\documentclass{IEEEtran}
\usepackage[latin9]{inputenc}
\usepackage{color}
\usepackage{units}
\usepackage{textcomp}
\usepackage{amsmath}
\usepackage{amsthm}
\usepackage{amssymb}
\usepackage{graphicx}
\usepackage{esint}
\makeatletter

\newcommand*\LyXThinSpace{\,\hspace{0pt}}

\theoremstyle{plain}
\newtheorem{thm}{\protect\theoremname}
\theoremstyle{definition}
\newtheorem{defn}[]{\protect\definitionname}
\theoremstyle{remark}
\newtheorem{rem}[]{\protect\remarkname}
\theoremstyle{plain}
\newtheorem{prop}[]{\protect\propositionname}
\theoremstyle{definition}
\newtheorem{example}[]{\protect\examplename}
\theoremstyle{plain}
\newtheorem{lem}[]{\protect\lemmaname}

\usepackage{bm}

\usepackage{subfigure}
\usepackage{epstopdf}
\usepackage{float}
\usepackage{subfloat}
\usepackage{array}
\usepackage{tabularx}
\usepackage{multirow}
\usepackage{tikz}
\usepackage{algorithmic}

\newtheorem{property}{Property}

\newcommand{\Ebb}{\mathbb{E}}

\newcommand{\Pbb}{\mathbb{P}}

\newcommand{\Ccal}{\mathcal{C}}

\newcommand{\Lcal}{\mathcal{L}}

\newcommand{\Scal}{\mathcal{S}}
\newcommand{\Tcal}{\mathcal{T}}

\newcommand{\Xcal}{\mathcal{X}}

\newcommand{\Zcal}{\mathcal{Z}}

\tikzstyle{arw}=[->,>=latex]
\tikzstyle{node}=[draw,rectangle,rounded corners, minimum width=1cm,minimum height =.75 cm]

\usepackage{color}%May be necessary if you want to color links

\usepackage{adjustbox}

\usepackage{algorithmic}
\catcode`~=11 \def\UrlSpecials{\do\~{\kern -.15em\lower .7ex\hbox{~}\kern .04em}} \catcode`~=13

\allowdisplaybreaks[1]

\newcommand{\nn}{\nonumber}

\newcommand{\calT}{\mathcal{T}}

\newcommand{\bB}{\mathbf{B}}

\newcommand{\bU}{\mathbf{U}}

\newcommand{\bV}{\mathbf{V}}

\newcommand{\bbE}{\mathbb{E}}

\newcommand{\bbN}{\mathbb{N}}

\newcommand{\eps}{\varepsilon}
\newcommand{\floor}[1]{\lfloor{#1}\rfloor}

\DeclareMathOperator*{\plimsup}{p-lim\, sup\,}
\DeclareMathOperator*{\pliminf}{p-lim\, inf\,}

\newcommand{\underI}{\underline{I}}
\newcommand{\overI}{\overline{I}}

\providecommand{\definitionname}{Definition}
\providecommand{\examplename}{Example}
\providecommand{\lemmaname}{Lemma}
\providecommand{\propositionname}{Proposition}
\providecommand{\theoremname}{Theorem}

\newcounter{mytempeqncnt}

\makeatother

\providecommand{\definitionname}{Definition}
\providecommand{\examplename}{Example}
\providecommand{\lemmaname}{Lemma}
\providecommand{\propositionname}{Proposition}
\providecommand{\remarkname}{Remark}
\providecommand{\theoremname}{Theorem}

\begin{document}

\title{Source-Channel Secrecy \\
for Shannon Cipher System}

\author{Lei Yu, Houqiang Li, \textit{Senior} \textit{Member, IEEE, }and Weiping
Li, \textit{Fellow, IEEE}
\thanks{Manuscript received August 22, 2016; revised December 23, 2016; accepted
February 1, 2017.  This work was supported by the 973 Program under
Grant 2013CB329004, and the NSFC under Grants 61325009, 61272316,
and 61631017. This paper was presented in part at the 2016 54th Annual
Allerton Conference on Communications, Control and Computing \cite{Yu-2}.}

\thanks{The authors are with the Department of Electronic Engineering and
Information Science, University of Science and Technology of China,
Hefei 230027, China (e-mail: yulei@ustc.edu.cn; lihq@ustc.edu.cn;
wpli@ustc.edu.cn).}
\thanks{Communicated by S. Watanabe, Associate Editor for
Shannon Theory.}
\thanks{Copyright (c) 2017 IEEE. Personal use of this material
is permitted. However, permission to use this material for any other
purposes must be obtained from the IEEE by sending a request to pubs-permissions@ieee.org.}}

\maketitle
\begin{abstract}
Recently, a secrecy measure based on list-reconstruction has been
proposed \cite{Schieler}, in which a wiretapper is allowed to produce
a list of $2^{mR_{L}}$ reconstruction sequences and the secrecy is
measured by the minimum distortion over the entire list. In this paper,
we show that this list secrecy problem is equivalent to the one with
secrecy measured by a new quantity \emph{lossy-equivocation}, which
is proven to be the minimum optimistic 1-achievable source coding
rate (the minimum coding rate needed to reconstruct the source within
target distortion with positive probability for\emph{ infinitely many
blocklengths}) of the source with the wiretapped signal as two-sided
information, and also can be seen as a lossy extension of conventional
equivocation. Upon this (or list) secrecy measure, we study source-channel
secrecy problem in the discrete memoryless Shannon cipher system with
\emph{noisy} wiretap channel. Two inner bounds and an outer bound
on the achievable region of secret key rate, list rate, wiretapper
distortion, and distortion of legitimate user are given. The inner
bounds are derived by using uncoded scheme and (operationally) separate
scheme, respectively. Thanks to the equivalence between lossy-equivocation
secrecy and list secrecy, information spectrum method is leveraged
to prove the outer bound. As special cases, the admissible region
for the case of degraded wiretap channel or lossless communication
for legitimate user has been characterized completely. For both these
two cases, separate scheme is proven to be optimal. Interestingly,
however, separation indeed suffers performance loss for other certain
cases. Besides, we also extend our results to characterize the achievable
region for Gaussian communication case. As a side product optimistic
lossy source coding has also been addressed.
\end{abstract}


\section{Introduction}

\label{sec:introduction}

Shannon cipher system (the one with noisy channel depicted in Fig.
\ref{fig:Shannon}) was first investigated by Shannon \cite{Shannon49},
in which a sender A communicates with a legitimate receiver B secretly
by exploiting a secret key that is shared by them. Shannon showed
that the perfect secrecy for this system is achievable if and only
if the rate of secret key is not smaller than the entropy of the source.
However in practice, the amount of key may be insufficient, and the
wiretapper might only want to reconstruct a lossy version of the source.
Recently, Schieler et al. \cite{Schieler} proposed a distortion-based
secrecy measure around the assumption that the wiretapper has ability
to conduct list decoding with fixed list size, and the induced distortion
is set to the minimum distortion over the entire list. They studied
it in the Shannon cipher system with noiseless channel, and characterized
the optimal tradeoff of secret key rate, list rate, wiretapper distortion,
and distortion of legitimate user. For this secrecy measure, the wiretapper
can be seen as a \textquotedblleft guessing wiretapper\textquotedblright{}
who produces a list of guesses to reconstruct the source within target
distortion (suppose some testing mechanism is available at the wiretapper).
From the perspective of \emph{computational security}, the list size
indicates the computational complexity that the wiretapper needs to
reconstruct the source within target distortion (different from \cite{Merhav}
the number of guesses for list secrecy is fixed during the guessing
process). On the other hand, from the perspective of \emph{information-theoretic
security}, the minimum list rate also indicates the minimum additional
information rate (received from an omniscient helper, named henchman)
for the wiretapper to reconstruct the source within target distortion
\cite{Schieler}. Besides, this secrecy measure is also compatible
with the conventional equivocation, and actually it can be characterized
by \emph{lossy-equivocation}, a lossy extension of the conventional
equivocation (this point will be shown in this paper). Furthermore,
this kind of measure could apply to  any secrecy system (not restricted
to lossy communication system or the secrecy system with testing mechanism
at the wiretapper), such as secure multimedia communication, communication
of personal verification information (password or bank account), and
communication of any other classified database that consists of sensitive
information. This is because on one hand, the results in \cite{Schieler}
still hold for lossless communication case; and on the other hand,
the interpretation from the perspective of information-theoretic security
is always valid for any secrecy system.

The secrecy measure in \cite{Schieler} is defined in sense of strong
secrecy, which requires that  there exists no sequence of codes achieving
target distortion for\emph{ all sufficiently large blocklengths.}
It implies the supremum of achievable list rate  equals the infimum
of all $R$ for which there exists a sequence of $R$-rate codes achieving
target distortion for\emph{ infinitely many blocklengths. }This kind
of codes is related to optimistic coding, which requires that good
codes exist for\emph{ infinitely many blocklengths} as opposed to
for \emph{all sufficiently large blocklengths}, required by the (pessimistic)
source coding or channel coding. \emph{ }Optimistic source coding
and optimistic channel coding are originated from the study of the
source-channel separation theorem and strong converse for general
sources and channels \cite{Vembu,Verdu}, and then systematically
investigated by Chen et al. as a separate problem \cite{Chen}. Recently,
Tan et al. \cite{Tan} applied the optimistic coding to the secure
communication over wiretap channels, and studied the optimistic capacity
of legitimate user under given secrecy constraint. However, in this
paper, it is used to study the secrecy performance, instead of the
communication performance of legitimate user.

Furthermore, some researchers investigated source-channel secrecy
problem. Yamamoto \cite{Yamamoto} studied it in Shannon cipher system
with secrecy measured by the minimum distortion that incurs in reconstructing
the source for a wiretapper. A different formulation of the problem
was considered in \cite{Wilson}, where the authors assumed there
is a fixed information leakage to the wiretapper and wish to minimize
the distortion at the legitimate receiver, while at the same time
providing a graceful distortion degradation when there is an SNR (Signal
Noise Ratio) mismatch. They showed that, for a positive leakage, this
can be achieved by combining vector quantization and scaling together.
This scenario was extended to consider side information at the receiver
in \cite{Bagherikaram} or side information at the sender in \cite{Bagherikaram13}.
Besides, joint source-channel secrecy with noncausal side information
available at both the legitimate user and the eavesdropper has been
studied in \cite{Villard}, and with causal source disclosure at the
eavesdropper has been studied in \cite{Song15}.

In this paper, the list secrecy measure proposed in \cite{Schieler}
is applied to the source-channel secrecy problem for Shannon cipher
system with a discrete memoryless source transmitted over a discrete
memoryless noisy wiretap channel (see Fig. \ref{fig:Shannon}), and
then we investigate the the achievable region of secret key rate,
list rate, wiretapper distortion, and distortion of legitimate user
for this system. The secrecy of the system is obtained by exploiting
both the secret key and  the wiretap channel. By information spectrum
analysis, we find that this problem (or an equivalent problem, henchman
problem) is equivalent to the one with secrecy measured by a new quantity,
\emph{lossy-equivocation}, which is defined as a conditional rate-distortion
function obtained by extending conventional equivocation (conditional
entropy) to the lossy case. From perspective of optimistic source
coding, the lossy-equivocation is proven to be the minimum optimistic
1-achievable source coding rate (the minimum coding rate to achieve
target distortion with positive probability for\emph{ infinitely many
blocklengths}) of the source with the wiretapped signal as two-sided
information. Thanks to the equivalence between lossy-equivocation
secrecy and list secrecy, it enables us to leverage the information
spectrum method to analyze these problems and obtain a converse result.
Note that this proof  method is different from the one used in \cite{Schieler}
which  mainly relies on the method of types. Besides, two achievability
schemes, uncoded scheme and (operationally) separate scheme, are analyzed.
When specialized to lossless communication case or degraded wiretap
channel case, the separate scheme is proven to be optimal. Hence for
these cases, the admissible regions are characterized completely.
 Interestingly, however, we observe that separate scheme indeed loses
the optimality for other certain cases. This implies separation is
not optimal in general for the source-channel secrecy problem.

Besides, we extend our result to characterize the achievable region
for the Gaussian communication case. Since the standard discretization
technique, usually used in proving the achievability for the continuous
source or continuous channel, is invalid in bounding the probability
of excess distortion, some other techniques including $\mathsf{d}-$tilted
information, weak typicality, and specified discretization, are exploited
in our proof.

In our work, optimistic source coding plays a key role in building
a bridge between the list secrecy problem and the lossy-equivocation
secrecy problem. Optimistic lossless source coding was investigated
by Chen et al. \cite{Chen}. As an extension to lossy case, optimistic
lossy source coding has been addressed in this paper as a side product.

The rest of this paper is organized as follows. Section II summarizes
basic notations, preliminaries, and formulation of the problem. Section
III and Section IV give the main results for the discrete memoryless
systems of lossless communication and lossy communication, respectively.
Section V extends the results to quadratic Gaussian communication
scenario. Finally, Section VI gives the concluding remarks.

\section{\label{sec:Problem-Formulation}Problem Formulation and Preliminaries}

\subsection{Notation and Preliminaries}

We use $P_{X}(x)$ to denote the probability distribution of random
variable $X$, which is also shortly denoted as $P_{X}$ or $P(x)$.
We also use $P_{X}$ and $Q_{X}$ to denote different probability
distribution with common alphabet $\mathcal{X}$.

The total variation distance between two probability measures $P$
and $Q$ with common alphabet is defined by
\begin{equation}
\lVert P-Q\rVert_{TV}\triangleq\sup_{A\in\mathcal{F}}|P(A)-Q(A)|,
\end{equation}
where $\mathcal{F}$ is the $\sigma$-algebra of the probability space.
The following properties of total variation distance hold.

\begin{property}\cite{Schieler} \label{tvproperties} Total variation
distance satisfies:
\begin{enumerate}
\item If the support of $P$ and $Q$ is a countable set $\Xcal$, then
\begin{equation}
\lVert P-Q\rVert_{TV}=\frac{1}{2}\sum_{x\in\Xcal}|P(\{x\})-Q(\{x\})|.
\end{equation}
\item Let $\eps>0$ and let $f(x)$ be a function with bounded range of
width $b>0$. Then
\begin{equation}
\lVert P-Q\rVert_{TV}<\eps\:\Longrightarrow\:\big|\Ebb_{P}f(X)-\Ebb_{Q}f(X)\big|<\eps b,\label{eq:tvcontinuous}
\end{equation}
where $\Ebb_{P}$ indicates that the expectation is taken with respect
to the distribution $P$. %\item Total variation satisfies the triangle inequality. For any $R \in \Delta_{\Xcal}$, %\begin{equation}%\lVert P - Q \rVert \leq \lVert P - R \rVert + \lVert R - Q \rVert.%\end{equation}
\item Let $P_{X}P_{Y|X}$ and $Q_{X}P_{Y|X}$ be two joint distributions
with common channel $P_{Y|X}$. Then
\begin{equation}
\lVert P_{X}P_{Y|X}-Q_{X}P_{Y|X}\rVert_{TV}=\lVert P_{X}-Q_{X}\rVert_{TV}.\label{eq:PP-QP}
\end{equation}
\end{enumerate}
\end{property}

Information spectrum analysis \cite{Han} will be used frequently
in this paper. For a general sequence of random variables $\bB=\{B_{n}\}_{n\in\bbN}$\footnote{Throughout this paper, we use the boldface to denote a sequence of
random variables, e.g., $\bB=\{B_{n}\}_{n\in\bbN}$. Note that the
alphabets of these random variables are not restricted to be Cartesian
product, and the information spectrum quantities defined in this subsection
will be applied for sequences of random variables of different length
in the latter sections.}, define
\begin{equation}
\pliminf_{n\to\infty}\!B_{n}\!\triangleq\!\sup\Big\{ r:\lim_{n\to\infty}\mathbb{P}(B_{n}<r)\!=\!0\Big\}\label{eqn:epliminf}
\end{equation}
and
\begin{equation}
\plimsup_{n\to\infty}\!B_{n}\!\triangleq\!\inf\Big\{ r:\lim_{n\to\infty}\mathbb{P}(B_{n}>r)\!=\!0\Big\}.\label{eqn:eplimsup}
\end{equation}
For any general pair of random variables $(\bU,\bV)$ with joint distribution
$P_{\bU\bV}\triangleq\{P_{U^{n}V^{n}}\}_{n\in\bbN}$, define, for
each $n$, the normalized information density random variables\footnote{For convenience, all $\log$\textquoteright s and $\exp$\textquoteright s
in this paper are with respect to base 2.}
\begin{equation}
\imath_{n}(U^{n};V^{n})\triangleq\frac{1}{n}\log\frac{P_{V^{n}|U^{n}}(V^{n}|U^{n})}{P_{V^{n}}(V^{n})}.\label{eq:infdens}
\end{equation}
Given $\{\imath_{n}(U^{n};V^{n})\}_{n\in\bbN}$, we may now define
\begin{align}
\underI(\bU;\bV) & \triangleq\pliminf_{n\to\infty}\imath_{n}(U^{n};V^{n})\label{eqn:underIeps}\\
\overI(\bU;\bV) & \triangleq\plimsup_{n\to\infty}\imath_{n}(U^{n};V^{n}).\label{eqn:overIeps}
\end{align}
In information spectrum analysis, $\underI(\bU;\bV)$ and $\overI(\bU;\bV)$
are termed the {\em spectral inf- and sup-mutual information rates}
respectively. They are respectively the $\pliminf$ and $\plimsup$
of the sequence of random variables $\{\imath_{n}(U^{n};V^{n})\}_{n\in\bbN}$.
As special cases, {\em spectral inf- and sup-entropy rates} are
defined as
\begin{align}
\underline{H}(\bU) & \triangleq\pliminf_{n\to\infty}\imath_{n}(U^{n};U^{n})\label{eqn:underIeps-1}\\
\overline{H}(\bU) & \triangleq\plimsup_{n\to\infty}\imath_{n}(U^{n};U^{n}).\label{eqn:overIeps-1}
\end{align}
The conditional versions of these quantities can be defined similarly,
which will be used in the latter sections as well. Furthermore, for
any sequence of distortion functions $d_{n}(u^{n},v^{n})$, define
\begin{align}
\underline{D}(\bU,\bV) & \triangleq\pliminf_{n\to\infty}d_{n}(U^{n},V^{n})\label{eqn:underIeps-1-1}\\
\overline{D}(\bU,\bV) & \triangleq\plimsup_{n\to\infty}d_{n}(U^{n},V^{n}).\label{eqn:overIeps-1-1}
\end{align}

\subsection{List Secrecy}

\begin{figure}
\centering\includegraphics[width=0.4\textwidth]{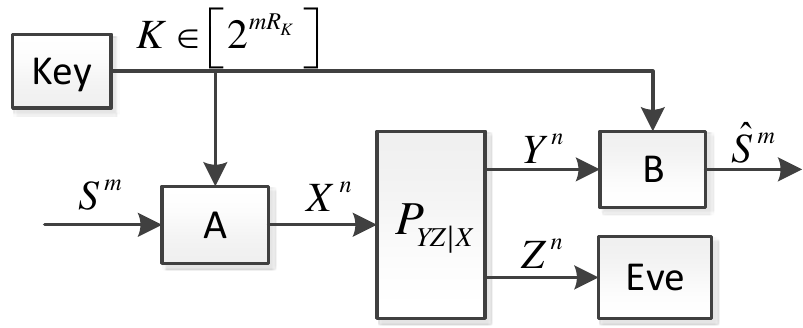} \protect\caption{\label{fig:Shannon} Shannon cipher system with noisy wiretap channel.}
\end{figure}

Consider Shannon cipher system with noisy wiretap channel shown in
Fig. \ref{fig:Shannon}, where a sender A and a legitimate receiver
B share a secret key $K$ that is uniformly distributed over $\left[2^{mR_{K}}\right]$\footnote{In this paper, the set $\left\{ 1,...,l\right\} $ is sometimes denoted
by $[l]$.} and independent of a source $S^{m}$. The sender A observes the discrete
memoryless (DM) source sequence $S^{m}$ with each element i.i.d.
(independent and identically distributed) according to $P_{S}$, and
then transmits it to the legitimate user B over a DM wiretap channel
$P_{YZ|X}$ confidentially by utilizing the secret key and the wiretap
channel. Finally, the legitimate user B produces a source reconstruction
$\hat{S}^{m}$ using the received sequence $Y^{n}$ and the secret
key $K$.

\begin{defn}
An $\left(m,n,R_{K}\right)$ block code consists of\footnote{Similar to \cite{Wyner}, it can be shown that one cannot benefit
from replacing the decoder with a stochastic one. }\\
 1) Stochastic encoder: $P_{X^{n}|S^{m}K}:\mathcal{S}^{m}\times\left[2^{mR_{K}}\right]\mapsto\mathcal{X}^{n}$;\\
 2) Decoder: $\psi:\mathcal{Y}^{n}\times\left[2^{mR_{K}}\right]\mapsto\mathcal{\hat{S}}^{m}$.
\end{defn}
Assume the source sample rate is $B_{s}$ and the channel sample rate
is $B_{c}$. Define the bandwidth mismatch factor as $\gamma\triangleq\frac{B_{c}}{B_{s}}.$
Then for any $\left(m,n,R_{K}\right)$ block code that can be implemented
in the system with bandwidth mismatch factor $\gamma$, it must hold
that $\frac{n}{m}\leq\gamma.$ Note that any $\left(m,n',R_{K}\right)$
code with $\frac{n'}{m}\leq\gamma$ can be seen as a special case
of $\left(m,n,R_{K}\right)$ codes with $n=\floor{m\gamma}$\footnote{Here $\floor{x}$ denotes the maximum integer not larger than $x$}
in which for each block, only $n'$ channel symbols are used. Hence
for the system with bandwidth mismatch factor $\gamma$, we only need
consider $\left(m,n,R_{K}\right)$ codes with $n=\floor{m\gamma}$.

Another output $Z^{n}$ of the channel is accessed by a wiretapper
Eve. Based on $Z^{n}$, the wiretapper produces a list $\Lcal(Z^{n})\subseteq\check{\mathcal{S}}^{m}$
and the induced distortion is set to the minimum one over the entire
list, i.e., $\min_{\check{s}^{m}\in\Lcal(Z^{n})}d_{E}(S^{m},\check{s}^{m}),$
where $d_{E}\left(s^{m},\check{s}^{m}\right)\triangleq\frac{1}{m}\sum_{i=1}^{m}d_{E}\left(s_{i},\check{s}_{i}\right)$
is a distortion measure for the wiretapper. For given distortion levels
$D_{B}$ and $D_{E}$, by exploiting the secret key and the wiretap
channel, Nodes A and B want to communicate the source within distortion
$D_{B}$, while ensuring that the wiretapper's strategy always suffers
distortion above $D_{E}$ with high probability.
\begin{defn}
\label{def:list}The tuple $\left(R_{K},R_{L},D_{B},D_{E}\right)$
is achievable if there exists a sequence of $(m,n,R_{K})$ codes with
$n=\floor{m\gamma}$ such that $\forall\delta>0$,

\noindent 1) Distortion constraint:
\begin{align}
\lim_{m\rightarrow\infty}\Pbb\Big[d_{B}(S^{m},\hat{S}^{m})\le D_{B}+\delta\Big] & =1,\label{eq:legitimatedistortion}
\end{align}
where $d_{B}\left(s^{m},\hat{s}^{m}\right)\triangleq\frac{1}{m}\sum_{i=1}^{m}d_{B}\left(s_{i},\hat{s}_{i}\right)$
is a distortion measure for the legitimate user B;

\noindent 2) Secrecy constraint:
\begin{equation}
\lim_{m\rightarrow\infty}\Pbb\Big[d_{E}(S^{m},\check{S}^{m})\geq D_{E}-\delta\Big]=1\label{eq:henchman}
\end{equation}
for any sequence of lists $\left\{ \Lcal_{m}\right\} $ such that
$\limsup_{m\rightarrow\infty}\frac{1}{m}\log|\Lcal_{m}|\leq R_{L}$,
where $\check{S}^{m}=\arg\min_{\check{s}^{m}\in\Lcal_{m}(Z^{n})}d_{E}(S^{m},\check{s}^{m})$.
\end{defn}
It is easy to verify that in Definition \ref{def:list}, the constraints
1) and 2) can be respectively replaced with $\overline{D_{B}}(\mathbf{S},\mathbf{\hat{S}})\leq D_{B}$,
and $\underline{D_{E}}(\mathbf{S},\mathbf{\check{S}})\geq D_{E}$
for any sequence of lists such that $\limsup_{m\rightarrow\infty}\frac{1}{m}\log|\Lcal_{m}|\leq R_{L}$.
\begin{defn}
\label{def:The-admissible-region}The admissible region $\mathcal{R}\triangleq\mathrm{Closure}\left\{ \textrm{Achievable }\left(R_{K},R_{L},D_{B},D_{E}\right)\right\} $\footnote{The admissible region $\mathcal{R}$ does not change if we replace
the constraint $\limsup_{m\rightarrow\infty}\frac{1}{m}\log|\Lcal_{m}|\leq R_{L}$
of Definition \ref{def:list} with $\frac{1}{m}\log|\Lcal_{m}|\leq R_{L}$
for all $m$. This is because it only affects the achievability of
the points on the boundary of $\mathcal{R}$, and however, $\mathcal{R}$
is defined as a closed set (all the boundary points are incorporated
into it). To keep consistent with the standard formulation of rate
constraint in \cite{Han}, here we write $\limsup_{m\rightarrow\infty}\frac{1}{m}\log|\Lcal_{m}|\leq R_{L}$,
instead of the other one.}.
\end{defn}
We assume all the alphabets of the source and its reconstructions
(at the legitimate user or wiretapper), as well as the alphabets of
channel input and outputs, are finite.

\subsection{Henchman Problem}

\begin{figure}
\centering\includegraphics[width=0.4\textwidth]{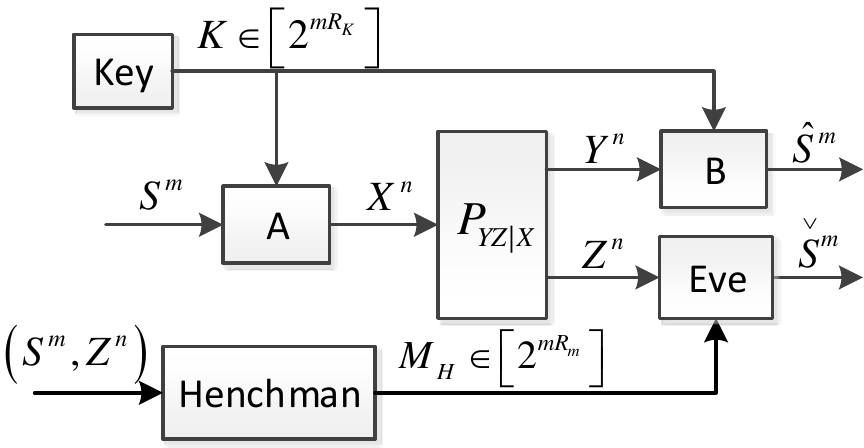} \protect\protect\protect\caption{\label{fig:henchman}Henchman problem, where $\limsup_{m\rightarrow\infty}R_{m}\leq R_{L}$.}
\end{figure}

The problem above is equivalent to a henchman problem \cite{Schieler},
in which a wiretapper reconstructs a single sequence with the help
of a rate-limited henchman who can access to both the source $S^{m}$
and the wiretapped signal $Z^{n}$. As depicted in Fig. \ref{fig:henchman},
the wiretapper receives the best possible $mR_{m}$ bits from the
henchman to assist in producing a reconstruction sequence $\check{S}^{m}$.

\begin{defn}
The $R_{m}$-rate henchman code of an $\left(m,n,R_{K}\right)$ block
code consists of \\
 1) Encoder: $\varphi_{H}:\mathcal{S}^{m}\times\mathcal{Z}^{n}\mapsto\left[2^{mR_{m}}\right]$;\\
 2) Decoder: $\psi_{H}:\left[2^{mR_{m}}\right]\times\mathcal{Z}^{n}\mapsto\mathcal{\check{S}}^{m}$.
\end{defn}
We assume that the wiretapper and henchman are aware of the $\left(m,n,R_{K}\right)$
block code adopted by Nodes A and B, and they cooperate to design
a henchman code based on the $\left(m,n,R_{K}\right)$ block code.
\begin{defn}
\label{def:henchman} The tuple $\left(R_{K},R_{L},D_{B},D_{E}\right)$
is achievable in the henchman problem if there exists a sequence of
$(m,n,R_{K})$ codes with $n=\floor{m\gamma}$ such that

\noindent 1) Distortion constraint: $\overline{D_{B}}(\mathbf{S},\mathbf{\hat{S}})\leq D_{B}$;

\noindent 2) Secrecy constraint: $\underline{D_{E}}(\mathbf{S},\mathbf{\check{S}})\geq D_{E}$
for any sequence of henchman codes such that $\limsup_{m\rightarrow\infty}R_{m}\leq R_{L}$,
where $R_{m}$'s are the rates of henchman codes.
\end{defn}

\subsection{\label{sub:Lossy-Equivocation-Secrecy}Lossy-Equivocation Secrecy}

Besides, the list secrecy problem and the henchman problem also can
be formulated as a secure communication problem with secrecy measured
by lossy-equivocation.
\begin{defn}
\label{def:LE} For a general source $\mathbf{S}$, general two-sided
information $\mathbf{Z}$, and a general distortion measure $d_{E}\left(s^{m},\check{s}^{m}\right)$,
the (spectral inf-) lossy-equivocation (or rate-distortion based spectral
inf-equivocation) $\underline{R_{\mathbf{S}|\mathbf{Z}}}(D_{E})$
is defined as
\begin{equation}
\underline{R_{\mathbf{S}|\mathbf{Z}}}(D_{E})\triangleq\inf_{P_{\mathbf{\check{S}}|\mathbf{S}\mathbf{Z}}:\underline{D_{E}}(\mathbf{S},\mathbf{\check{S}})\leq D_{E}}\overline{I}(\mathbf{S};\mathbf{\check{S}}|\mathbf{Z})\label{eq:ratechannel-2-2}
\end{equation}
for $D_{E}\geq D_{E,\min}\triangleq\inf_{\mathbf{g}\left(\mathbf{s}\right)}\underline{D_{E}}(\mathbf{S},\mathbf{g}\left(\mathbf{S}\right))$,
where $\mathbf{g}\left(\mathbf{s}\right)$ denotes a sequence of functions
$\left\{ g^{m}\left(s^{m}\right)\right\} _{m\in\bbN}$; and $\underline{R_{\mathbf{S}|\mathbf{Z}}}(D_{E})\triangleq+\infty$
for $D_{E}<D_{E,\min}$.
\end{defn}
In the lossy-equivocation secrecy problem, a sender and a legitimate
user jointly design an $\left(m,n,R_{K}\right)$ block code to maximize
the lossy-equivocation.
\begin{defn}
\label{def:lossyequiv} The tuple $\left(R_{K},R_{L},D_{B},D_{E}\right)$
is achievable if there exists a sequence of $(m,n,R_{K})$ codes with
$n=\floor{m\gamma}$ such that

\noindent 1) Distortion constraint: $\overline{D_{B}}(\mathbf{S},\mathbf{\hat{S}})\leq D_{B}$;

\noindent 2) Secrecy constraint: $R_{L}\leq\underline{R_{\mathbf{S}|\mathbf{Z}}}(D_{E}).$
\end{defn}
Besides, by setting $d_{B}(s^{m},\hat{s}^{m})=1\left\{ s^{m}\neq\hat{s}^{m}\right\} $
and $D_{B}=0$, the achievable tuples and admissible region for lossless
communication  can be defined similarly\footnote{Furthermore, by setting $d_{E}(s^{m},\check{s}^{m})=1\left\{ s^{m}\neq\check{s}^{m}\right\} $
and $D_{E}=0$, the admissible region for losslessly reconstructing
the source at the wiretapper can be defined as well. By checking our
proof, it can be verified that the admissible region for this case
is the same as that obtained by specializing our result to the case
of $d_{E}(s^{m},\check{s}^{m})=\frac{1}{m}\sum_{i=1}^{m}1\left\{ s_{i}\neq\check{s}_{i}\right\} $
and $D_{E}=0$. }.
\begin{defn}
The tuple $\left(R_{K},R_{L},D_{E}\right)$ is achievable for lossless
communication if there exists a sequence of $(m,n,R_{K})$ codes with
$n=\floor{m\gamma}$ such that $\lim_{m\rightarrow\infty}\Pbb(S^{m}\neq\hat{S}^{m})=0,$
and the secrecy constraint in Definition \ref{def:list}, \ref{def:henchman}
or \ref{def:lossyequiv} holds. The admissible region $\mathcal{R}\triangleq\mathrm{Closure}\left\{ \textrm{Achievable} \left(R_{K},R_{L},D_{E}\right)\right\} $
for this case.
\end{defn}
Observe that the secrecy constraint in Definition \ref{def:henchman}
 requires that  $D_{E}$ cannot be achieved (with any positive probability)
by any sequence of $R_{L}$-rate henchman codes for \emph{all sufficiently
large blocklengths. }By contrapositive, the supremum of all $R_{L}$
satisfying the secrecy constraint equals the infimum of all $R$ for
which there exists a sequence of $R$-rate codes achieving $D_{E}$
(with some positive probability) for\emph{ infinitely many blocklengths.
}This is related to optimistic source coding, which requires that
good codes exist for\emph{ infinitely many blocklengths} as opposed
to for \emph{all sufficiently large blocklengths}, required by the
(pessimistic) source coding. Hence the supremum of $R_{L}$ also equals
the infimum of all $R$ for which there exists a sequence of $R$-rate
optimistic source codes achieving $D_{E}$.  This enables us to
apply optimistic source coding to analyzing the henchman problem.

The information spectrum characterization of optimistic lossy source
coding is given in the following theorem, the proof of which is presented
in Appendix \ref{sec:Proof-of-Theorem}. To state our results concisely,
for a general sequence of random variables $\mathbf{U},\mathbf{V}$
and a sequence of distortion measures $d_{m}(u^{m},v^{m})$, we define
\begin{equation}
\underline{D^{(\varepsilon)}}(\mathbf{U},\mathbf{V})\triangleq\!\inf\left\{ d:\liminf_{m\rightarrow\infty}\Pbb\Big[d_{m}(U^{m},V^{m})\geq d\Big]<\varepsilon\right\} .\label{eqn:eplimsup-1}
\end{equation}
Note that $\underline{D^{(\varepsilon)}}(\mathbf{U},\mathbf{V})$
is consistent with $\underline{D}(\mathbf{U},\mathbf{V})$ when $\varepsilon=1$,
i.e., $\underline{D^{(1)}}(\mathbf{U},\mathbf{V})=\underline{D}(\mathbf{U},\mathbf{V})$.
\begin{thm}[Optimistic $\varepsilon$-Achievable Source Coding]
\label{thm:optimistic} Given a general source $\mathbf{S}$ and
general two-sided information $\mathbf{Z}$, and a general distortion
measure $d_{E}\left(s^{m},\check{s}^{m}\right)$, there exists a sequence
of $R_{m}$-rate fixed-length source codes $\left(f_{m},g_{m}\right)_{m=1}^{\infty}$
such that
\begin{align}
\limsup_{m\rightarrow\infty}R_{m} & \leq R,\label{eq:-8}\\
\underline{D_{E}^{(\varepsilon)}}(\mathbf{S},\mathbf{\check{S}}) & \leq D_{E},\label{eq:-84-1}
\end{align}
(these codes are named optimistic $\varepsilon$-achievable source
codes; see \cite{Chen}) with $\check{S}^{m}\triangleq g_{m}(f_{m}(S^{m},Z^{n}),Z^{n})$
denoting the reconstruction, if and only if
\begin{equation}
R\geq\underline{R_{\mathbf{S}|\mathbf{Z}}^{(\varepsilon)}}(D_{E})\triangleq\inf_{P_{\mathbf{\widetilde{S}}|\mathbf{S}\mathbf{Z}}:\underline{D_{E}^{(\varepsilon)}}(\mathbf{S},\mathbf{\widetilde{S}})\leq D_{E}}\overline{I}(\mathbf{S};\mathbf{\mathbf{\widetilde{S}}}|\mathbf{Z}).\label{eq:optimisticrate}
\end{equation}
\end{thm}
\begin{rem}
\label{rem:When-,-.}When $\varepsilon=1$, $\underline{D_{E}^{(1)}}(\mathbf{S},\mathbf{\check{S}})=\underline{D_{E}}(\mathbf{S},\mathbf{\check{S}})$.
Hence $\underline{R_{\mathbf{S}|\mathbf{Z}}}(D_{E})=\underline{R_{\mathbf{S}|\mathbf{Z}}^{(1)}}(D_{E})$,
i.e., $\underline{R_{\mathbf{S}|\mathbf{Z}}}(D_{E})$ is the minimum
optimistic 1-achievable source coding rate (the minimum coding rate
such that $\underline{D_{E}}(\mathbf{S},\mathbf{\check{S}})\leq D_{E}$).
\end{rem}
Chen et al. have shown that $\underline{H}(\mathbf{S}|\mathbf{Z})$
is the minimum optimistic 1-achievable source coding rate for lossless
case (minimum source coding rate for the source $\mathbf{S}$ and
two-sided information $\mathbf{Z}$ such that $\liminf_{m\rightarrow\infty}\mathbb{P}(S^{m}\neq\check{S}^{m})<1$)
\cite{Chen}. By Remark \ref{rem:When-,-.}, and the operational
definitions of optimistic 1-achievability for both lossless and lossy
cases, we have the following proposition. Besides, to understand
this proposition more intuitively, a direct proof is also given in
Appendix \ref{sec:Proof-of-Theorem-lossyequ}.
\begin{prop}
\label{prop:lossyequivocation}For a general source $\mathbf{S}$,
general two-sided information $\mathbf{Z}$, and a general distortion
measure $d_{E}\left(s^{m},\check{s}^{m}\right)$, $\underline{R_{\mathbf{S}|\mathbf{Z}}}(D_{E})\leq\underline{H}(\mathbf{S}|\mathbf{Z})$
for $D_{E}\geq D_{E,\min}^{\prime}\triangleq\inf_{\mathbf{g}\left(\mathbf{s}\right)}\overline{D_{E}}(\mathbf{S},\mathbf{g}\left(\mathbf{S}\right))$.
Moreover, $\underline{R_{\mathbf{S}|\mathbf{Z}}}(D_{E})=\underline{H}(\mathbf{S}|\mathbf{Z})$
for $d_{E}(s^{m},\check{s}^{m})=1\left\{ s^{m}\neq\check{s}^{m}\right\} $
and $D_{E}=0$.
\end{prop}
\begin{rem}
\label{rem:LE}For a DM source $\mathbf{S}$ with finite alphabet
and an additive distortion measure $d_{E}\left(s^{m},\check{s}^{m}\right)=\frac{1}{m}\sum_{i=1}^{m}d_{E}\left(s_{i},\check{s}_{i}\right)$,
we have $D_{E,\min}=D_{E,\min}^{\prime}=\inf_{\check{s}\left(s\right)}\mathbb{E}d_{E}(S,\check{s}\left(S\right))$,
where $D_{E,\min}$ is given in Definition \ref{def:LE}. Hence for
this case, $\underline{R_{\mathbf{S}|\mathbf{Z}}}(D_{E})\leq\underline{H}(\mathbf{S}|\mathbf{Z})$
for any $D_{E}\geq D_{E,\min}$.
\end{rem}
It seems somewhat counterintuitive that as the optimistic 1-achievable
source coding rate, the lossy equivocation $\underline{R_{\mathbf{S}|\mathbf{Z}}}(D_{E})$
(and also  $\underline{R_{\mathbf{S}|\mathbf{Z}}^{(\varepsilon)}}(D_{E})$),
is defined via sup-conditional mutual information rate, but in \cite{Chen},
the lossless optimistic 1-achievable rate $\underline{H}(\mathbf{S}|\mathbf{Z})$
is characterized by inf-entropy rate (instead of sup-entropy rate).
For the case of $d_{E}(s^{m},\check{s}^{m})=1\left\{ s^{m}\neq\check{s}^{m}\right\} $
and $D_{E}=0$, $\underline{D_{E}}(\mathbf{S},\mathbf{\check{S}})\leq D_{E}$
is equivalent to $\liminf_{m\rightarrow\infty}\mathbb{P}(S^{m}\neq\check{S}^{m})<1$.
Hence for this case, Proposition \ref{prop:lossyequivocation} implies
$\underline{R_{\mathbf{S}|\mathbf{Z}}}(D_{E})=\underline{H}(\mathbf{S}|\mathbf{Z})=\inf_{P_{\mathbf{\check{S}}|\mathbf{S}\mathbf{Z}}:\liminf_{m\rightarrow\infty}\mathbb{P}(S^{m}\neq\check{S}^{m})<1}\overline{I}(\mathbf{S};\mathbf{\check{S}}|\mathbf{Z})$.
That is, $\underline{H}(\mathbf{S}|\mathbf{Z})$ can be characterized
by sup-conditional mutual information rate as well.

Due to Proposition \ref{prop:lossyequivocation}, the quantity $\underline{H}(\mathbf{S}|\mathbf{Z})$
can be also named as \emph{spectral inf-equivocation} (or \emph{conditional
spectral inf-entropy}). This term is closely related to the conventional
equivocation $\liminf_{m\to\infty}H(S^{m}|Z^{n})$. Both of them denote
conditional entropies, but the former is defined in sense of limit
inferior in probability, and the latter is defined in sense of expectation.
Moreover, $\underline{H}(\mathbf{S}|\mathbf{Z})\leq\liminf_{m\to\infty}H(S^{m}|Z^{n})$
holds in general.

Comparing the secrecy constraint of Definition \ref{def:henchman}
with \eqref{eq:-8} and \eqref{eq:-84-1} tells us that given source
$S^{m}$ and two-sided information $Z^{n}$, the minimum optimistic
1-achievable source coding rate equals the supremum of all $R_{L}$
satisfying the secrecy constraint. Hence Theorem \ref{thm:optimistic}
combined with Remark \ref{rem:When-,-.} implies the equivalence between
the lossy-equivocation secrecy problem and the henchman problem.On
the other hand, the equivalence between the list secrecy problem and
the henchman problem has been proven by Schieler and Cuff \cite[Prop. 1]{Schieler}.
Hence these three problems are equivalent to each other, as stated
in the following proposition.
\begin{prop}
\label{prop:The-tuple-} The tuple $\left(R_{K},R_{L}+\delta_{1},D_{B},D_{E}+\delta_{1}\right)$
for some $\delta_{1}>0$ is achievable for lossy communication (or
$\left(R_{K},R_{L}+\delta_{1},D_{E}+\delta_{1}\right)$ for lossless
communication) in the list secrecy problem if and only if $\left(R_{K},R_{L}+\delta_{2},D_{B},D_{E}+\delta_{2}\right)$
for some $\delta_{2}>0$ is achievable in the henchman problem, and
also if and only if $\left(R_{K},R_{L}+\delta_{3},D_{B},D_{E}+\delta_{3}\right)$
for some $\delta_{3}>0$ is achievable in the lossy-equivocation secrecy
problem. In other words, the admissible region $\mathcal{R}$ remains
the same for these three problems.
\end{prop}
\begin{IEEEproof}
Schieler and Cuff \cite[Prop. 1]{Schieler} have shown that for any
tuple $\left(R_{K},R_{L},D_{B},D_{E}\right)$, the achievability is
equivalent for the list secrecy problem and the henchman problem.
Hence we only need to show that $\left(R_{K},R_{L}+\delta_{2},D_{B},D_{E}+\delta_{2}\right)$
for some $\delta_{2}>0$ is achievable in the henchman problem, if
and only if $\left(R_{K},R_{L}+\delta_{3},D_{B},D_{E}+\delta_{3}\right)$
for some $\delta_{3}>0$ is achievable in the lossy-equivocation secrecy
problem.

Suppose that the tuple $\left(R_{K},R_{L}+\delta_{2},D_{B},D_{E}+\delta_{2}\right)$
satisfies the secrecy constraint: $\underline{D_{E}}(\mathbf{S},\mathbf{\check{S}})\geq D_{E}+\delta_{2}$
for any sequence of henchman codes such that $\limsup_{m\rightarrow\infty}R_{m}\leq R_{L}+\delta_{2}$.
It immediately gives us $\underline{D_{E}}(\mathbf{S},\mathbf{\check{S}})>D_{E}+\frac{\delta_{2}}{2}$
for any sequence of codes such that $\limsup_{m\rightarrow\infty}R_{m}\leq R_{L}+\delta_{2}$.
By Theorem \ref{thm:optimistic} together with Remark \ref{rem:When-,-.},
we have $R_{L}+\delta_{2}<\underline{R_{\mathbf{S}|\mathbf{Z}}}(D_{E}+\frac{\delta_{2}}{2})$.
Hence $\left(R_{K},R_{L}+\delta_{3},D_{B},D_{E}+\delta_{3}\right)$
with $\delta_{3}=\frac{\delta_{2}}{2}>0$ is achievable in the lossy-equivocation
secrecy problem.

Suppose that the tuple $\left(R_{K},R_{L}+\delta_{3},D_{B},D_{E}+\delta_{3}\right)$
satisfies the secrecy constraint: $R_{L}+\delta_{3}\leq\underline{R_{\mathbf{S}|\mathbf{Z}}}(D_{E}+\delta_{3}).$
It immediately gives us $R_{L}+\frac{\delta_{3}}{2}<\underline{R_{\mathbf{S}|\mathbf{Z}}}(D_{E}+\delta_{3}).$
Again, by Theorem \ref{thm:optimistic} together with Remark \ref{rem:When-,-.},
we have $\underline{D_{E}}(\mathbf{S},\mathbf{\check{S}})\geq D_{E}+\delta_{3}$
for any sequence of codes such that $\limsup_{m\rightarrow\infty}R_{m}\leq R_{L}+\frac{\delta_{3}}{2}$.
Hence $\left(R_{K},R_{L}+\delta_{2},D_{B},D_{E}+\delta_{2}\right)$
with $\delta_{2}=\frac{\delta_{3}}{2}>0$ is achievable in the henchman
problem.

Therefore, the achievability of $\left(R_{K},R_{L}+\delta_{2},D_{B},D_{E}+\delta_{2}\right)$
for  the henchman problem is equivalent to the achievability of $\left(R_{K},R_{L}+\delta_{3},D_{B},D_{E}+\delta_{3}\right)$
for  the lossy-equivocation secrecy problem.

So far, we have shown that the achievability of the interior points
of $\mathcal{R}$ is equivalent for these three problems. On the other
hand, $\mathcal{R}$ is defined as a closed set. So the admissible
region $\mathcal{R}$ remains the same for these three problems.
\end{IEEEproof}
From this proposition and Definition \ref{def:lossyequiv}, the admissible
region for these problems can be characterized using information spectrum
quantities as
\begin{equation}
\mathcal{R}=\mathrm{Closure}\underset{P_{\mathbf{X}|\mathbf{S}\mathbf{K}},P_{\mathbf{\hat{S}}|\mathbf{Y}\mathbf{K}}}{\bigcup}\left\{ \begin{array}{l}
\left(R_{K},R_{L},D_{B},D_{E}\right):\\
\overline{D_{B}}(\mathbf{S},\mathbf{\hat{S}})\leq D_{B},\\
R_{L}\leq\underline{R_{\mathbf{S}|\mathbf{Z}}}(D_{E})
\end{array}\right\} .\label{eq:-120}
\end{equation}

In addition to the DM system, we also consider the Shannon cipher
system of communicating Gaussian source over power-constrained Gaussian
wiretap channel. For this case, the channel input cost constraint
\begin{equation}
\lim_{n\rightarrow\infty}\mathbb{P}\bigl[\rho\left(X^{n}\right)\le P+\delta\bigr]=1,\forall\delta>0,\label{eq:powerconstraint}
\end{equation}
should be added to Definitions \ref{def:list}, \ref{def:henchman}
and \ref{def:lossyequiv}, where $\rho\left(x^{n}\right)\triangleq\frac{1}{n}\sum_{i=1}^{n}\rho\left(x_{i}\right)$
is cost function of the wiretap channel, and for power-constrained
Gaussian wiretap channel $\rho\left(x\right)=x^{2}$. For the system
involving channel cost constraint, Proposition \ref{prop:The-tuple-}
still holds.

%\begin{equation} \label{9)}%\mathop{\lim }\limits_{n\to \infty } E\left\{\mathop{\min }\limits_{i\in \left[2^{nlR_{L} } \right]} \tilde{d}\left(S^{nl} ,g_{L,i} \left(M^{l} \right)\right)\right\}\le D_{0} -\varepsilon%\end{equation}

\section{Lossless Communication }

When lossless communication is required for the legitimate user, the
admissible region $\mathcal{R}$ is given by the following theorem.
\begin{thm}[Lossless Communication]
\label{thm:lossless} For the lossless DM Shannon cipher system,
we have
\begin{equation}
\mathcal{R}=\left\{ \begin{array}{l}
\left(R_{K},R_{L},D_{E}\right):\gamma C_{B}\geq H\left(S\right),\\
R_{L}\leq\min\left\{ R_{K}+\gamma\Gamma_{1}(\frac{1}{\gamma}H(S)),R_{S}(D_{E})\right\}
\end{array}\right\} ,\label{eq:-45}
\end{equation}
where
\begin{equation}
C_{B}=\max_{P_{X}}I\left(X;Y\right)\label{eq:capacity}
\end{equation}
denotes the capacity of the channel $P_{Y|X}$,
\begin{align}
\Gamma_{1}\left(R\right) & \triangleq\max_{\substack{P_{X}P_{V|X}P_{U|V}:\\
I\left(V;Y\right)\geq R,\\
I(U;Y)\leq I(U;Z)
}
}\left[I\left(V;Y|U\right)-I\left(V;Z|U\right)\right]\label{eq:gamma1}
\end{align}
denotes the equivocation-capacity function of the wiretap channel
which was first defined by Csiszár et al. \cite{Csiszar} and
\begin{align}
R_{S}\left(D\right) & =\mathop{\min}\limits _{P_{\check{S}|S}:\mathbb{E}d_{E}(S,\check{S})\le D}I(S;\check{S})\label{eq:rd}
\end{align}
denotes the rate-distortion function of $S$.
\end{thm}
\begin{rem}
More precisely, the equivocation-capacity function of the wiretap
channel \cite[Cor. 2]{Csiszar} was defined as
\begin{align}
\Gamma_{1}^{\prime}\left(R\right) & =\min\bigl\{ R,\Gamma_{1}\left(R\right)\bigr\}.\label{eq:gamma1-4}
\end{align}
However, for our problem, it does not affect the admissible region
$\mathcal{R}$ by replacing $\Gamma_{1}\left(R\right)$ with $\Gamma_{1}^{\prime}\left(R\right)$,
since if $\Gamma_{1}\left(R\right)>R$ then $\min\left\{ R_{K}+\gamma\Gamma_{1}^{\prime}(\frac{1}{\gamma}H(S)),R_{S}(D_{E})\right\} =\min\left\{ R_{K}+\gamma\Gamma_{1}(\frac{1}{\gamma}H(S)),R_{S}(D_{E})\right\} =R_{S}(D_{E})$.
\end{rem}
\begin{rem}
\label{rem:If-the-secrecy}If the secrecy measure is replaced with
the conventional equivocation $\liminf_{m\to\infty}H(S^{m}|Z^{n})$,
then the admissible region becomes the $\mathcal{R}$ in \eqref{eq:-45}
with the term $R_{S}(D_{E})$ replaced by $H\left(S\right)$. The
achievability part of this claim follows from the fact $\underline{R_{\mathbf{S}|\mathbf{Z}}}(D_{E})\leq\underline{H}(\mathbf{S}|\mathbf{Z})\leq\liminf_{m\to\infty}H(S^{m}|Z^{n})$
and Theorem~\ref{thm:lossless} with the setting  $d_{E}(s^{m},\check{s}^{m})=\frac{1}{m}\sum_{i=1}^{m}1\left\{ s_{i}\neq\check{s}_{i}\right\} $
and $D_{E}=0$; the converse part of this claim can be shown by following
similar steps to the proof of the converse part of Theorem~\ref{thm:lossless}.
Note that this claim with $R_{K}=0$ is a restatement of the result
of \cite{Csiszar} with common message rate $R_{0}=0$, or the result
of \cite{Wyner}.
\end{rem}
\begin{IEEEproof}
The proof of converse part is given in Appendix \ref{sec:Proof-of-Theorem-lossless}.
As for the achievability part, lossless communication can be roughly
considered as a special case of lossy communication, hence the achievability
part  can be obtained by following similar steps to the proof of
the achievability part of Theorem \ref{thm:lossy}, which is stated
in Section \ref{sec:lossy}.
\end{IEEEproof}
Note that the result for the case with no henchman (or only a single
reconstruction allowed) is obtained by setting $R_{L}=0$ in the region
of Theorem~\ref{thm:lossless}. For this case, any positive rate
of secret key or any positive $\Gamma_{1}(\frac{1}{\gamma}H(S))$
results in the maximum expected distortion that can occur. This observation
coincides with that for noiseless wiretap channel case \cite{Schieler}.

The first constraint of $\mathcal{R}$ is consistent with the source-channel
separation theorem, and the second constraint of $\mathcal{R}$, roughly
speaking, follows from the following argument. On one hand, the henchman
and the wiretapper can always ignore the signal $Z^{n}$ and use a
point-to-point code to achieve $R_{S}(D_{E})$. On the other hand,
the optimal strategy of the sender and legitimate user is an operationally
separate coding scheme, in which the source is first compressed by
an optimal source code with rate $H\left(S\right)$, then a part ($R_{K}$
rate) of the resulting bitstream is encrypted by one-time pad using
the secret key, finally all the bitstream is transmitted losslessly
to the legitimate user using an optimal secrecy-channel code \cite{Csiszar}.
The optimal secrecy-channel code consists of two layers $U^{n}$ and
$V^{n}$, and the secrecy is obtained only from the second layer,
i.e., $V^{n}$. For such optimal strategy, upon $Z^{n}$ the wiretapper
is able to reconstruct the first layer $U^{n}$ directly, and further
reconstruct $V^{n}$ upon both $Z^{n}$ and $U^{n}$ by using rate
$\gamma\left(I\left(V;Y|U\right)-I\left(V;Z|U\right)\right)$. Then
the wiretapper uses $R_{K}$ rate to decrypt the secret key, and finally
reconstructs the source losslessly by using the secret key and the
messages $\left(U^{n},V^{n}\right)$ as the legitimate user does.

Furthermore, if the legitimate user's channel is \emph{less noisy
}than the wiretapper's (i.e., for every $V\rightarrow X\rightarrow YZ$,
$I\left(V;Y\right)\geq I\left(V;Z\right)$ holds; see \cite{Gamal}),
then Csiszár et al. \cite[Thm. 3]{Csiszar} showed
\begin{align}
\Gamma_{1}\left(R\right) & =\max_{P_{X}:I\left(X;Y\right)\geq R}\left[I\left(X;Y\right)-I\left(X;Z\right)\right].\label{eq:gamma1-3}
\end{align}

\section{Lossy Communication \label{sec:lossy}}

Now, we consider lossy communication case. Define
\[
\mathcal{R}_{\textrm{unc}}^{(i)}=\underset{P_{X^{n}|S^{m}K}}{\bigcup}\left\{ \begin{array}{l}
\left(R_{K},R_{L},D_{B},D_{E}\right):\\
D_{B}\geq\min_{\hat{s}^{m}\left(y^{n},k\right)}\mathbb{E}d_{B}\left(S^{m},\hat{S}^{m}\right),\\
R_{L}\leq\frac{1}{m}R_{S^{m}|Z^{n}}(D_{E})
\end{array}\right\} ,
\]
\[
\mathcal{R}_{\textrm{sep}}^{(i)}=\underset{P_{\hat{S}|S}}{\bigcup}\left\{ \begin{array}{l}
\left(R_{K},R_{L},D_{B},D_{E}\right):\gamma C_{B}\geq I(S;\hat{S}),\\
D_{B}\geq\mathbb{E}d_{B}(S,\hat{S}),\\
R_{L}\leq\min\Bigl\{ R_{K}+\gamma\Gamma_{1}\bigl(\frac{1}{\gamma}I(S;\hat{S})\bigr)\\
\qquad+R_{S|\hat{S}}(D_{E}),R_{S}(D_{E})\Bigr\}
\end{array}\right\} ,
\]
and
\[
\mathcal{R}^{(o)}=\underset{P_{\hat{S}|S}}{\bigcup}\left\{ \begin{array}{l}
\left(R_{K},R_{L},D_{B},D_{E}\right):\gamma C_{B}\geq I(S;\hat{S}),\\
D_{B}\geq\mathbb{E}d_{B}(S,\hat{S}),\\
R_{L}\leq\min\Bigl\{ R_{K}+\gamma\Gamma_{2}\bigl(\frac{1}{\gamma}I(S;\hat{S})\bigr)\\
\qquad+R_{S|\hat{S}}(D_{E}),R_{S}(D_{E})\Bigr\}
\end{array}\right\} ,
\]
where $C_{B}$, $\Gamma_{1}\left(\cdot\right)$ and $R_{S}\left(\cdot\right)$
are respectively defined by \eqref{eq:capacity}, \eqref{eq:gamma1}
and \eqref{eq:rd},
\begin{align}
\Gamma_{2}\left(R\right) & \triangleq\min_{\substack{Q_{YZ|X}:Q_{Y|X}=P_{Y|X},\\
Q_{Z|X}=P_{Z|X}
}
}\max_{Q_{X}:I_{Q}\left(X;Y\right)\geq R}I_{Q}\left(X;Y|Z\right)
\end{align}
with $I_{Q}\left(\cdot\right)$ denoting the mutual information under
distribution $Q_{X}Q_{YZ|X}$, is another function specified by the
wiretap channel, and
\begin{align}
R_{S|\hat{S}}\left(D\right) & =\mathop{\min}\limits _{P_{\check{S}|S\hat{S}}:\mathbb{E}d_{E}(S,\check{S})\le D}I(S;\check{S}|\hat{S})\label{eq:rd-si}
\end{align}
denotes the rate-distortion function of $S$ with the two-sided information
$\hat{S}$. Then we have the following theorem for lossy communication
case.
\begin{thm}[Lossy Communication]
\label{thm:lossy} For the lossy DM Shannon cipher system, we have
\begin{equation}
\mathcal{R}_{\textrm{unc}}^{(i)}\cup\mathcal{R}_{\textrm{sep}}^{(i)}\subseteq\mathcal{R}\subseteq\mathcal{R}^{(o)}.
\end{equation}
\end{thm}
\begin{rem}
As shown in the following, $\mathcal{R}_{\textrm{unc}}^{(i)}$ is
achieved by an uncoded scheme. In this scheme, the secret key is used
in a symbol-by-symbol way. However, applying the key to a sufficiently
large blocklength will result in a better secrecy performance. Based
on this idea, an improved version of this scheme is proposed in our
another paper \cite{Yu}, which cascades a random permutation (according
to the secret key) with a symbol-by-symbol mapping. We refer the interested
reader to \cite{Yu} for the details.
\end{rem}
\begin{rem}
\label{rem:Similar-to-Remark}Similar to Remark \eqref{rem:If-the-secrecy},
if the secrecy measure is replaced with the conventional equivocation
$\liminf_{m\to\infty}H(S^{m}|Z^{n})$, then by replacing the terms
$\frac{1}{m}R_{S^{m}|Z^{n}}(D_{E})$, $R_{S|\hat{S}}(D_{E})$ and
$R_{S}(D_{E})$ with $\frac{1}{m}H(S^{m}|Z^{n})$, $H(S|\hat{S})$
and $H\left(S\right)$, Theorem~\ref{thm:lossy} still holds. The
proof of this claim follows similar argument to that of Remark \eqref{rem:If-the-secrecy}.
Note that this claim with $R_{K}=0$ is a lossy extension of the result
of \cite{Csiszar} with common message rate $R_{0}=0$ and the result
of \cite{Wyner}.
\end{rem}
\begin{IEEEproof}
The proofs of the outer bound and the inner bound $\mathcal{R}_{\textrm{sep}}^{(i)}$
are given in Appendices \ref{sec:Outer-Bound} and \ref{sec:Inner-Bound},
respectively. Now we give a proof of the inner bound $\mathcal{R}_{\textrm{unc}}^{(i)}$.
For simplicity, we only consider the case of $m=n=1$. For given $P_{X|SK}$,
suppose $\hat{s}\left(y,k\right)$ achieves $\min_{\hat{s}\left(y,k\right)}\mathbb{E}d_{B}(S,\hat{S})$.
Then consider the following uncoded scheme.

{\textbf{Encoder}}: Upon $\left(S,K\right)$, the sender produces
$X$ stochastically according to probability distribution $P_{X|SK}$.

{\textbf{Decoder}}: The legitimate user produces $\hat{S}=\hat{s}\left(Y,K\right)$.

Observe that $(S_{i},K_{i},X_{i},Y_{i},Z_{i},\hat{S}_{i})_{i=1}^{\infty}$
are i.i.d. Then by law of large numbers,
\begin{align}
\Pbb\Big[d_{B}(S^{l},\hat{S}^{l})\le\mathbb{E}d_{B}(S^{l},\hat{S}^{l})+\epsilon\Big] & \xrightarrow{l\to\infty}1,\label{eq:legitimatedistortion-1}
\end{align}
for any $\epsilon>0$. Hence the distortion constraint for legitimate
user is satisfied.

Next we prove the secrecy constraint is also satisfied if $R_{L}<R_{S|Z}(D_{E})$.
That is, if $R_{L}<R_{S|Z}(D_{E})$ then $\lim_{l\rightarrow\infty}\mathbb{P}\bigl(d_{E}(S^{l},\check{S}^{l})<D_{E}-\epsilon\bigr)=0$
for any $\epsilon>0$. This is equivalent to the strong converse for
lossy source coding with two-sided information.

Define $\mathcal{A}\triangleq\left\{ \left(S^{l},Z^{l}\right)\in\mathcal{T}_{\delta}^{l}\right\} .$
From the property of typicality, we have $\mathbb{P}\left(\mathcal{A}\right)\rightarrow1$
as $l\to\infty$. Denote the codebook of henchman code as $\mathcal{C}$.
Given $Z^{l}$, $\mathcal{C}$ has $2^{lR_{l}}$ elements at most.
Denote them as $\check{s}^{l}\left(i,z^{l}\right),i\in[2^{lR_{l}}]$.
Consider
\begin{align}
 & \mathbb{P}\left(d_{E}(S^{l},\check{S}^{l})<D_{E}-\epsilon\right)\nonumber \\
\leq & \mathbb{P}\left(d_{E}(S^{l},\check{S}^{l})<D_{E}-\epsilon,\mathcal{A}\right)+\mathbb{P}\left(\mathcal{A}^{c}\right)\\
\leq & \mathbb{P}\left(d_{E}(S^{l},\check{S}^{l})<D_{E}-\epsilon,\mathcal{A}\right)+\epsilon_{l}\\
= & \sum_{z^{l}}P_{Z^{n},\mathcal{C}}\left(z^{l},c\right)\mathbb{P}\left(d_{E}(S^{l},\check{S}^{l})<D_{E}-\epsilon,\mathcal{A}|z^{l},c\right)+\epsilon_{l},\label{eq:-86}
\end{align}
where $\epsilon_{l}$ is a term that vanishes as $l\to\infty$. Utilizing
union bound, we have
\begin{align}
 & \mathbb{P}\left(d_{E}(S^{l},\check{S}^{l})<D_{E}-\epsilon,\mathcal{A}|z^{l},c\right)\nonumber \\
\leq & \sum_{i\in[2^{lR_{l}}]}\mathbb{P}\left(d_{E}(S^{m},\check{s}^{m}\left(i,z^{l}\right))<D_{E}-\epsilon,\mathcal{A}|z^{l},c\right).\label{eq:-85}
\end{align}
On the other hand, Lemma \ref{lem:typebound-rd-si} (given in Appendix
\ref{sec:Inner-Bound}) implies $\mathbb{P}\left(d_{E}(S^{m},\check{s}^{m}\left(i,z^{l}\right))<D_{E}-\epsilon,\mathcal{A}|z^{l},c\right)\leq2^{-l(R_{S|Z}(D_{E}-\epsilon)-o(1))}\leq2^{-l(R_{S|Z}(D_{E})-o(1))}$.
Then combining it with \eqref{eq:-86} and \eqref{eq:-85}, we have
\begin{align}
 & \mathbb{P}\left(d_{E}(S^{l},\check{S}^{l})<D_{E}-\epsilon\right)\nonumber \\
\leq & 2^{-l(R_{S|Z}(D_{E})-R_{l}-o(1))}+\epsilon_{l}.
\end{align}
If $R_{L}<R_{S|Z}(D_{E})$, i.e., $\limsup_{l\rightarrow\infty}R_{l}<R_{S|Z}(D_{E})$,
then
\begin{align}
 & \mathbb{P}\left(d_{E}(S^{l},\check{S}^{l})<D_{E}-\epsilon\right)\xrightarrow{l\to\infty}0,
\end{align}
for any $\epsilon>0$. Hence $\left(R_{K},R_{L},D_{B},D_{E}\right)$
is achievable.
\end{IEEEproof}
Note that $\mathcal{R}_{\textrm{sep}}^{(i)}$ is a generalization
of the achievability part of Theorem \ref{thm:lossless} to the lossy
case, and obtained by an operationally separate scheme as well. $\mathcal{R}_{\textrm{unc}}^{(i)}$
is achieved by the uncoded scheme above in which both the encoder
and decoder are symbol-by-symbol mappings. Different from the lossless
case, for the lossy communication the source may be transmitted using
uncoded scheme or other lossy joint source-channel secrecy code, and
hence there may be no message (digital information) transmitted over
the channel. For this case the wiretapper cannot decrypt the source
through decrypting the digital information. This leads to the difficulty
for proving the outer bound part. Here we leverage information spectrum
method to derive the outer bound. Instead of reconstructing the source
directly, an indirect decryption strategy is considered in our proof,
which can be roughly considered as follows: the wiretapper first reconstructs
$\hat{S}^{n}$ using rate $\gamma\Gamma_{2}\bigl(\frac{1}{\gamma}I(S;\hat{S})\bigr)$,
next decrypts the secret key using rate $R_{K}$, then upon $Y^{n}$
and secret key, produces the legitimate user's reconstruction $\hat{S}^{n}$,
and finally upon $\hat{S}^{n}$ produces a final reconstruction $\check{S}^{n}$
using rate $R_{S|\hat{S}}(D_{E})$. This leads to the outer bound
$\mathcal{R}^{(o)}$.

Note that $\mathcal{R}^{(o)}$ and $\mathcal{R}_{\textrm{sep}}^{(i)}$
differ only in the gamma functions $\Gamma_{1}\left(\cdot\right)$
and $\Gamma_{2}\left(\cdot\right)$. Obviously, $\Gamma_{1}\left(\cdot\right)$
and $\Gamma_{2}\left(\cdot\right)$ both only depend on the margin
distributions of the wiretap channel, and $\Gamma_{1}\left(R\right)\leq\Gamma_{2}\left(R\right)$,
or equivalently, $\mathcal{R}_{\textrm{sep}}^{(i)}\subseteq\mathcal{R}^{(o)}$.
Moreover for (stochastically) degraded wiretap channel, it is easy
to verify that $\Gamma_{1}\left(R\right)=\Gamma_{2}\left(R\right)$.
Hence for this case, $\mathcal{R}_{\textrm{sep}}^{(i)}$ and $\mathcal{R}^{(o)}$
coincide.
\begin{thm}[DM System with Degraded Wiretap Channel]
\label{thm:Cor1-1} For lossy DM Shannon cipher system with a degraded
wiretap channel ($X\rightarrow Y\rightarrow Z$ or $X\rightarrow Z\rightarrow Y$),
we have
\begin{equation}
\mathcal{R}=\mathcal{R}_{\textrm{sep}}^{(i)}=\mathcal{R}^{(o)}.\label{eq:Thm1-2-2-1}
\end{equation}
\end{thm}
Theorem \ref{thm:Cor1-1} implies the separate coding is optimal for
the degraded channel case. However, this is not true in general. In
fact, as illustrated by the following example, there exists some source-channel
pair such that uncoded scheme strictly outperforms separate scheme.
This interesting observation is very different from the case with
no secrecy constraint, since without secrecy constraint separation
is proven to be optimal for source-channel coding problem \cite{Gamal}.
\begin{example}
(Uncoded scheme may strictly outperform separate scheme). Assume there
is no secret key, i.e., $R_{K}=0$. Consider the case of transmitting
a Bernoulli source $S\sim\textrm{Bern}(\frac{1}{2})$ over a bandwidth-matched
($\gamma=1$) wiretap channel $P_{YZ|X}$ with $X\in\left\{ 0,1\right\} ,$
$Y\in\left\{ 0,1,e\right\} $, and $Z\in\left\{ 0,1\right\} $, where
the channel from $X$ to $Y$ is a binary erasure channel $\textrm{BEC}(\epsilon),\epsilon\in\left(0,1\right)$,
and the channel from $X$ to $Z$ is a binary symmetric channel $\textrm{BSC}(p),p\in\left(0,\frac{1}{2}\right)$.
Assume $2p<\epsilon\leq4p(1\text{\textminus}p)$. Then $Y$ is less
noisy than $Z$, but $Z$ is not a degraded version of $Y$ (see \cite[Example 5.4]{Gamal}).
Assume both the legitimate user's reconstruction and wiretapper's
reconstruction satisfy $\hat{S},\check{S}\in\left\{ 0,1,e\right\} $,
and the distortion measures are set to the erasure distortion functions:
\[
d_{B}\left(s,\hat{s}\right)=d_{E}\left(s,\hat{s}\right)=\begin{cases}
0, & \textrm{if }\left(s,\hat{s}\right)=\left(0,0\right)\textrm{ or }\left(1,1\right);\\
1, & \textrm{if }\left(s,\hat{s}\right)=\left(0,e\right)\textrm{ or }\left(1,e\right);\\
\infty, & \textrm{if }\left(s,\hat{s}\right)=\left(0,1\right)\textrm{ or }\left(1,0\right).
\end{cases}
\]
Observe that $C_{B}=R_{S}(\epsilon)$, hence $D_{B}\geq\epsilon$.
Consider the case of $D_{B}=D_{E}=\epsilon$. Then to achieve capacity,
$X\sim\textrm{Bern}(\frac{1}{2})$, and to achieve rate-distortion
function, $P_{\hat{S}|S}$ forms a $\textrm{BEC}(\epsilon)$\footnote{Both the distributions $P_{X}$ and $P_{\hat{S}|S}$ are unique; see
\cite[Thm. 2.7.4]{Cover}.}. Hence $P_{S\hat{S}}=P_{XY}$, and $R_{S|\hat{S}}(\epsilon)=0$.
For less noisy wiretap channel, $\Gamma_{1}\left(R\right)$ is given
by \eqref{eq:gamma1-3}. Hence the maximum achievable $R_{L}$ by
separate scheme is $R_{L,\textrm{sep}}=I\left(X;Y\right)-I\left(X;Z\right)=1-\epsilon-\left(1-H_{2}(p)\right)=H_{2}(p)-\epsilon$,
where $H_{2}$ denotes the binary entropy function, i.e., $H_{2}(p)=-p\log p-(1-p)\log(1-p).$
On the other hand, for uncoded scheme, the legitimate user's distortion
$D_{B}=\epsilon$ can be achieved by setting $X=S$. Then the maximum
achievable $R_{L}$ by uncoded scheme is $R_{L,\textrm{unc}}=R_{S|Z}(\epsilon)=H_{2}(p)-\epsilon H_{2}(\frac{p}{\epsilon})$.
From the assumption $\frac{p}{\epsilon}<\frac{1}{2}$, we have $H_{2}(\frac{p}{\epsilon})<1$.
It implies $R_{L,\textrm{unc}}>R_{L,\textrm{sep}}$, i.e., uncoded
scheme provides stronger secrecy. Moreover, for this case,
\begin{align}
\Gamma_{2}\left(R\right) & =\min_{\substack{Q_{XYZ}:Q_{X}=P_{X},Q_{Y|X}=P_{Y|X},\\
Q_{Z|X}=P_{Z|X}
}
}I_{Q}\left(X;Y|Z\right)\\
 & =\min_{\substack{P_{U|XZ}:P_{U|X}=P_{Y|X}}
}I\left(X;U|Z\right)\\
 & =H_{2}(p)-\epsilon H_{2}(\frac{p}{\epsilon}).
\end{align}
Substitute this into $\mathcal{R}^{(o)}$, then we have the the upper
bound $R_{L}^{(o)}$ of the achievable $R_{L}$ satisfies $R_{L}^{(o)}=R_{L,\textrm{unc}}>R_{L,\textrm{sep}}$.
This implies that for the specified setting above, uncoded scheme
is optimal, and meanwhile, separate coding is strictly suboptimal.
\end{example}
For the above example, if set $d_{E}\left(s,\hat{s}\right)$ to the
Hamming distortion, $D_{E}=0$, and remain other settings unchanged,
then $R_{L}^{(o)}=R_{L,\textrm{unc}}>R_{L,\textrm{sep}}$ still holds.
Therefore, from Remark \ref{rem:Similar-to-Remark}, separate scheme
is not optimal in general not only for the list secrecy problem, but
also for the conventional equivocation secrecy problem. In \cite{Wyner,Csiszar},
Wyner, Csiszár and Körner have not found this interesting point, since
on one hand, only the case of lossless communication for legitimate
user was considered by them, and on the other hand, as shown by Theorem
\ref{thm:lossless}, for lossless communication case separate coding
is optimal.
\begin{prop}
Separate coding is neither optimal in general for the list secrecy
(or lossy-equivocation secrecy) problem, nor optimal in general for
the conventional equivocation secrecy problem.
\end{prop}
Besides, when specialized to the communication over noiseless wiretap
channel (with channel capacity $C_{B}$), i.e., $Y=Z=X$ and $H(X)=C_{B}$,
the problem turns into the one considered by Schieler and Cuff \cite{Schieler}.
For this case, $\Gamma_{1}\left(R\right)=\Gamma_{2}\left(R\right)=0$
for any $R\geq0$. Hence Theorem \ref{thm:lossy} recovers the admissible
region given in \cite[Thm.3]{Schieler}, i.e.,
\[
\mathcal{R}=\underset{P_{\hat{S}|S}}{\bigcup}\left\{ \begin{array}{l}
\left(R_{K},R_{L},D_{B},D_{E}\right):\gamma C_{B}\geq I(S;\hat{S}),\\
D_{B}\geq\mathbb{E}d_{B}(S,\hat{S}),\\
R_{L}\leq\min\left\{ R_{K}+R_{S|\hat{S}}(D_{E}),R_{S}(D_{E})\right\}
\end{array}\right\} .
\]

\begin{figure}
\centering \includegraphics[width=0.5\textwidth]{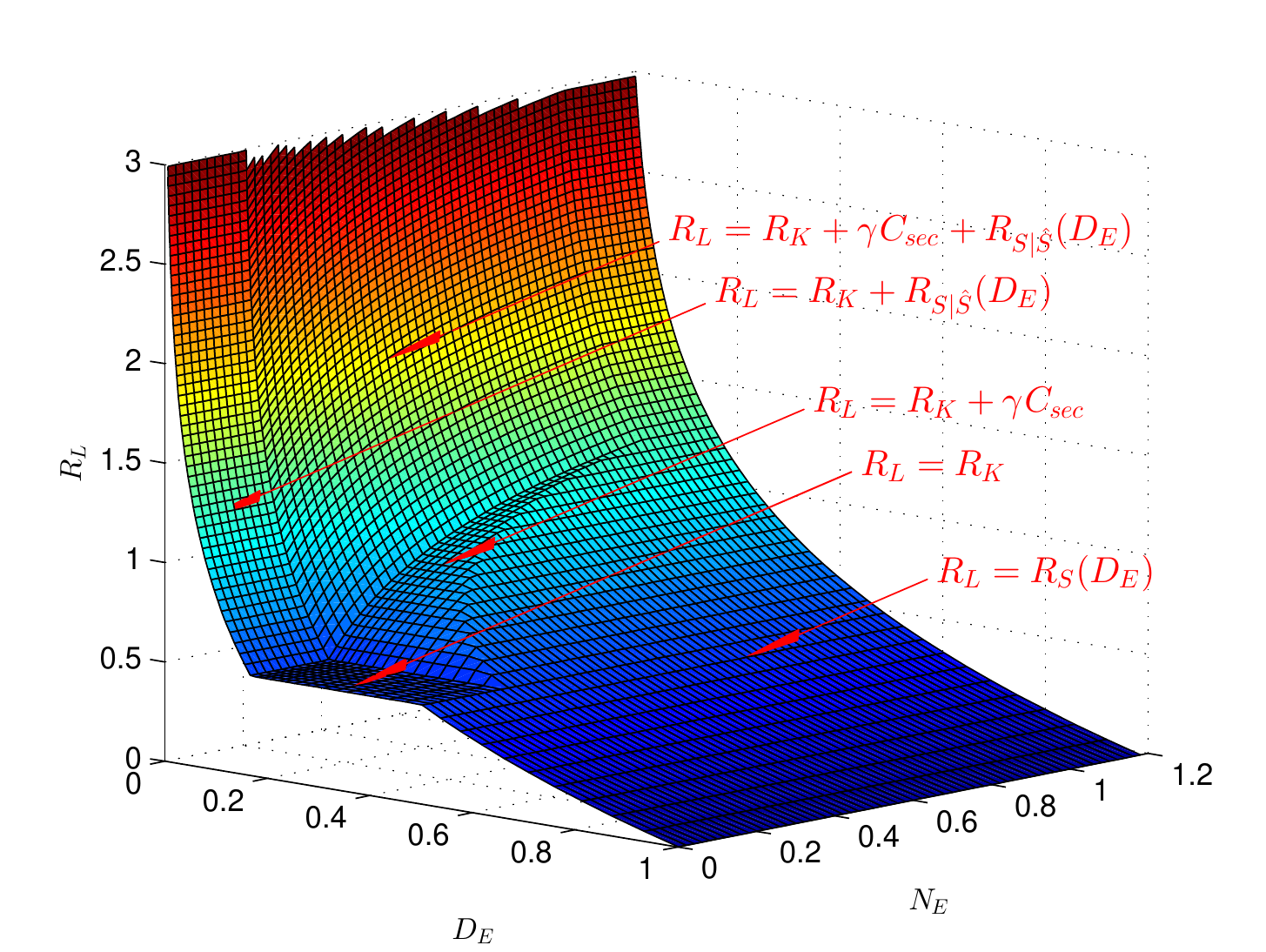}

\protect\protect\protect\caption{\label{fig:Gaussian}The region in Theorem \ref{thm:gaussian} for
$\gamma=1,R_{K}=0.5,N_{S}=1,P=1,N_{B}=0.2$ and $D_{B}=\frac{N_{S}N_{B}}{P+N_{B}}$.}
\end{figure}

\section{Gaussian Communication }

The results given in previous section can be extended to Gaussian
communication scenario. Consider the case of communicating a Gaussian
source $S\sim\mathcal{N}\left(0,N_{S}\right)$ over a Gaussian wiretap
channel, $Y=X+W_{B},Z=Y+W_{E}^{\prime}$, where $W_{B}\sim\mathcal{N}\left(0,N_{B}\right)$
and $W_{E}^{\prime}\sim\mathcal{N}\left(0,N_{E}-N_{B}\right)$ are
independent, and transmitting power is constrained as $\lim_{n\rightarrow\infty}\Pbb\bigl[\frac{1}{n}\sum_{i=1}^{n}X_{i}^{2}\le P+\delta\bigr]=1,\forall\delta>0$.
Then we have the following theorem. The proofs of the converse part
and the achievability part are given in Appendices \ref{sec:Proof-of-Theorem-gaussian}
and \ref{sec:Proof-of-Theorem-gaussian-1}, respectively.
\begin{thm}[Gaussian Communication]
\label{thm:gaussian} For the Gaussian Shannon cipher system, the
admissible region is
\[
\mathcal{R}=\left\{ \begin{array}{l}
\left(R_{K},R_{L},D_{B},D_{E}\right):D_{B}\geq\frac{N_{S}}{\left(1+\nicefrac{P}{N_{B}}\right)^{\gamma}},\\
R_{L}\leq\min\biggl\{ R_{K}+\frac{\gamma}{2}\log^{+}\left(\frac{1+\nicefrac{P}{N_{B}}}{1+\nicefrac{P}{N_{E}}}\right)\\
\qquad+\frac{1}{2}\log^{+}\frac{D_{B}}{D_{E}},\frac{1}{2}\log^{+}\frac{N_{S}}{D_{E}}\biggr\}
\end{array}\right\} ,
\]
where $\log^{+}x\triangleq\max\left\{ \log x,0\right\} $.
\end{thm}
The region in Theorem \ref{thm:gaussian} is illustrated in Fig. \ref{fig:Gaussian}.
If the wiretapper's channel satisfies $N_{E}\leq N_{B}$, then $R_{L}\leq\min\biggl\{ R_{K}+\frac{1}{2}\log^{+}\frac{D_{B}}{D_{E}},\frac{1}{2}\log^{+}\frac{N_{S}}{D_{E}}\biggr\}$,
i.e., no secrecy can be obtained from exploiting the wiretap channel.
On the contrary, if the wiretapper's channel satisfies $N_{E}>N_{B}$,
then $R_{L}\leq\min\biggl\{ R_{K}+\frac{\gamma}{2}\log^{+}\left(\frac{1+\nicefrac{P}{N_{B}}}{1+\nicefrac{P}{N_{E}}}\right)+\frac{1}{2}\log^{+}\frac{D_{B}}{D_{E}},\frac{1}{2}\log^{+}\frac{N_{S}}{D_{E}}\biggr\}$,
i.e., the secrecy of the system is obtained from exploiting the secret
key and the wiretap channel at the same time. Moreover, if the wiretapper's
channel is bad enough such that $N_{E}\geq\frac{P}{2^{2R_{K}}-1}$,
then $R_{L}\leq\frac{1}{2}\log^{+}\frac{N_{S}}{D_{E}}$, i.e., the
received signal $Z^{n}$ does not help the wiretapper to produce a
better reconstruction.

The achievability part of Theorem \ref{thm:gaussian} is proven by
using a separate coding scheme. Apart from separate coding, two uncoded
schemes, permutation based scheme and orthogonal-transform based scheme,
have been proposed in \cite{Yu} for secure Gaussian \emph{broadcast}
communication (two or more legitimate users are involved in the system)
with matched bandwidth (i.e., $\gamma=1$). In these two uncoded schemes,
the source is transmitted after random permutation or random orthogonal
transform (according to the secret key) and (symbol-by-symbol) scaling
operation. As shown in \cite{Yu}, they achieve a same region.  Compared
with the case of no secrecy constraint, they do not incur any  performance
loss in aspect of legitimate users' distortions; and meanwhile,
achieve the optimal secrecy performance under some certain regimes.
However,  the separate coding proposed in this paper cannot achieve
the optimal distortion performance for broadcast scenarios. Hence
for certain cases, the uncoded schemes in \cite{Yu} will outperform
 the separate coding of this paper for the bandwidth-matched Gaussian
broadcast communication.

When specialized to  \emph{point-to-point} communication (the system
considered in this paper), the performance of such uncoded schemes
is characterized by the following theorem.
\begin{thm}[Uncoded Schemes]
\cite{Yu} \label{thm:Thm1} The uncoded schemes above could achieve
the region
\begin{align*}
\mathcal{R}_{\textrm{unc}}^{(i)} & =\left\{ \begin{array}{l}
\left(R_{K},R_{L},D_{B},D_{E}\right):0\leq P'\leq P,\\
D_{B}\geq\frac{N_{S}N_{B}}{P'+N_{B}},\\
R_{L}\leq\min\biggl\{ R_{K}+\frac{1}{2}\log^{+}\left(\frac{N_{S}N_{E}}{D_{E}\left(P'+N_{E}\right)}\right),\\
\qquad\frac{1}{2}\log^{+}\frac{N_{S}}{D_{E}}\biggr\}
\end{array}\right\} .
\end{align*}
\end{thm}
Obviously, $\mathcal{R}_{\textrm{unc}}^{(i)}\subseteq\mathcal{R}$.
Moreover, it can be verified that under condition that $N_{E}\leq N_{B},D_{B}\leq D_{E}$
or $N_{E}\geq N_{B},D_{B}=\frac{\lambda N_{B}}{P+N_{B}}\geq D_{E}$,
i.e., the wiretapper has a better channel and wants to produce a
worse reconstruction, or the legitimate user's distortion is restricted
to be the Shannon limit and meanwhile the wiretapper has a worse channel
 and wants to produce a better reconstruction, the uncoded schemes
above are optimal as well.

\section{Concluding Remarks}

\label{sec:conclusion}

In this paper, we investigated the source-channel secrecy problem
with list secrecy measure for Shannon cipher system. By associating
it with optimistic source coding, we established the equivalence between
this problem and the one with secrecy measured by a new quantity,
lossy-equivocation. Utilizing information spectrum method to bound
the lossy-equivocation, we obtained several converse results for the
systems with various classes of source-channel pairs. Some special
cases including lossless communication, communication over degraded
wiretap channel, and Gaussian communication, were completely resolved.
For these cases, separate coding was proven to be optimal. But this
does not hold in general.

The equivalence between the henchman problem and the lossy-equivocation
secrecy problem implies that the quantity, lossy-equivocation, indicates
the minimum additional information rate needed for the wiretapper
to reconstruct the source within a target distortion. On the other
hand, the lossy-equivocation can be considered as a lossy extension
of the conventional equivocation. Therefore, an explicit meaning
of the conventional equivocation is obtained: roughly speaking, the
conventional equivocation indicates the minimum additional rate needed
for the wiretapper to reconstruct the source losslessly. From this
perspective, the conventional equivocation, however, is not suitable
for lossy communication systems, especially for the communication
of continuous sources, since for these systems the wiretapper usually
only want to reconstruct a lossy version of the source. Consequently,
measuring secrecy by the lossy-equivocation is more reasonable for
these systems. Besides, as a generation of the conventional equivocation,
the lossy-equivocation also applies to lossless communication systems.
For these reasons, the lossy-equivocation secrecy is expected to have
widespread applications  in general secure network communications.

\appendices{}

\section{\label{sec:Proof-of-Theorem}Proof of Theorem \ref{thm:optimistic}}

\subsection{Sufficiency}

Suppose $\mathbf{\widetilde{S}}=\{\widetilde{S}^{m}\}_{m=1}^{\infty}$
achieves  $\underline{R_{\mathbf{S}|\mathbf{Z}}^{(\varepsilon)}}(D_{E})$,
and for $\delta>0$, denote
\begin{align}
R & =\overline{I}(\mathbf{S};\mathbf{\widetilde{S}}|\mathbf{Z})+\delta,\label{eq:-7}\\
D_{E} & =\underline{D_{E}^{(\varepsilon)}}(\mathbf{S},\mathbf{\widetilde{S}})+\delta.\label{eq:-84}
\end{align}
Then consider the following coding scheme.

{\textbf{Codebook Generation}}: For each $z^{n}$, we independently
generate sequences $\widetilde{s}^{m}\left(i,z^{n}\right),i\in[2^{mR}]$
according to $P_{\widetilde{S}^{m}|Z^{n}=z^{n}}$. Denote the subcodebook
as $\Ccal_{z^{n}}$, and the whole codebook as $\Ccal=\left\{ \Ccal_{z^{n}}\right\} _{z^{n}\in\mathcal{Z}^{n}}$.

{\textbf{Encoder}}: Upon $\left(S^{m},Z^{n}\right)$, the encoder
$f_{m}(S^{m},Z^{n})=M$, where $M$ is determined by $m=\min_{i\in[2^{mR}]}d_{E}(s^{m},\widetilde{s}^{m}\left(i,z^{n}\right)).$

{\textbf{Decoder}}: Upon $\left(M,Z^{n}\right)$, the decoder $g_{m}(M,Z^{n})=\widetilde{s}^{m}\left(M,Z^{n}\right)$.

Denote
\begin{align}
\calT_{1}^{m} & =\left\{ (\mathbf{s},\mathbf{\widetilde{s}},\mathbf{z}):\frac{1}{m}\log\frac{P_{\widetilde{S}^{m}|S^{m}Z^{n}}(\mathbf{\widetilde{s}}|\mathbf{s},\mathbf{z})}{P_{\widetilde{S}^{m}|Z^{n}}(\mathbf{\widetilde{s}}|\mathbf{z})}<R\right\} ,\label{eqn:Tgamma}\\
\calT_{2}^{m} & =\left\{ (\mathbf{s},\mathbf{\widetilde{s}}):\frac{1}{m}d_{E}(\mathbf{s},\mathbf{\widetilde{s}})<D_{E}\right\} \times\mathcal{Z}^{n}.
\end{align}
and $\calT^{m}=\calT_{1}^{m}\cap\calT_{2}^{m}$. If define
\begin{equation}
P_{e}^{m}=\mathbb{P}\big(d_{E}(S^{m},g_{m}(f_{m}(S^{m},Z^{n}),Z^{n}))>D_{E}\big),\label{eqn:fein-1}
\end{equation}
then following from the argument in \cite[proof 1) of Thm. 5.2.1]{Han},
we have the following lemma. The proof is omitted here.
\begin{lem}
\label{lem:}
\begin{equation}
P_{e}^{m}\le\mathbb{P}\big((S^{m},\widetilde{S}^{m},Z^{n})\notin\calT^{m}\big)+e^{e^{-m\delta}}.\label{eqn:fein}
\end{equation}
\end{lem}
On the other hand, according to the definitions of $\overline{I}(\cdot)$
and $\underline{D_{E}^{(\varepsilon)}}(\cdot)$, and from \eqref{eq:-7}
and \eqref{eq:-84}, we have
\begin{align}
\lim_{m\rightarrow\infty}\mathbb{P}\big((S^{m},\widetilde{S}^{m},Z^{n})\notin\calT_{1}^{m}\big) & =0,\label{eqn:fein-2}\\
\liminf_{m\rightarrow\infty}\mathbb{P}\big((S^{m},\widetilde{S}^{m},Z^{n})\notin\calT_{2}^{m}\big) & <\varepsilon.
\end{align}
Hence $\liminf_{m\rightarrow\infty}\mathbb{P}\big((S^{m},\widetilde{S}^{m},Z^{n})\notin\calT^{m}\big)<\varepsilon.$
Combining this with Lemma \ref{lem:} gives us $\liminf_{m\rightarrow\infty}P_{e}^{m}<\varepsilon.$
Therefore, $\left(R,D_{E}\right)$ is optimistically $\varepsilon$-achievable.
Letting $\delta\rightarrow0$ completes the proof of the sufficiency.

\subsection{Necessity}

Assume there exists a sequence of $R_{m}$-rate fixed-length codes
$\left(f_{m},g_{m}\right)_{m=1}^{\infty}$ with reconstructions $\mathbf{\check{S}}=\{\check{S}^{m}\}_{m=1}^{\infty}$
such that
\begin{align}
\limsup_{m\rightarrow\infty}R_{m} & \leq R,\\
\underline{D_{E}^{(\varepsilon)}}(\mathbf{S},\mathbf{\check{S}}) & \leq D_{E}.\label{eq:-87}
\end{align}
Set $\mathbf{\widetilde{S}}=\mathbf{\check{S}}$, then \eqref{eq:-87}
immediately yields
\begin{equation}
\underline{D_{E}^{(\varepsilon)}}(\mathbf{S},\mathbf{\widetilde{S}})\leq D_{E}.\label{eq:-82}
\end{equation}
Hence we only need prove $\overline{I}(\mathbf{S};\mathbf{\widetilde{S}}|\mathbf{Z})\leq R$
or $\overline{I}(\mathbf{S};\mathbf{\check{S}}|\mathbf{Z})\leq R.$
Notice that given $Z^{n}=z^{n}$, $\check{S}^{m}=g_{m}(f_{m}(S^{m},z^{n}),z^{n}))$
cannot take more than $2^{mR_{m}}$ values. Then the following lemma
holds.
\begin{lem}
\cite[Lem. 2.6.2]{Han}\label{lem:-1}
\begin{equation}
\mathbb{P}\left\{ \frac{1}{m}\log\frac{1}{P_{\check{S}^{m}|Z^{n}}(\check{S}^{m}|Z^{n})}\geq R_{m}+\delta\right\} \le e^{-m\delta}.\label{eqn:fein-3}
\end{equation}
\end{lem}
On the other hand,
\begin{equation}
\frac{1}{m}\log\frac{P_{\check{S}^{m}|S^{m}Z^{n}}(\check{S}^{m}|S^{m},Z^{n})}{P_{\check{S}^{m}|Z^{n}}(\check{S}^{m}|Z^{n})}\leq\frac{1}{m}\log\frac{1}{P_{\check{S}^{m}|Z^{n}}(\check{S}^{m}|Z^{n})},
\end{equation}
and $\limsup_{m\rightarrow\infty}R_{m}\leq R$ for any $z^{n}$. Hence
it holds that
\begin{equation}
\mathbb{P}\left\{ \frac{1}{m}\log\frac{P_{\check{S}^{m}|S^{m}Z^{n}}(\check{S}^{m}|S^{m},Z^{n})}{P_{\check{S}^{m}|Z^{n}}(\check{S}^{m}|Z^{n})}\geq R+2\delta\right\} \le e^{-m\delta},\label{eqn:fein-3-1}
\end{equation}
which further implies $\overline{I}(\mathbf{S};\mathbf{\check{S}}|\mathbf{Z})\leq R+2\delta.$
This means $\overline{I}(\mathbf{S};\mathbf{\check{S}}|\mathbf{Z})\leq R$
because $\delta>0$ is arbitrary. Combining it with \eqref{eq:-82}
completes the proof of $\underline{R_{\mathbf{S}|\mathbf{Z}}^{(\varepsilon)}}(D_{E})\leq R.$

\section{\label{sec:Proof-of-Theorem-lossyequ}Proof of Proposition \ref{prop:lossyequivocation}}

Actually, $\underline{R_{\mathbf{S}|\mathbf{Z}}}(D_{E})$ equals
the minimum source coding rate for the source $\mathbf{S}$ and two-sided
information $\mathbf{Z}$ such that $\underline{D_{E}}(\mathbf{S},\mathbf{\check{S}})\leq D_{E}$
 (see Theorem \ref{thm:optimistic} and Remark \ref{rem:When-,-.});
while $\underline{H}(\mathbf{S}|\mathbf{Z})$ equals the minimum source
coding rate for the source $\mathbf{S}$ and two-sided information
$\mathbf{Z}$ such that $\limsup_{m\rightarrow\infty}\mathbb{P}\left(S^{m}=\check{S}^{m}\right)>0$
\cite{Chen}.  Therefore, there exists a source code with rate $\underline{H}(\mathbf{S}|\mathbf{Z})+\delta,\delta>0,$
such that $\limsup_{m\rightarrow\infty}\mathbb{P}\left(S^{m}=\check{S}^{m}\right)>0$
which further implies $\underline{D_{E}}(\mathbf{S},\mathbf{\check{S}})\leq D_{E}$
for any $D_{E}\geq D_{E,\min}^{\prime}$ or for $d_{E}(s^{m},\check{s}^{m})=1\left\{ s^{m}\neq\check{s}^{m}\right\} $
and $D_{E}=0$. Then we have $\underline{R_{\mathbf{S}|\mathbf{Z}}}(D_{E})\leq\underline{H}(\mathbf{S}|\mathbf{Z})+\delta$,
since $\underline{R_{\mathbf{S}|\mathbf{Z}}}(D_{E})$ equals the minimum
coding rate satisfying $\underline{D_{E}}(\mathbf{S},\mathbf{\check{S}})\leq D_{E}$.
Observe that $\delta>0$ is arbitrary, hence $\underline{R_{\mathbf{S}|\mathbf{Z}}}(D_{E})\leq\underline{H}(\mathbf{S}|\mathbf{Z})$.
Similarly, we can also have $\underline{R_{\mathbf{S}|\mathbf{Z}}}(D_{E})\geq\underline{H}(\mathbf{S}|\mathbf{Z})$
for $d_{E}(s^{m},\check{s}^{m})=1\left\{ s^{m}\neq\check{s}^{m}\right\} $
and $D_{E}=0$.

The argument above is rather superficial. To understand the proposition
more intuitively, we provide a direct proof in the following. Note
that the following proof is essentially the same as the argument above.

For $\delta>0$, denote
\begin{align}
\calT^{m} & =\left\{ (\mathbf{s},\mathbf{z}):\frac{1}{m}\log\frac{1}{P_{S^{m}|Z^{n}}(\mathbf{s}|\mathbf{z})}<\underline{H}(\mathbf{S}|\mathbf{Z})+\delta\right\} \label{eqn:Tgamma-1-1}
\end{align}
and
\begin{align}
\calT^{m}\left(\mathbf{z}\right) & =\left\{ \mathbf{s}:(\mathbf{s},\mathbf{z})\in\calT^{m}\right\} .\label{eqn:Tgamma-1-1-4}
\end{align}
Then, we have $P_{S^{m}|Z^{n}}(\mathbf{s}|\mathbf{z})>2^{-m\left(\underline{H}(\mathbf{S}|\mathbf{Z})+\delta\right)}$
for $(\mathbf{s},\mathbf{z})\in\calT^{m}$. Since it holds that
\begin{align}
1 & \geq\sum_{\mathbf{s}\in\calT^{m}\left(\mathbf{z}\right)}P_{S^{m}|Z^{n}}(\mathbf{s}|\mathbf{z})\label{eqn:Tgamma-1-1-2}\\
 & >\sum_{\mathbf{s}\in\calT^{m}\left(\mathbf{z}\right)}2^{-m\left(\underline{H}(\mathbf{S}|\mathbf{Z})+\delta\right)}\\
 & \geq\left|\calT^{m}\left(\mathbf{z}\right)\right|2^{-m\left(\underline{H}(\mathbf{S}|\mathbf{Z})+\delta\right)},
\end{align}
we have $\left|\calT^{m}\left(\mathbf{z}\right)\right|<2^{m\left(\underline{H}(\mathbf{S}|\mathbf{Z})+\delta\right)}$
for any $\mathbf{z}$.

For $\delta'>0$, suppose $\mathbf{g}\left(\mathbf{s}\right)$ is
a sequence of functions such that $\overline{D_{E}}(\mathbf{S},\mathbf{g}\left(\mathbf{S}\right))=D_{E,\min}^{\prime}+\delta'$.
Set $\check{S}^{m}=\begin{cases}
S^{m}, & \textrm{if }\left(S^{m},Z^{n}\right)\in\calT^{m}\\
s_{0}^{m}, & \textrm{otherwise}
\end{cases}$ and $\check{S}^{*m}=g^{m}(\check{S}^{m})$, where $s_{0}^{m}\in\mathcal{S}^{m}$
is a constant vector. Then it immediately yields
\begin{align}
\overline{I}(\mathbf{S};\mathbf{\check{S}}^{*}|\mathbf{Z}) & \leq\overline{I}(\mathbf{S};\mathbf{\check{S}}^{*}\mathbf{\check{S}}|\mathbf{Z})\label{eq:-33}\\
 & =\overline{I}(\mathbf{S};\mathbf{\check{S}}|\mathbf{Z})\label{eq:-51}\\
 & \leq\overline{H}(\mathbf{\check{S}}|\mathbf{Z}),
\end{align}
where \eqref{eq:-51} follows from that $\check{S}^{*m}$ is a function
of $\check{S}^{m}$.

It has been shown that given $Z^{n}=z^{n}$, $\check{S}^{m}$ cannot
take more than $2^{m\left(\underline{H}(\mathbf{S}|\mathbf{Z})+\delta\right)}$
values. Hence we have the following lemma.
\begin{lem}
\cite[Lem. 2.6.2]{Han}\label{lem:-1-1} For any $\delta''>0$,
\begin{equation}
\mathbb{P}\left\{ \frac{1}{m}\log\frac{1}{P_{\check{S}^{m}|Z^{n}}(\check{S}^{m}|Z^{n})}\geq\underline{H}(\mathbf{S}|\mathbf{Z})+\delta+\delta''\right\} \le e^{-m\delta''}.\label{eqn:fein-3-2}
\end{equation}
\end{lem}
It implies $\overline{H}(\mathbf{\check{S}}|\mathbf{Z})\leq\underline{H}(\mathbf{S}|\mathbf{Z})+\delta$.
Therefore,
\begin{equation}
\overline{I}(\mathbf{S};\mathbf{\check{S}}^{*}|\mathbf{Z})\leq\underline{H}(\mathbf{S}|\mathbf{Z})+\delta.\label{eq:-44}
\end{equation}

Next we will show $\underline{D_{E}}(\mathbf{S},\mathbf{\check{S}}^{*})\leq D_{E}$.
First we have
\begin{align}
 & \lVert P_{S^{m}Z^{n}\check{S}^{m}}-P_{S^{m}Z^{n}S^{m}}\rVert_{TV}\nonumber \\
 & =\frac{1}{2}\sum_{\left(s_{1}^{m},z^{n},s_{2}^{m}\right)\in\mathcal{S}^{m}\times\mathcal{Z}^{n}\times\mathcal{S}^{m}}|P_{S^{m}Z^{n}\check{S}^{m}}\left(s_{1}^{m},z^{n},s_{2}^{m}\right)\nonumber \\
 & \qquad\qquad\qquad-P_{S^{m}Z^{n}S^{m}}\left(s_{1}^{m},z^{n},s_{2}^{m}\right)|\\
 & =\frac{1}{2}\sum_{\left(s_{1}^{m},z^{n}\right)\in\calT^{m}}\sum_{s_{2}^{m}\in\mathcal{S}^{m}}|P_{S^{m}Z^{n}\check{S}^{m}}\left(s_{1}^{m},z^{n},s_{2}^{m}\right)\nonumber \\
 & \qquad-P_{S^{m}Z^{n}S^{m}}\left(s_{1}^{m},z^{n},s_{2}^{m}\right)|\nonumber \\
 & \qquad+\frac{1}{2}\sum_{\left(s_{1}^{m},z^{n}\right)\notin\calT^{m}}\sum_{s_{2}^{m}\in\mathcal{S}^{m}}|P_{S^{m}Z^{n}\check{S}^{m}}\left(s_{1}^{m},z^{n},s_{2}^{m}\right)\nonumber \\
 & \qquad-P_{S^{m}Z^{n}S^{m}}\left(s_{1}^{m},z^{n},s_{2}^{m}\right)|\\
 & =\frac{1}{2}\sum_{\left(s_{1}^{m},z^{n}\right)\notin\calT^{m}}\sum_{s_{2}^{m}\in\mathcal{S}^{m}}|P_{S^{m}Z^{n}\check{S}^{m}}\left(s_{1}^{m},z^{n},s_{2}^{m}\right)\nonumber \\
 & \qquad-P_{S^{m}Z^{n}S^{m}}\left(s_{1}^{m},z^{n},s_{2}^{m}\right)|\\
 & \leq\sum_{\left(s_{1}^{m},z^{n}\right)\notin\calT^{m}}P_{S^{m}Z^{n}}\left(s_{1}^{m},z^{n}\right)\\
 & =\mathbb{P}\bigl[\left(S^{m},Z^{n}\right)\notin\calT^{m}\bigr].\label{eq:-12}
\end{align}
On the other hand,
\begin{align}
 & \mathbb{P}\bigl[\left(S^{m},Z^{n}\right)\notin\calT^{m}\bigr]\nonumber \\
 & =1-\mathbb{P}\bigl[\frac{1}{m}\log\frac{1}{P_{S^{m}|Z^{n}}(S^{m}|Z^{n})}<\underline{H}(\mathbf{S}|\mathbf{Z})+\delta\bigr]\label{eq:-13}
\end{align}
and according to the definition of $\underline{H}(\mathbf{S}|\mathbf{Z})$,
\begin{equation}
\limsup_{m\rightarrow\infty}\mathbb{P}\bigl[\frac{1}{m}\log\frac{1}{P_{S^{m}|Z^{n}}(S^{m}|Z^{n})}<\underline{H}(\mathbf{S}|\mathbf{Z})+\delta\bigr]>0.\label{eq:-30}
\end{equation}
Combining \eqref{eq:-12}, \eqref{eq:-13} and \eqref{eq:-30} gives
us
\begin{equation}
\liminf_{m\rightarrow\infty}\lVert P_{S^{m}Z^{n}\check{S}^{m}}-P_{S^{m}Z^{n}S^{m}}\rVert_{TV}<1,
\end{equation}
which further yields
\begin{equation}
\liminf_{m\rightarrow\infty}\lVert P_{S^{m}\check{S}^{m}}-P_{S^{m}S^{m}}\rVert_{TV}<1.\label{eq:-32}
\end{equation}

Furthermore, for any $D_{E}\geq D_{E,\min}^{\prime}+\delta'$, we
have $\overline{D_{E}}(\mathbf{S},\mathbf{g}\left(\mathbf{S}\right))\leq D_{E}$,
i.e., $\lim_{m\rightarrow\infty}\mathbb{P}\left[d_{E}(S^{m},g^{m}(S^{m}))\geq D_{E}+\delta''\right]=0,\forall\delta''>0$.
Combining it with \eqref{eq:-32}, and according to the definition
of total variation distance, we have
\begin{equation}
\liminf_{m\rightarrow\infty}\mathbb{P}\left[d_{E}(S^{m},g^{m}(\check{S}^{m}))\geq D_{E}+\delta''\right]<1,
\end{equation}
 i.e.,
\begin{equation}
\limsup_{m\rightarrow\infty}\mathbb{P}\left[d_{E}(S^{m},\check{S}^{*m})<D_{E}+\delta''\right]>0.\label{eq:-46}
\end{equation}
Since $\delta''>0$ is arbitrary, it must hold that
\begin{equation}
\underline{D_{E}}(\mathbf{S},\mathbf{\check{S}}^{*})\leq D_{E}.\label{eq:-47}
\end{equation}

Combining \eqref{eq:-44} with \eqref{eq:-47} and by the definition
of $\underline{R_{\mathbf{S}|\mathbf{Z}}}(D_{E})$, we have
\begin{align}
\underline{R_{\mathbf{S}|\mathbf{Z}}}(D_{E}) & \leq\overline{I}(\mathbf{S};\mathbf{\check{S}}^{*}|\mathbf{Z})\leq\underline{H}(\mathbf{S}|\mathbf{Z})+\delta.\label{eq:-88-1-1}
\end{align}
Since $\delta>0$ is arbitrary, it holds that
\begin{align}
\underline{R_{\mathbf{S}|\mathbf{Z}}}(D_{E}) & \leq\underline{H}(\mathbf{S}|\mathbf{Z}).\label{eq:-88-1-1-1}
\end{align}

Similarly, it can be shown that both $\underline{R_{\mathbf{S}|\mathbf{Z}}}(D_{E})\leq\underline{H}(\mathbf{S}|\mathbf{Z})$
and $\underline{R_{\mathbf{S}|\mathbf{Z}}}(D_{E})\geq\underline{H}(\mathbf{S}|\mathbf{Z})$
hold for the case of $d_{E}(s^{m},\check{s}^{m})=1\left\{ s^{m}\neq\check{s}^{m}\right\} $
and $D_{E}=0$.

\section{\label{sec:Proof-of-Theorem-lossless}Proof of Converse of Theorem
\ref{thm:lossless}}

From source-channel coding theorem \cite{Gamal}, we have $\gamma C_{B}\geq H\left(S\right)$.
Furthermore, if $R_{L}>R_{S}\left(D_{E}\right)$, then no matter what
scheme Nodes A and B use, the henchman and the wiretapper can ignore
$Z^{n}$ altogether and simply use a point-to-point rate-distortion
code to describe $S^{m}$ within distortion $D_{E}$ (with probability
1). Hence, to prevent the wiretapper from achieving this, it should
hold that $R_{L}\leq R_{S}\left(D_{E}\right)$. Furthermore, to show
the converse part, we only need to prove if $\left(R_{K},R_{L},D_{E}\right)$
is achievable, then
\begin{align}
R_{L} & \leq R_{K}+\gamma\Gamma_{1}\bigl(\frac{1}{\gamma}H\left(S\right)\bigr),\label{eq:-29-3-2}
\end{align}
or equivalently to show
\begin{align}
\underline{R_{\mathbf{S}|\mathbf{Z}}}(D_{E}) & \leq R_{K}+\gamma\Gamma_{1}\bigl(\frac{1}{\gamma}H\left(S\right)\bigr).\label{eq:-29-3}
\end{align}

To that end, we borrow the information spectrum method \cite{Han}.
By Proposition \ref{prop:lossyequivocation} together with Remark
\ref{rem:LE}, we have
\begin{align}
\underline{R_{\mathbf{S}|\mathbf{Z}}}(D_{E}) & \leq\underline{H}(\mathbf{S}|\mathbf{Z}).\label{eq:-88-1}
\end{align}
Next we prove $\underline{H}(\mathbf{S}|\mathbf{Z})\leq R_{K}+\gamma\Gamma_{1}\bigl(\frac{1}{\gamma}H\left(S\right)\bigr).$

Consider
\begin{align}
\underline{H}(\mathbf{S}|\mathbf{Z}) & \leq\underline{H}(\mathbf{S}\mathbf{K}|\mathbf{Z})\\
 & \leq\overline{H}(\mathbf{K}|\mathbf{Z})+\underline{H}(\mathbf{S}|\mathbf{Z}\mathbf{K})\\
 & \leq\overline{H}(\mathbf{K})+\underline{H}(\mathbf{S}|\mathbf{Z}\mathbf{K})\\
 & =R_{K}+\underline{H}(\mathbf{S}|\mathbf{Z}\mathbf{K}).\label{eq:-88}
\end{align}

To upper bound $\underline{H}(\mathbf{S}|\mathbf{Z}\mathbf{K})$,
the following lemma is needed.
\begin{lem}
\cite[Thm.1.7.2]{Han} For any $P_{\mathbf{K}\mathbf{S}\mathbf{Z}}$,
\begin{equation}
\underline{H}(\mathbf{S}|\mathbf{Z}\mathbf{K})\leq\liminf_{m\rightarrow\infty}\frac{1}{m}H(S^{m}|Z^{n}K).\label{eq:-54}
\end{equation}
\end{lem}
Observe that if $K=\emptyset$, the term $\frac{1}{m}H(S^{m}|Z^{n})$
is just the conventional equivocation defined in \cite{Wyner,Csiszar}.
Csiszár et al. \cite{Csiszar} proved that when there is no the secret
key $K$, $\frac{1}{m}H(S^{m}|Z^{n})\leq\gamma\Gamma_{1}\left(\frac{1}{\gamma}H\left(S\right)\right)$
holds. Here we follow similar steps to their proof, and show that
for the case with secret key, $\frac{1}{m}H(S^{m}|Z^{n}K)$ is also
upper bounded by $\gamma\Gamma_{1}\left(\frac{1}{\gamma}H\left(S\right)\right)$.

Denote $\epsilon_{m}=\Pbb(S^{m}\neq\hat{S}^{m})$, then $\lim_{m\rightarrow\infty}\epsilon_{m}=0$.
By Fano's inequality,
\begin{equation}
H(S^{m}|Y^{n}K)\leq1+m\epsilon_{m}\left|\mathcal{S}\right|\leq m\epsilon_{m}^{\prime},
\end{equation}
where $\epsilon_{m}^{\prime}$ is a term such that $\lim_{m\rightarrow\infty}\epsilon_{m}^{\prime}=0$.
Then we have
\begin{align}
H(S^{m}|Z^{n}K) & \leq H(S^{m}|Z^{n}K)-H(S^{m}|Y^{n}K)+m\epsilon_{m}^{\prime}\\
 & =I(S^{m};Y^{n}|K)-I(S^{m};Z^{n}|K)+m\epsilon_{m}^{\prime}.
\end{align}
Observe that
\begin{align}
I(S^{m};Y^{n}|K) & =\sum_{i=1}^{n}I(S^{m};Y_{i}|Y^{i-1}Z_{i+1}^{n}K)+\Sigma_{1}-\Sigma_{2},\\
I(S^{m};Z^{n}|K) & =\sum_{i=1}^{n}I(S^{m};Z_{i}|Y^{i-1}Z_{i+1}^{n}K)+\Sigma_{1}^{*}-\Sigma_{2}^{*},
\end{align}
where
\begin{align}
\Sigma_{1} & =\sum_{i=1}^{n}I(Z_{i+1}^{n};Y_{i}|Y^{i-1}K),\\
\Sigma_{1}^{*} & =\sum_{i=1}^{n}I(Y^{i-1};Z_{i}|Z_{i+1}^{n}K),
\end{align}
and $\Sigma_{2},\Sigma_{2}^{*}$ are the analogous sums with $S^{m}K$
instead of $K$. By Csiszár Sum Identity \cite{Csiszar,Gamal}, $\Sigma_{1}=\Sigma_{1}^{*}$
and $\Sigma_{2}=\Sigma_{2}^{*}$. Therefore,
\begin{align}
 & \frac{1}{m}H(S^{m}|Z^{n}K)\nonumber \\
 & \leq\frac{1}{m}\sum_{i=1}^{n}I(S^{m};Y_{i}|Y^{i-1}Z_{i+1}^{n}K)\nonumber \\
 & \qquad-\frac{1}{m}\sum_{i=1}^{n}I(S^{m};Z_{i}|Y^{i-1}Z_{i+1}^{n}K)+\epsilon_{m}^{\prime}\\
 & =\gamma I(S^{m};Y_{Q}|Y^{Q-1}Z_{Q+1}^{n}KQ)\nonumber \\
 & \qquad-\gamma I(S^{m};Z_{Q}|Y^{Q-1}Z_{Q+1}^{n}KQ)+\epsilon_{m}^{\prime}\\
 & =\gamma I(V;Y|U)-\gamma I(V;Z|U)+\epsilon_{m}^{\prime},\label{eq:-31-2}
\end{align}
where $Q$ denotes a timesharing random variable uniformly distributed
over $[n]$, and $U\triangleq Y^{Q-1}Z_{Q+1}^{n}KQ,V\triangleq S^{m}U,Y\triangleq Y_{Q},Z\triangleq Z_{Q}$.

Furthermore,
\begin{equation}
H\left(S\right)\leq\frac{1}{m}I(S^{m};Y^{n}|K)+\epsilon_{m}^{\prime},\label{eq:-89-1}
\end{equation}
\begin{align}
\frac{1}{m}I(S^{m};Y^{n}|K) & =\frac{1}{m}\sum_{i=1}^{n}I(S^{m};Y_{i}|Y^{i-1}K)\\
 & \leq\frac{1}{m}\sum_{i=1}^{n}I(S^{m}Y^{i-1}Z_{i+1}^{n}K;Y_{i})\\
 & =\gamma I(V;Y),\label{eq:-49-1}
\end{align}
and
\begin{align}
 & \frac{1}{m}I(S^{m};Y^{n}|K)\nonumber \\
 & =\frac{1}{m}\sum_{i=1}^{n}I(S^{m};Y_{i}|Y^{i-1}Z_{i+1}^{n}K)+\frac{1}{m}\Sigma_{1}-\frac{1}{m}\Sigma_{2}\\
 & \leq\frac{1}{m}\sum_{i=1}^{n}I(S^{m};Y_{i}|Y^{i-1}Z_{i+1}^{n}K)+\frac{1}{m}\Sigma_{1}\\
 & =\frac{1}{m}\sum_{i=1}^{n}I(S^{m};Y_{i}|Y^{i-1}Z_{i+1}^{n}K)+\frac{1}{m}\Sigma_{1}^{*}\\
 & =\gamma I(S^{m};Y_{Q}|Y^{Q-1}Z_{Q+1}^{n}KQ)+\gamma I(Y^{Q-1};Z_{Q}|Z_{Q+1}^{n}KQ)\\
 & \leq\gamma I(S^{m};Y_{Q}|Y^{Q-1}Z_{Q+1}^{n}KQ)+\gamma I(Y^{Q-1}Z_{Q+1}^{n}KQ;Z_{Q})\\
 & =\gamma I(V;Y|U)+\gamma I(U;Z).\label{eq:-81-1}
\end{align}

Combining \eqref{eq:-88-1}, \eqref{eq:-88}, \eqref{eq:-54}, \eqref{eq:-31-2},
\eqref{eq:-89-1}, \eqref{eq:-49-1} and \eqref{eq:-81-1}, we have
\begin{align}
R_{L} & \leq R_{K}+\sup_{P_{\mathbf{X}|\mathbf{K}\mathbf{S}}}\liminf_{m\rightarrow\infty}\left(\gamma\widetilde{\Gamma}_{1}\left(\frac{1}{\gamma}H\left(S\right)-\epsilon_{m}^{\prime}\right)+\epsilon_{m}^{\prime}\right)\label{eq:-28-1-1}\\
 & \leq R_{K}+\gamma\widetilde{\Gamma}_{1}\left(\frac{1}{\gamma}H\left(S\right)\right),
\end{align}
where
\begin{align}
\widetilde{\Gamma}_{1}\left(R\right) & \triangleq\max_{\substack{P_{X}P_{V|X}P_{U|V}:\\
I\left(V;Y\right)\geq R,\\
I\left(V;Y|U\right)+I(U;Z)\geq R
}
}\left[I\left(V;Y|U\right)-I\left(V;Z|U\right)\right].\label{eq:gamma1-1}
\end{align}
Actually $\widetilde{\Gamma}_{1}\left(R\right)=\Gamma_{1}\left(R\right)$
for any $R\geq0$ (see \cite[Proof of Cor.2 ]{Csiszar}), hence
\begin{align}
R_{L} & \leq R_{K}+\gamma\Gamma_{1}\left(\frac{1}{\gamma}H\left(S\right)\right).
\end{align}
This completes the proof.

\section{\label{sec:Outer-Bound}Proof of Outer Bound of Theorem \ref{thm:lossy}}

Suppose $Q_{YZ|X}$ is the pmf achieving the minimization in the definition
of $\Gamma_{2}\left(R\right)$. Hence it satisfies $Q_{Y|X}=P_{Y|X},Q_{Z|X}=P_{Z|X}$.
Furthermore, given the code adopted by Nodes A and B, the achievability
of  $\left(R_{K},R_{L},D_{B},D_{E}\right)$ only depends on the margin
distributions of the wiretap channel. Hence the admissible region
$\mathcal{R}$ does not change if we replace the channel $P_{YZ|X}$
with $Q_{YZ|X}$. In the following, without loss of generality, we
only need consider the case of $P_{YZ|X}=Q_{YZ|X}$, i.e., $P_{YZ|X}$
achieves  $\Gamma_{2}\left(R\right)$.

Following the argument for lossless communication case, it should
hold that $R_{L}\leq R_{S}\left(D_{E}\right)$. Next, we prove that
if $\left(R_{K},R_{L},D_{B},D_{E}\right)$ is achievable, then there
exists a conditional pmf $P_{\hat{S}|S}$ such that
\begin{align}
\gamma C_{B} & \geq I(S;\hat{S}),\label{eq:-1}\\
D_{B} & \geq\mathbb{E}d_{B}(S,\hat{S}),\label{eq:-22}\\
\underline{R_{\mathbf{S}|\mathbf{Z}}}(D_{E}) & \leq R_{K}+\gamma\Gamma_{2}\bigl(\frac{1}{\gamma}I(S;\hat{S})\bigr)+R_{S|\hat{S}}(D_{E}).\label{eq:-29}
\end{align}

Since $\overline{I}(\mathbf{S};\mathbf{\check{S}}|\mathbf{Z})\leq\overline{I}(\mathbf{S};\mathbf{\check{S}YK}|\mathbf{Z})$,
we have
\begin{align}
 & \underline{R_{\mathbf{S}|\mathbf{Z}}}(D_{E})\nonumber \\
\leq & \inf_{P_{\mathbf{\check{S}}|\mathbf{S}\mathbf{ZYK}}:\underline{D_{E}}(\mathbf{S},\mathbf{\check{S}})\leq D_{E}}\overline{I}(\mathbf{S};\mathbf{\check{S}YK}|\mathbf{Z})\\
\leq & \overline{I}(\mathbf{S};\mathbf{Y}|\mathbf{Z})+\overline{I}(\mathbf{S};\mathbf{K}|\mathbf{YZ})\nonumber \\
 & \qquad+\inf_{P_{\mathbf{\check{S}}|\mathbf{S}\mathbf{ZYK}}:\underline{D_{E}}(\mathbf{S},\mathbf{\check{S}})\leq D_{E}}\overline{I}(\mathbf{S};\mathbf{\check{S}}|\mathbf{KYZ}).\label{eq:-90}
\end{align}

By data processing inequality,
\begin{align}
\overline{I}(\mathbf{S};\mathbf{Y}|\mathbf{Z}) & \leq\gamma\overline{I}(\mathbf{X};\mathbf{Y}|\mathbf{Z}).\label{eq:-91}
\end{align}
The second term in \eqref{eq:-90} is bounded by
\begin{equation}
\overline{I}(\mathbf{S};\mathbf{K}|\mathbf{YZ})\leq\overline{H}(\mathbf{K}|\mathbf{YZ})\leq\overline{H}(\mathbf{K})=R_{K},\label{eq:-96}
\end{equation}
and the last term in \eqref{eq:-90} is bounded by
\begin{align}
 & \inf_{P_{\mathbf{\check{S}}|\mathbf{S}\mathbf{ZYK}}:\underline{D_{E}}(\mathbf{S},\mathbf{\check{S}})\leq D_{E}}\overline{I}(\mathbf{S};\mathbf{\check{S}}|\mathbf{KYZ})\nonumber \\
= & \inf_{P_{\mathbf{\check{S}}|\mathbf{S}\mathbf{ZYK}\mathbf{\hat{S}}}:\underline{D_{E}}(\mathbf{S},\mathbf{\check{S}})\leq D_{E}}\overline{I}(\mathbf{S};\mathbf{\check{S}}|\mathbf{KYZ}\mathbf{\hat{S}})\label{eq:-92}\\
\leq & \inf_{P_{\mathbf{\check{S}}|\mathbf{S}\mathbf{\hat{S}}}:\underline{D_{E}}(\mathbf{S},\mathbf{\check{S}})\leq D_{E}}\overline{I}(\mathbf{S};\mathbf{\check{S}}|\mathbf{KYZ}\mathbf{\hat{S}})\label{eq:-93}\\
\leq & \inf_{P_{\mathbf{\check{S}}|\mathbf{S}\mathbf{\hat{S}}}:\underline{D_{E}}(\mathbf{S},\mathbf{\check{S}})\leq D_{E}}\overline{I}(\mathbf{S};\mathbf{\check{S}}|\mathbf{\hat{S}})\label{eq:-94}\\
\triangleq & \underline{R_{\mathbf{S}|\mathbf{\hat{S}}}}(D_{E}),\label{eq:-97}
\end{align}
where \eqref{eq:-92} follows from $\mathbf{S}\mathbf{\check{S}}\mathbf{Z}\rightarrow\mathbf{KY}\rightarrow\mathbf{\hat{S}}$,
and \eqref{eq:-94} follows from that for any $\mathbf{\check{S}}$
such that $\mathbf{KYZ}\rightarrow\mathbf{S}\mathbf{\hat{S}}\rightarrow\mathbf{\check{S}}$,
\begin{align}
 & \overline{I}(\mathbf{S};\mathbf{\check{S}}|\mathbf{KYZ}\mathbf{\hat{S}})\nonumber \\
= & \plimsup_{m\to\infty}\frac{1}{m}\log\frac{P_{\check{S}^{m}|S^{m}\hat{S}^{m}}(\check{S}^{m}|S^{m}\hat{S}^{m})}{P_{\check{S}^{m}|\hat{S}^{m}Y^{n}Z^{n}K}(\check{S}^{m}|S^{m}\hat{S}^{m}Y^{n}Z^{n}K)}\\
= & \plimsup_{m\to\infty}\biggl(\frac{1}{m}\log\frac{P_{\check{S}^{m}|S^{m}\hat{S}^{m}}(\check{S}^{m}|S^{m}\hat{S}^{m})}{P_{\check{S}^{m}|\hat{S}^{m}}(\check{S}^{m}|\hat{S}^{m})}\nn\nonumber \\
 & \qquad-\frac{1}{m}\log\frac{P_{\check{S}^{m}|\hat{S}^{m}Y^{n}Z^{n}K}(\check{S}^{m}|S^{m}\hat{S}^{m}Y^{n}Z^{n}K)}{P_{\check{S}^{m}|\hat{S}^{m}}(\check{S}^{m}|\hat{S}^{m})}\biggr)\\
\le & \overline{I}(\mathbf{S};\mathbf{\check{S}}|\mathbf{\hat{S}})-\underline{I}(\mathbf{KYZ};\mathbf{\check{S}}|\mathbf{\hat{S}})\\
\le & \overline{I}(\mathbf{S};\mathbf{\check{S}}|\mathbf{\hat{S}}).\label{eq:-95}
\end{align}
The inequality \eqref{eq:-95} follows from $\underline{I}(\mathbf{KYZ};\mathbf{\check{S}}|\mathbf{\hat{S}})\geq0$,
which is a conditional version of \cite[Eqn. (3.2.3)]{Han}.

Combining \eqref{eq:-90}, \eqref{eq:-91}, \eqref{eq:-96} and \eqref{eq:-97}
gives us
\begin{equation}
\underline{R_{\mathbf{S}|\mathbf{Z}}}(D_{E})\leq R_{K}+\gamma\overline{I}(\mathbf{X};\mathbf{Y}|\mathbf{Z})+\underline{R_{\mathbf{S}|\mathbf{\hat{S}}}}(D_{E}).\label{eq:-5}
\end{equation}
Hence to show \eqref{eq:-1}-\eqref{eq:-29} we only need to prove
there exists a conditional pmf $P_{\hat{S}|S}$ such that \eqref{eq:-1},
\eqref{eq:-22}, and
\begin{align}
\overline{I}(\mathbf{X};\mathbf{Y}|\mathbf{Z}) & \leq\Gamma_{2}\bigl(\frac{1}{\gamma}I(S;\hat{S})\bigr),\label{eq:-98}\\
\underline{R_{\mathbf{S}|\mathbf{\hat{S}}}}(D_{E}) & \leq R_{S|\hat{S}}(D_{E}).\label{eq:-107}
\end{align}

We first consider \eqref{eq:-98}, and prove $\overline{I}(\mathbf{X};\mathbf{Y}|\mathbf{Z})\leq\Gamma_{2}\left(\frac{1}{\gamma}\overline{I}(\mathbf{S};\mathbf{\hat{S}})\right)$.
Denote $X^{n}$, and $Y^{n}$ and $Z^{n}$ to be output variables
of the channel corresponding to $X^{n}$. We also denote $\overline{X}^{n}$
to be a sequence of independent random variables with pmf $P_{\overline{X}^{n}}=\prod_{i=1}^{n}P_{X_{i}}$,
and $\overline{Y}^{n}$ and $\overline{Z}^{n}$ to be output variables
of the channel corresponding to $\overline{X}^{n}$. Obviously, $P_{\overline{X}_{i}\overline{Y}_{i}\overline{Z}_{i}}=P_{X_{i}Y_{i}Z_{i}}$
for $1\leq i\leq n$. Consider the sequence of random variables
\begin{align}
U_{n} & \triangleq\frac{1}{n}\log\frac{P_{Y^{n}|X^{n}Z^{n}}(Y^{n}|X^{n},Z^{n})}{P_{\overline{Y}^{n}|\overline{Z}^{n}}(Y^{n}|Z^{n})}\\
 & =\frac{1}{n}\sum_{i=1}^{n}\log\frac{P_{Y_{i}Z_{i}|X_{i}}(Y_{i},Z_{i}|X_{i})}{P_{\overline{Y}_{i}|\overline{Z}_{i}}(Y_{i}|Z_{i})P_{Z_{i}|X_{i}}(Z_{i}|X_{i})}\label{eq:-42}\\
 & =\frac{1}{n}\sum_{i=1}^{n}\log\frac{P_{Y_{i}Z_{i}|X_{i}}(Y_{i},Z_{i}|X_{i})}{P_{Y_{i}|Z_{i}}(Y_{i}|Z_{i})P_{Z_{i}|X_{i}}(Z_{i}|X_{i})}\label{eq:-43}
\end{align}
where \eqref{eq:-42} follows from that $P_{\overline{Y}^{n}|\overline{Z}^{n}}$,
$P_{Y^{n}Z^{n}|X^{n}}$ and $P_{Z^{n}|X^{n}}$ are memoryless.

From the sub-additivity of $\plimsup$ \cite[Sec. 1.3]{Han}, and
by introducing the product distribution $P_{\overline{Y}^{n}|\overline{Z}^{n}}$,
we obtain
\begin{align}
\overline{I}(\mathbf{Y};\mathbf{X}|\mathbf{Z}) & =\plimsup_{n\to\infty}\biggl(\frac{1}{n}\log\frac{P_{Y^{n}|X^{n}Z^{n}}(Y^{n}|X^{n},Z^{n})}{P_{\overline{Y}^{n}|\overline{Z}^{n}}(Y^{n}|Z^{n})}\nonumber \\*
 & \qquad\qquad\qquad-\frac{1}{n}\log\frac{P_{Y^{n}|Z^{n}}(Y^{n}|Z^{n})}{P_{\overline{Y}^{n}|\overline{Z}^{n}}(Y^{n}|Z^{n})}\biggr)\label{eq:-25}\\
 & \le\plimsup_{n\to\infty}\frac{1}{n}\log\frac{P_{Y^{n}|X^{n}Z^{n}}(Y^{n}|X^{n},Z^{n})}{P_{\overline{Y}^{n}|\overline{Z}^{n}}(Y^{n}|Z^{n})}\nonumber \\*
 & \qquad-\pliminf_{n\to\infty}\frac{1}{n}\log\frac{P_{Y^{n}|Z^{n}}(Y^{n}|Z^{n})}{P_{\overline{Y}^{n}|\overline{Z}^{n}}(Y^{n}|Z^{n})}.\label{eq:-99}
\end{align}
The final term is non-negative following~\cite[Lem. 3.2.1]{Han}
and hence
\begin{align}
\overline{I}(\mathbf{Y};\mathbf{X}|\mathbf{Z})\le\plimsup_{n\to\infty}U_{n}.\label{eq:u}
\end{align}
Now we bound $\plimsup_{n\to\infty}U_{n}$. First define the information
density random variables $L_{i}$ as
\begin{align}
L_{i}\triangleq\log\frac{P_{Y_{i}Z_{i}|X_{i}}(Y_{i},Z_{i}|X_{i})}{P_{Y_{i}|Z_{i}}(Y_{i}|Z_{i})P_{Z_{i}|X_{i}}(Z_{i}|X_{i})}.\label{eq:l}
\end{align}
Then
\begin{equation}
U_{n}=\frac{1}{n}\sum_{i=1}^{n}L_{i}.\label{eq:u2}
\end{equation}
Obviously,
\begin{equation}
\bbE[L_{i}]=I\left(Y_{i};X_{i}|Z_{i}\right).
\end{equation}
Now define
\begin{equation}
P_{\tilde{X}}\left(x\right)=\frac{1}{n}\sum_{i=1}^{n}P_{X_{i}}\left(x\right)
\end{equation}
and the induced distribution
\begin{equation}
P_{\tilde{X}\tilde{Y}\tilde{Z}}\left(x,y,z\right)=P_{\tilde{X}}\left(x\right)P_{YZ|X}\left(y,z|x\right).
\end{equation}
Since $I\left(Y;X|Z\right)$ is a concave-$\cap$ function of the
input probability distribution $P_{X}$ \cite[Thm. 2]{Dijk}, we have
\begin{equation}
\frac{1}{n}\sum_{i=1}^{n}I\left(Y_{i};X_{i}|Z_{i}\right)\leq I\left(\tilde{X};\tilde{Y}|\tilde{Z}\right).
\end{equation}
It means that
\begin{equation}
\mu\triangleq\bbE\Bigl[\frac{1}{n}\sum_{i=1}^{n}L_{i}\Bigr]\leq I\left(\tilde{X};\tilde{Y}|\tilde{Z}\right)\le\min\left\{ \log\left|\mathcal{X}\right|,\log\left|\mathcal{Y}\right|\right\} .\label{eq:bound}
\end{equation}
Furthermore, by \cite[Rmk. 3.1.1]{Han}
\begin{equation}
\mathrm{Var}\left[L_{i}\right]\leq\log\frac{8\min\left\{ \left|\mathcal{X}\right|,\left|\mathcal{Y}\right|\right\} }{e^{2}}\triangleq\sigma_{0}^{2},
\end{equation}
and
\begin{equation}
\mathrm{Var}\left[\frac{1}{n}\sum_{i=1}^{n}L_{i}\right]=\frac{1}{n^{2}}\sum_{i=1}^{n}\mathrm{Var}\left[L_{i}\right]\leq\frac{\sigma_{0}^{2}}{n}.
\end{equation}
Hence by the Chebyshev inequality, for any $\alpha>0$,
\begin{equation}
\mathbb{P}\left[\left|\frac{1}{n}\sum_{i=1}^{n}L_{i}-\mu\right|\geq\alpha\right]\leq\frac{\sigma_{0}^{2}}{n\alpha^{2}}.
\end{equation}
The upper bound $\frac{\sigma_{0}^{2}}{n\alpha^{2}}$ clearly tends
to zero as $n\rightarrow\infty$. From the definition of $\plimsup$,
we have for any $\alpha>0$,
\begin{equation}
\plimsup_{n\to\infty}\frac{1}{n}\sum_{i=1}^{n}L_{i}\leq\limsup_{n\to\infty}\mu+\alpha.\label{eq:-14}
\end{equation}
Finally combine \eqref{eq:u}, \eqref{eq:u2}, \eqref{eq:bound} and
\eqref{eq:-14}, then we have
\begin{equation}
\overline{I}(\mathbf{Y};\mathbf{X}|\mathbf{Z})\leq I\left(\tilde{X};\tilde{Y}|\tilde{Z}\right)+\alpha.\label{eq:-15}
\end{equation}

In the same way, we have
\begin{equation}
\frac{1}{\gamma}\overline{I}(\mathbf{S};\mathbf{\hat{S}})\leq\overline{I}(\mathbf{X};\mathbf{Y})\leq I\left(\tilde{X};\tilde{Y}\right)+\alpha\leq C_{B}+\alpha.\label{eq:-16}
\end{equation}

Combining \eqref{eq:-15} and \eqref{eq:-16}, we have
\begin{align}
\overline{I}(\mathbf{Y};\mathbf{X}|\mathbf{Z}) & \leq I\left(\tilde{X};\tilde{Y}|\tilde{Z}\right)+\alpha\\
 & \leq\max_{P_{X}:I\left(X;Y\right)\geq\frac{1}{\gamma}\overline{I}(\mathbf{S};\mathbf{\hat{S}})-\alpha}I\left(X;Y|Z\right)+\alpha\label{eq:-19}\\
 & =\Gamma_{2}\left(\frac{1}{\gamma}\overline{I}(\mathbf{S};\mathbf{\hat{S}})-\alpha\right)+\alpha,\label{eq:-17}
\end{align}
where \eqref{eq:-17} follows from the assumption that the considered
channel $P_{YZ|X}$ achieves $\Gamma_{2}\left(R\right)$.

Since \eqref{eq:-16} and \eqref{eq:-17} hold for any $\alpha>0$
and $\Gamma_{2}\left(R\right)$ is continuous in $R$ \cite{Wyner},
by letting $\alpha\rightarrow0$, we have
\begin{align}
\frac{1}{\gamma}\overline{I}(\mathbf{S};\mathbf{\hat{S}}) & \leq C_{B},\label{eq:-18}\\
\overline{I}(\mathbf{Y};\mathbf{X}|\mathbf{Z}) & \leq\Gamma_{2}\left(\frac{1}{\gamma}\overline{I}(\mathbf{S};\mathbf{\hat{S}})\right).
\end{align}
Furthermore, from the assumption, we have
\begin{equation}
\overline{D_{B}}(\mathbf{S},\mathbf{\hat{S}})\leq D_{B}.\label{eq:}
\end{equation}
From \eqref{eq:-18}-\eqref{eq:}, and the fact that $\Gamma_{2}\left(R\right)$
is a decreasing function, we have that to complete the proof, we only
need prove there exists a conditional pmf $P_{\hat{S}|S}$ such that
\begin{align}
\overline{I}(\mathbf{S};\mathbf{\hat{S}}) & \geq I(S;\hat{S}),\\
\overline{D_{B}}(\mathbf{S},\mathbf{\hat{S}}) & \geq\mathbb{E}d_{B}(S,\hat{S}),\\
\underline{R_{\mathbf{S}|\mathbf{\hat{S}}}}(D_{E}) & \leq R_{S|\hat{S}}(D_{E}).
\end{align}

For $\left(S^{m},\hat{S}^{m}\right)$, denote $\check{V}^{m}$ to
be a sequence of conditionally independent random variables with pmf
$P_{\check{V}^{m}|S^{m}\hat{S}^{m}}=\prod_{i=1}^{m}P_{\check{V}_{i}|S_{i}\hat{S}_{i}}$.
Then we have
\begin{align}
\underline{R_{\mathbf{S}|\mathbf{\hat{S}}}}(D_{E}) & \leq\inf_{P_{\mathbf{\check{V}}|\mathbf{S}\mathbf{\hat{S}}}:\underline{D_{E}}(\mathbf{S},\mathbf{\check{V}})\leq D_{E}}\overline{I}(\mathbf{S};\mathbf{\check{V}}|\mathbf{\hat{S}}).\label{eq:-100}
\end{align}
For $\left(S^{m},\hat{S}^{m},\check{V}^{m}\right)$, denote $\left(U^{m},V^{m}\right)$
to be a sequence of independent random variables with pmf $P_{S^{m}U^{m}V^{m}}=\prod P_{S_{i}U_{i}V_{i}}$,
and $P_{S_{i}U_{i}V_{i}}=P_{S_{i}\hat{S}_{i}\check{V}_{i}}$ for $1\leq i\leq m$.
Following similar steps to \eqref{eq:-25}-\eqref{eq:-14}, we can
get
\begin{align}
\overline{I}(\mathbf{S};\mathbf{\check{V}}|\mathbf{\hat{S}}) & \leq\limsup_{m\to\infty}\bbE\left[\frac{1}{m}\sum_{i=1}^{m}\log\frac{P_{V_{i}|S_{i}U_{i}}(\check{V}_{i}|S_{i}\hat{S}_{i})}{P_{V_{i}|U_{i}}(\check{V}_{i}|\hat{S}_{i})}\right]\label{eq:-15-1}\\
 & =\limsup_{m\to\infty}I\left(S_{Q};V_{Q}|U_{Q}Q\right),\label{eq:-102}
\end{align}
where $Q$ is a timesharing random variable and uniformly distributed
over $[m]$, independent of other random variables.

By \cite[Lem. 5.8.1]{Han} and \cite[Lem. 5.8.2]{Han} we have
\begin{align}
\overline{D_{B}}(\mathbf{S},\mathbf{\hat{S}}) & \geq\overline{D_{B}}(\mathbf{S},\mathbf{U})=\limsup_{m\rightarrow\infty}\mathbb{E}d_{B}\left(S_{Q},U_{Q}\right),\label{eq:-105}
\end{align}
and
\begin{align}
\overline{I}(\mathbf{S};\mathbf{\hat{S}}) & \geq\overline{I}(\mathbf{S};\mathbf{U})\\
 & =\limsup_{m\rightarrow\infty}I\left(S_{Q};U_{Q}|Q\right)\\
 & \geq\limsup_{m\rightarrow\infty}I\left(S_{Q};U_{Q}\right).\label{eq:-106}
\end{align}
Following similar steps to the proof of the first inequality of \cite[Thm. 1.7.2]{Han}
and the proof of \cite[Lem. 5.8.1]{Han}, we have
\begin{align}
\underline{D_{E}}(\mathbf{S},\mathbf{\check{V}}) & \leq\underline{D_{E}}(\mathbf{S},\mathbf{V})=\liminf_{m\rightarrow\infty}\mathbb{E}d_{E}\left(S_{Q},V_{Q}\right).\label{eq:-103}
\end{align}

According to the definitions of liminf and limsup, for any arbitrarily
small $\epsilon>0$, there always exists a sufficiently large $m_{0}$
such that
\begin{align}
\limsup_{m\to\infty}I\left(S_{Q};V_{Q}|U_{Q}Q\right) & \leq I\left(S_{Q};V_{Q}|U_{Q}Q\right)+\epsilon,\exists m\geq m_{0},\label{eq:-115}\\
\liminf_{m\rightarrow\infty}\mathbb{E}d_{E}\left(S_{Q},V_{Q}\right) & \leq\mathbb{E}d_{E}\left(S_{Q},V_{Q}\right)+\epsilon,\forall m\geq m_{0},\label{eq:-117}\\
\limsup_{m\rightarrow\infty}\mathbb{E}d_{B}\left(S_{Q},U_{Q}\right) & \geq\mathbb{E}d_{B}\left(S_{Q},U_{Q}\right)-\epsilon,\forall m\geq m_{0},\label{eq:-118}\\
\limsup_{m\rightarrow\infty}I\left(S_{Q};U_{Q}\right) & \geq I\left(S_{Q};U_{Q}\right)-\epsilon,\forall m\geq m_{0}.\label{eq:-116}
\end{align}
Hence there exists some $m$ (or equivalently there exists some $\left(V_{Q},U_{Q},Q\right)$)
satisfying the inequalities \eqref{eq:-115}-\eqref{eq:-116} simultaneously.
Combining \eqref{eq:-100}, \eqref{eq:-102}, \eqref{eq:-103}, \eqref{eq:-115}
and \eqref{eq:-117} gives us (the subscript $Q$ is omitted)
\begin{align}
\underline{R_{\mathbf{S}|\mathbf{\hat{S}}}}(D_{E}) & \leq\inf_{P_{V|SUQ}:\mathbb{E}d_{E}\left(S,V\right)+\epsilon\leq D_{E}}I\left(S;V|UQ\right)+\epsilon\\
 & =R_{S|UQ}(D_{E}-\epsilon)+\epsilon\\
 & \leq R_{S|U}(D_{E}-\epsilon)+\epsilon,\label{eq:-104}
\end{align}
where \eqref{eq:-104} follows from the fact that introducing side
information reduces the minimum rate of source coding. Combining \eqref{eq:-105},
\eqref{eq:-106}, \eqref{eq:-118} and \eqref{eq:-116} gives us (the
subscript $Q$ is omitted)
\begin{align}
\overline{D_{B}}(\mathbf{S},\mathbf{\hat{S}}) & \geq\mathbb{E}d_{B}\left(S,U\right)-\epsilon,\label{eq:-103-2}\\
\overline{I}(\mathbf{S};\mathbf{\hat{S}}) & \geq I\left(S;U\right)-\epsilon.\label{eq:-105-2}
\end{align}
Since $\epsilon>0$ is arbitrary and rate-distortion function $R_{S|U}(\cdot)$
is continuous, combining \eqref{eq:-104}, \eqref{eq:-103-2} and
\eqref{eq:-105-2} gives us
\begin{align}
\overline{D_{B}}(\mathbf{S},\mathbf{\hat{S}}) & \geq\mathbb{E}d_{B}\left(S,U\right),\label{eq:-103-2-1}\\
\overline{I}(\mathbf{S};\mathbf{\hat{S}}) & \geq I\left(S;U\right),\label{eq:-105-2-1}\\
\underline{R_{\mathbf{S}|\mathbf{\hat{S}}}}(D_{E}) & \leq R_{S|U}(D_{E}).
\end{align}
Hence $P_{U|S}$ is the desired distribution. This completes the proof
of the converse part.

\section{\label{sec:Inner-Bound}Proof of Inner Bound $\mathcal{R}_{\textrm{sep}}^{(i)}$
of Theorem \ref{thm:lossy}}

It is hard to obtain any satisfactory inner bound through bounding
the lossy-equivocation directly, but it is feasible by analyzing
 the henchman problem instead. Next we prove $\mathcal{R}_{\textrm{sep}}^{(i)}$
by following similar steps to the proof of the achievability part
in \cite{Schieler}.\emph{ }It is worth noting that the complications
introduced by a source-channel setting compared to \cite{Schieler}
mainly comes from that, for the following equivalent subproblem, the
chosen index is not uniformly or nearly-uniformly distributed in general
(which is dependent on the codebook) given the wiretapper's received
signal. This is different from noiseless channel case \cite{Schieler}.
For that case, the chosen index naturally has a uniform conditional
distribution. Hence the proof given in \cite{Schieler} cannot be
applied to our case directly. To address this difficulty, we first
need to find a high-probability set of codebooks on which the chosen
index with a nearly-uniform conditional distribution (given the wiretapper's
received signal) is generated. Then following similar steps to the
proof given in \cite{Schieler}, we show that for any codebook in
this high-probability set, the probability that the wiretapper achieves
the target distortion vanishes. This leads to our result. Furthermore,
applying \emph{two-layered }secrecy-channel code is necessary to achieve
the desired performance. This makes our problem more complicated.

Before proving $\mathcal{R}_{\textrm{sep}}^{(i)}$ of Theorem \ref{thm:lossy},
we first consider the following equivalent problem (the proof of the
equivalence will be given in Appendix \ref{sub:Likelihood-encoder}).

\subsection{\label{sub:Interlude}Subproblem: lossy compression of a codeword
drawn uniformly from a random codebook with side information }

Let $\epsilon>0$. Assume $X$ and $Y$ are the variables achieving
the maximum in the definition of $\Gamma_{1}\left(R\right)$ with
$R=\frac{1}{\gamma}\left(I(S;\hat{S})+3\epsilon\right)$. Then $I\left(V;Y|U\right)-I\left(V;Z|U\right)=\Gamma_{1}\left(\frac{1}{\gamma}\left(I(S;\hat{S})+3\epsilon\right)\right)$
and $I\left(V;Y\right)\geq\frac{1}{\gamma}\left(I(S;\hat{S})+3\epsilon\right)$.
Fix
\begin{eqnarray}
R_{0} & = & \gamma I\left(U;Y\right)-\epsilon,\label{eq:-23}\\
R_{1} & = & \gamma I\left(V;Y|U\right)-\epsilon,\\
R_{1}^{\prime} & = & \gamma I\left(V;Z|U\right)+\epsilon,\label{eq:-24}
\end{eqnarray}
and define\footnote{Assume $R_{K}\leq R_{t}-R_{c}$. If $R_{K}>R_{t}-R_{c}$, then only
$R_{t}-R_{c}$ rate of key is used in our scheme.}
\begin{eqnarray}
R_{t} & = & R_{0}+R_{1},\label{eq:-23-1}\\
R_{c} & = & R_{1}-R_{1}^{\prime},\\
R_{p} & = & R_{t}-R_{c}-R_{K}=R_{0}+R_{1}^{\prime}-R_{K}.\label{eq:-24-1}
\end{eqnarray}
Hence $R_{t}\geq I(S;\hat{S})+\epsilon$. Besides, for simplicity
we assume $m\gamma$ is an integer, hence $n=m\gamma$.

{\textbf{Codebook Generation}}: Randomly and independently generate
sequences $\hat{s}^{m}(j_{k},j_{p},j_{c}),(j_{k},j_{p},j_{c})\in[2^{mR_{K}}][2^{mR_{p}}][2^{mR_{c}}]$
with each according to $\prod_{i=1}^{m}P_{\hat{S}}\left(\hat{s}_{i}\right)$.
Randomly and independently generate sequences $u^{n}(m_{0}),m_{0}\in[2^{mR_{0}}]$
with each according to $\prod_{i=1}^{n}P_{U}\left(u_{i}\right)$,
and for each message $m_{0}\in[2^{mR_{0}}]$, randomly and independently
generate sequences $v^{n}(m_{0},m_{1}),m_{1}\in[2^{mR_{1}}]$ with
each according to $\prod_{i=1}^{n}P_{V|U}\left(v_{i}|u_{i}(m_{0})\right)$.
The codebook
\begin{align*}
\mathcal{C}= & \Bigl\{\hat{s}^{m}(j_{k},j_{p},j_{c}),(j_{k},j_{p},j_{c})\in[2^{mR_{K}}][2^{mR_{p}}][2^{mR_{c}}],\\
 & \left(u^{n}(m_{0}),v^{n}(m_{0},m_{1})\right),(m_{0},m_{1})\in[2^{mR_{0}}][2^{mR_{1}}]\Bigr\},
\end{align*}
is revealed to all parties including the wiretapper.

\textbf{Subproblem}: \label{Subproblem:--Denote} Denote $j=(j_{k},j_{p},j_{c})$\footnote{The subscripts means ``key'', ``public'', and ``channel'', respectively.},
then choose an index $J$ uniformly at random from $[2^{mR_{K}}][2^{mR_{p}}][2^{mR_{c}}]$
and generate $M_{k}=J_{k}\wedge K$, where $\land$ is the one-time
pad operation. Map $(M_{k},J_{p})$ into $(M_{0},M_{1}^{\prime})\in[2^{mR_{0}}][2^{mR_{1}^{\prime}}]$
through an arbitrary bijective function $(m_{0},m_{1}^{\prime})=g(m_{k},j_{p})$.
Denote $M\triangleq(M_{0},M_{1}),M_{1}\triangleq(M_{1}^{\prime},M_{c}),M_{c}\triangleq J_{c}$.
Given $K=k$, the mapping between $M$ and $J$, denoted as $j(m,k)$
or $m(j,k)$, is also bijective. Moreover, since $J$ follows a uniform
distribution, $M$ follows a uniform distribution as well.

\begin{figure}
\centering \includegraphics[width=0.45\textwidth]{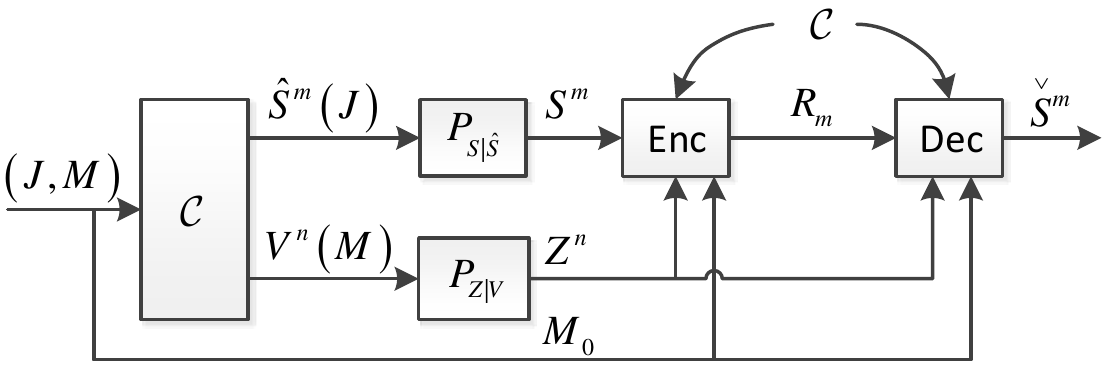}

\protect\protect\protect\caption{\label{fig:lossycompression}Lossy compression of a codeword drawn
uniformly from a random codebook with side information.}
\end{figure}

Based on the codebook above, on one hand, pass $\hat{s}^{m}\left(J\right)$
through a memoryless channel $\prod P_{S|\hat{S}}$ to generate a
sequence $S^{m}$; on the other hand, pass $v^{n}\left(M\right)$
through a memoryless channel $\prod P_{Z|V}$ to generate a sequence
$Z^{n}$ (this is equivalent to applying a stochastic channel encoder
$P_{X^{n}|M,\mathcal{C}}=\prod_{i=1}^{n}P_{X|V}(x_{i}|v_{i}\left(M\right))$
for the channel $\prod P_{Z|X}$). Finally, transmit $S^{m}$ over
a $R_{m}$ rate noiseless channel with the help of two-sided information
$\left(Z^{n},M_{0}\right)$. See Fig. \ref{fig:lossycompression}.

The distribution $Q_{KS^{m}JM\hat{S}^{m}U^{n}V^{n}Y^{n}Z^{n}}$ incurred
by the subproblem above can be expressed as
\begin{align}
 & Q_{KS^{m}JM\hat{S}^{m}U^{n}V^{n}Y^{n}Z^{n}}\nonumber \\
\triangleq & P_{K}Q_{S^{m}J}P_{\hat{S}^{m}|J}P_{M|KJ}P_{U^{n}V^{n}|M}P_{Y^{n}Z^{n}|V^{n}}\nonumber \\
= & 2^{-mR_{K}}Q_{S^{m}J}1\{\hat{s}^{m}=\hat{S}^{m}(j)\}P_{M|KJ}\nonumber \\
 & \qquad1\left\{ u^{n}=U^{n}(m_{0}),v^{n}=V^{n}(m)\right\} \prod P_{YZ|V},\label{eq:idealdist}
\end{align}
where $P_{M|KJ}$ is the distribution induced by the one-time pad
operation, and
\begin{equation}
Q_{S^{m}J}(s^{m},j)\triangleq2^{-mR_{t}}\prod_{i=1}^{m}P_{S|\hat{S}}(s_{i}|\hat{s}_{i}(j)).
\end{equation}
It is easy to verify
\begin{equation}
\Ebb_{\mathcal{C}}\left[Q_{S^{m}\hat{S}^{m}}\left(s^{m},\hat{s}^{m}\right)\right]=\prod_{i=1}^{m}P_{S\hat{S}}(s_{i},\hat{s}_{i}),\label{eq:Qsshat}
\end{equation}
and
\begin{equation}
\Ebb_{\mathcal{C}}\left[Q_{U^{n}V^{n}Y^{n}Z^{n}}\left(u^{n},v^{n},y^{n},z^{n}\right)\right]=\prod_{i=1}^{n}P_{UVYZ}(u_{i},v_{i},y_{i},z_{i}).\label{eq:-78}
\end{equation}

For this subproblem, we have the following theorem.
\begin{thm}
\label{thm:lossy2} If
\begin{align}
 & \limsup_{m\rightarrow\infty}R_{m}\leq\min\Bigl\{ R_{K}+\gamma\Gamma_{1}\bigl(\frac{1}{\gamma}(I(S;\hat{S})+3\epsilon)\bigr)\nonumber \\
 & \qquad\qquad\qquad\qquad+R_{S|\hat{S}}(D_{E}),R_{S}(D_{E})\Bigr\}-5\epsilon,\label{eq:RLopt}
\end{align}
then
\begin{equation}
\mathop{\lim}\limits _{m\to\infty}\mathbb{E}_{\mathcal{C}Z^{n}M_{0}}\Bigl[\max_{R_{m}codes}\mathbb{P}\bigl[d_{E}(S^{m},\check{S}^{m})\le D_{E}|\mathcal{C}Z^{n}M_{0}\bigr]\Bigr]=0.\label{eq:-3}
\end{equation}
\end{thm}
\begin{IEEEproof}[Proof of Theorem \ref{thm:lossy2}]
To prove Theorem \ref{thm:lossy2} we need the following lemma, the
proof of which is given in Appendix \ref{sec:Proof-of-Lemma-prob}.
\begin{lem}
\noindent \label{lem:probability} For any sequence of random variables
$\left\{ X_{n}\right\} $ and any sequence of events $\left\{ \mathcal{A}_{n}\right\} $,
$\lim_{n\to\infty}\mathbb{P}\left(\mathcal{A}_{n}\right)=0$, if and
only if $\lim_{n\to\infty}\mathbb{P}\left[\mathbb{P}\left(\mathcal{A}_{n}|X_{n}\right)>\tau_{n}\right]=0$
for some sequence $\left\{ \tau_{n}\right\} $ such that $\tau_{n}>0$
and $\lim_{n\to\infty}\tau_{n}=0$.
\end{lem}
From Lemma \ref{lem:probability}, to prove Theorem \ref{thm:lossy2}
we only need to show that if $R_{m}$ satisfies \eqref{eq:RLopt},
then
\begin{align}
 & \mathop{\lim}\limits _{m\to\infty}\mathbb{P}_{\mathcal{C}Z^{n}M_{0}}\Bigl[\max_{R_{m}codes}\mathbb{P}\bigl[d_{E}(S^{m},\check{S}^{m})\le D_{E}|\mathcal{C}Z^{n}M_{0}\bigr]>\tau_{m}\Bigr]\nonumber \\
 & =0,\label{eq:-2}
\end{align}
for some sequence $\left\{ \tau_{m}\right\} $ with $\tau_{m}>0$
and $\lim_{m\to\infty}\tau_{m}=0$. Next we prove this.

First define several events
\begin{align}
\mathcal{A}_{1} & \triangleq\left\{ \bigl(S^{m},\hat{S}^{m}\left(J\right)\bigr)\in\mathcal{T}_{\delta}^{m}\right\} ,\\
\mathcal{A}_{2} & \triangleq\left\{ \left(U^{n}\left(M_{0}\right),Z^{n}\right)\in\mathcal{T}_{\delta'}^{n}\right\} ,\\
\mathcal{A}_{3} & \triangleq\left\{ \left(U^{n}\left(M_{0}\right),V^{n}\left(M_{0},M_{1}\right),Z^{n}\right)\in\mathcal{T}_{\delta}^{n}\right\} ,\\
\mathcal{A}_{123} & \triangleq\mathcal{A}_{1}\mathcal{A}_{2}\mathcal{A}_{3},\\
\mathcal{A}_{23} & \triangleq\mathcal{A}_{2}\mathcal{A}_{3},
\end{align}
\begin{equation}
\mathcal{B}\triangleq\left\{ \begin{array}{l}
\min\limits _{(m_{0},m_{c},z^{n}):\left(U^{n}\left(m_{0}\right),z^{n}\right)\in\mathcal{T}_{\delta'}^{n}}\phi_{{\mathcal{C}}}\left(m_{0},m_{c},z^{n}\right)\geq1,\\
\mathop{\max}\limits _{(m_{0},m_{c},z^{n}):\left(U^{n}\left(m_{0}\right),z^{n}\right)\in\mathcal{T}_{\delta'}^{n}}\phi_{{\mathcal{C}}}\left(m_{0},m_{c},z^{n}\right)\leq2^{2m\epsilon}
\end{array}\right\} ,
\end{equation}
where $0<\delta'<\delta$,
\begin{align}
 & \qquad\phi_{{\mathcal{C}}}\left(m_{0},m_{c},z^{n}\right)\triangleq\nonumber \\
 & \sum_{m_{1}^{\prime}\in[2^{mR_{1}^{\prime}}]}1\left\{ \left(U^{n}\left(m_{0}\right),V^{n}\left(m_{0},m_{1}^{\prime},m_{c}\right),z^{n}\right)\in\mathcal{T}_{\delta}^{n}\right\}
\end{align}
and $\epsilon$ is the same to that in \eqref{eq:RLopt}. Observe
that if the codebook satisfies $\mathcal{B}$, then for any $(m_{0},m_{c},z^{n})\in\left[2^{mR_{0}}\right]\left[2^{mR_{c}}\right]\Zcal^{n}$
such that $\left(U^{n}\left(m_{0}\right),z^{n}\right)\in\mathcal{T}_{\delta'}^{n}$,
it holds that
\begin{align}
2^{mR_{c}} & \leq\sum_{m_{1}\in\left[2^{mR_{1}}\right]}1\left\{ \left(U^{n}\left(m_{0}\right),V^{n}\left(m_{0},m_{1}\right),z^{n}\right)\in\mathcal{T}_{\delta}^{n}\right\} \nonumber \\
 & \leq2^{m\left(R_{c}+2\epsilon\right)}.\label{eq:-80}
\end{align}
The $\delta$-typical set is defined according to the notion of strong
typicality, see \cite{Gamal}:
\begin{equation}
\Tcal_{\delta}^{m}(S)\triangleq\{s^{m}\in\Scal^{m}:\left|T_{s^{m}}-P_{S}\right|<\delta P_{S}\},\label{eq:typicality}
\end{equation}
where $T_{s^{m}}$ denotes the type (or empirical distribution) of
$s^{m}$. For simplicity, $\Tcal_{\delta}^{m}(S)$ is also shortly
denoted as $\Tcal_{\delta}^{m}$.

Then we have the following lemmas.
\begin{lem}
\noindent \label{lem:A123} $\lim_{m\to\infty}\Pbb\left[\mathcal{A}_{123}\right]=1$.
\end{lem}
\begin{lem}
\label{lem:B} $\lim_{m\to\infty}\Pbb\left[\mathcal{B}\right]=1$.
\end{lem}
\begin{lem}
\label{lem:B-1} For any codebook $c$ satisfying $\mathcal{B}$,
\begin{align*}
 & \Pbb\left(M_{1}=m_{1},\mathcal{A}_{3}|\mathcal{C}=c,Z^{n}=z^{n},M_{0}=m_{0}\right)\\
\leq & 2^{-m\left(R_{c}-\epsilon_{\delta}\right)}1\left\{ \left(U^{n}\left(m_{0}\right),V^{n}\left(m_{0},m_{1}\right),z^{n}\right)\in\mathcal{T}_{\delta}^{n}\right\} ,
\end{align*}
where $\epsilon_{\delta}$ is a term such that $\epsilon_{\delta}\rightarrow0$
as $\delta\to0$.
\end{lem}
The proofs of Lemma \ref{lem:B} and Lemma \ref{lem:B-1} are given
in Appendices \ref{sec:Proof-of-LemmaB} and \ref{sec:Proof-of-LemmaB-1},
respectively. Furthermore, from \eqref{eq:Qsshat} and \eqref{eq:-78},
and the fact typical set has total probability close to one \cite{Gamal},
we can easily establish Lemma \ref{lem:A123}.

Consider that the optimal $R_{m}$-rate code that maximizes $\mathbb{P}\left[d_{E}(S^{m},\check{S}^{m})\le D_{E}|\mathcal{C}Z^{n}M_{0}\right]$
is adopted. Then we only need to show $\mathop{\lim}\limits _{m\to\infty}\mathbb{P}_{\mathcal{C}Z^{n}M_{0}}\left[\mathbb{P}\left[d_{E}(S^{m},\check{S}^{m})\le D_{E}|\mathcal{C}Z^{n}M_{0}\right]>\tau_{m}\right]=0$
for this code. By utilizing Lemmas \ref{lem:probability}, \ref{lem:A123}
and \ref{lem:B}, we have
\begin{align}
 & \mathbb{P}_{\mathcal{C}Z^{n}M_{0}}\left[\mathbb{P}\left[d_{E}(S^{m},\check{S}^{m})\le D_{E}|\mathcal{C}Z^{n}M_{0}\right]>\tau_{m}\right]\nonumber \\
\leq & \mathbb{P}_{\mathcal{C}Z^{n}M_{0}}\Bigl[\mathbb{P}\left[d_{E}(S^{m},\check{S}^{m})\le D_{E}|\mathcal{C}Z^{n}M_{0}\right]>\tau_{m},\mathcal{B},\nonumber \\
 & \qquad\mathbb{P}\left[\mathcal{A}_{123}^{c}|\mathcal{C}Z^{n}M_{0}\right]\leq\epsilon_{m}\Bigr]+\Pbb\left[\mathcal{B}^{c}\right]\nonumber \\
 & \qquad+\Pbb\left[\mathbb{P}\left[\mathcal{A}_{123}^{c}|\mathcal{C}Z^{n}M_{0}\right]>\epsilon_{m}\right]\\
\leq & \mathbb{P}_{\mathcal{C}Z^{n}M_{0}}\Bigl[\mathbb{P}\left[d_{E}(S^{m},\check{S}^{m})\le D_{E}|\mathcal{C}Z^{n}M_{0}\right]>\tau_{m},\mathcal{B},\nonumber \\
 & \qquad\mathbb{P}\left[\mathcal{A}_{123}^{c}|\mathcal{C}Z^{n}M_{0}\right]\leq\epsilon_{m}\Bigr]+\epsilon_{m}^{\prime},\label{eq:-34}
\end{align}
for some $\epsilon_{m}$ and $\epsilon_{m}^{\prime}$ such that $\epsilon_{m}\rightarrow0$
and $\epsilon_{m}^{\prime}\rightarrow0$ as $m\to\infty$. Furthermore,
\begin{align}
 & \mathbb{P}\left[d_{E}(S^{m},\check{S}^{m})\le D_{E}|\mathcal{C}Z^{n}M_{0}\right]\nonumber \\
\leq & \mathbb{P}\left[d_{E}(S^{m},\check{S}^{m})\le D_{E},\mathcal{A}_{123}|\mathcal{C}Z^{n}M_{0}\right]\nonumber \\
 & \qquad+\mathbb{P}\left[\mathcal{A}_{123}^{c}|\mathcal{C}Z^{n}M_{0}\right]\\
\leq & \mathbb{P}\left[d_{E}(S^{m},\check{S}^{m})\le D_{E},\mathcal{A}_{123}|\mathcal{C}Z^{n}M_{0}\right]+\epsilon_{m},\label{eq:-63}
\end{align}
where \eqref{eq:-63} follows from the event $\mathbb{P}\left[\mathcal{A}_{123}^{c}|\mathcal{C}Z^{n}M_{0}\right]\leq\epsilon_{m}$
in \eqref{eq:-34}.

Owing to the rate constraint, given $(\Ccal,Z^{n},M_{0})$, the reconstruction
$\check{S}^{m}$ cannot take more than $R_{m}$ values. Denote the
set of possible values as $c(\Ccal,Z^{n},M_{0})$. Then
\begin{align}
 & \mathbb{P}\left[d_{E}(S^{m},\check{S}^{m})\le D_{E},\mathcal{A}_{123}|\mathcal{C}Z^{n}M_{0}\right]\nonumber \\
= & \Pbb\Big[\min_{\check{s}^{m}\in c(\Ccal,Z^{n},M_{0})}d_{E}(S^{m},\check{s}^{m})\leq D_{E},\mathcal{A}_{123}|\mathcal{C}Z^{n}M_{0}\Big].\label{eq:boundMax-1}
\end{align}
Now we apply a union bound to \eqref{eq:boundMax-1} and write
\begin{align}
 & \Pbb\Big[\min_{\check{s}^{m}\in c(\Ccal,Z^{n},M_{0})}d_{E}(S^{m},\check{s}^{m})\leq D_{E},\mathcal{A}_{123}|\mathcal{C}Z^{n}M_{0}\Big]\nonumber \\
\leq & \sum_{\check{s}^{m}\in c(\Ccal,Z^{n},M_{0})}\Pbb\Big[d_{E}(S^{m},\check{s}^{m})\leq D_{E},\mathcal{A}_{123}|\mathcal{C}Z^{n}M_{0}\Big]\label{eq:-9}\\
\leq & 2^{mR_{m}}\max_{\check{s}^{m}\in c(\Ccal,Z^{n},M_{0})}\Pbb\Big[d_{E}(S^{m},\check{s}^{m})\leq D_{E},\mathcal{A}_{123}|\mathcal{C}Z^{n}M_{0}\Big]\\
\leq & 2^{mR_{m}}\max_{\check{s}^{m}\in\check{\mathcal{S}}{}^{m}}\Pbb\Big[d_{E}(S^{m},\check{s}^{m})\leq D_{E},\mathcal{A}_{123}|\mathcal{C}Z^{n}M_{0}\Big]\label{eq:-35}\\
= & 2^{mR_{m}}\max_{\check{s}^{m}\in\check{\mathcal{S}}{}^{m}}\sum_{m_{1}=1}^{2^{mR_{1}}}\sum_{k=1}^{2^{mR_{K}}}\Pbb\left[M_{1}=m_{1},\mathcal{A}_{23}|\mathcal{C}Z^{n}M_{0}\right]\nonumber \\
 & \qquad\times\Pbb\left[K=k|\mathcal{A}_{23}\mathcal{C}Z^{n}M\right]\nonumber \\
 & \qquad\times\Pbb\left[d_{E}(S^{m},\check{s}^{m})\leq D_{E},\mathcal{A}_{1}|\mathcal{A}_{23}\mathcal{C}Z^{n}MK\right]\\
\leq & 2^{m\left(R_{m}-R_{K}\right)}\max_{\check{s}^{m}\in\check{\mathcal{S}}{}^{m}}\sum_{m_{1}=1}^{2^{mR_{1}}}\sum_{k=1}^{2^{mR_{K}}}\Pbb\left[M_{1}=m_{1},\mathcal{A}_{3}|\mathcal{C}Z^{n}M_{0}\right]\nonumber \\
 & \qquad\times\Pbb\left[d_{E}(S^{m},\check{s}^{m})\leq D_{E},\mathcal{A}_{1}|\hat{S}^{m}\left(j\left(m,k\right)\right)\right],\label{eq:-40}
\end{align}
where \eqref{eq:-9} follows from a union bound, and \eqref{eq:-40}
follows from $\Pbb\left[K=k|\mathcal{A}_{23}\mathcal{C}Z^{n}M\right]=\Pbb\left[K=k|M\right]=2^{-mR_{K}}$
and $\mathcal{A}_{23}\mathcal{C}Z^{n}MK\rightarrow\hat{S}^{m}\left(J\right)\rightarrow S^{m}\mathcal{A}_{1}$.

Furthermore, for any codebook $c$ satisfying $\mathcal{B}$ and for
any $(m_{0},m_{c},z^{n})\in\left[2^{mR_{0}}\right]\left[2^{mR_{c}}\right]\Zcal^{n}$
such that $\left(U^{n}\left(m_{0}\right),z^{n}\right)\in\mathcal{T}_{\delta'}^{n}$,
there exists at most $2^{2m\epsilon}$ codewords $V^{n}\left(m_{0},m_{1}^{\prime},m_{c}\right)$
such that $\left(U^{n}\left(m_{0}\right),V^{n}\left(m_{0},m_{1}^{\prime},m_{c}\right),z^{n}\right)\in\mathcal{T}_{\delta}^{n}$.
Define
\[
\mathcal{A}_{m_{0},m_{c},z^{n}}\triangleq\left\{ m_{1}^{\prime}:\left(U^{n}\left(m_{0}\right),V^{n}\left(m_{0},m_{1}^{\prime},m_{c}\right),z^{n}\right)\in\mathcal{T}_{\delta}^{n}\right\} .
\]
Then we can bound
\begin{eqnarray}
\left|\mathcal{A}_{m_{0},m_{c},z^{n}}\right| & \leq & 2^{2m\epsilon},\label{eq:-39}
\end{eqnarray}
for any $(m_{0},m_{c},z^{n})\in\left[2^{mR_{0}}\right]\left[2^{mR_{c}}\right]\Zcal^{n}$
such that $\left(U^{n}\left(m_{0}\right),z^{n}\right)\in\mathcal{T}_{\delta'}^{n}$.

Combining \eqref{eq:boundMax-1}, \eqref{eq:-40} and Lemma \ref{lem:B-1},
we have
\begin{align}
 & \mathbb{P}\left[d_{E}(S^{m},\check{S}^{m})\le D_{E},\mathcal{A}_{123}|\mathcal{C}Z^{n}M_{0}\right]\nonumber \\
\leq & 2^{m\lambda}\max_{\check{s}^{m}\in\check{\mathcal{S}}{}^{m}}\sum_{k=1}^{2^{mR_{K}}}\sum_{m_{c}=1}^{2^{mR_{c}}}\sum_{m_{1}^{\prime}\in\mathcal{A}_{m_{0},m_{c},z^{n}}}\nonumber \\
 & \qquad\Pbb\left[d_{E}(S^{m},\check{s}^{m})\leq D_{E},\mathcal{A}_{1}|\hat{S}^{m}\left(j\left(m,k\right)\right)\right],\label{eq:-36}
\end{align}
where $\lambda=R_{m}-R_{K}-R_{c}+\epsilon_{\delta}$. Combining \eqref{eq:-34},
\eqref{eq:-63} and \eqref{eq:-36}, we further have
\begin{align}
 & \mathbb{P}_{\mathcal{C}Z^{n}M_{0}}\left[\mathbb{P}\left[d_{E}(S^{m},\check{S}^{m})\le D_{E}\right]>\tau_{m}\right]\nonumber \\
 & \leq\mathbb{P}_{\mathcal{C}Z^{n}M_{0}}\Big[\max_{\check{s}^{m}\in\check{\mathcal{S}}{}^{m}}\sum_{k=1}^{2^{mR_{K}}}\sum_{m_{c}=1}^{2^{mR_{c}}}\sum_{m_{1}^{\prime}\in\mathcal{A}_{m_{0},m_{c},z^{n}}}\nonumber \\
 & \qquad\qquad\qquad\qquad\qquad\eta_{m,k,\check{s}^{m}}>\tau_{m}^{\prime}2^{-m\lambda}\Big]+\epsilon_{m}^{\prime}\label{eq:-67}\\
 & \leq\mathbb{P}_{\mathcal{C}Z^{n}M_{0}}\Big[\max_{\check{s}^{m}\in\check{\mathcal{S}}{}^{m},m_{1}^{\prime}\in[2^{mR_{1}^{\prime}}]}\sum_{k=1}^{2^{mR_{K}}}\sum_{m_{c}=1}^{2^{mR_{c}}}\nonumber \\
 & \qquad\qquad\qquad\qquad\qquad\eta_{m,k,\check{s}^{m}}>\tau_{m}^{\prime}2^{-m\lambda'}\Big]+\epsilon_{m}^{\prime}\label{eq:-4}\\
 & \leq2^{mR_{1}^{\prime}}|\check{\mathcal{S}}|^{m}\max_{\check{s}^{m}\in\check{\mathcal{S}}{}^{m},m_{1}^{\prime}\in[2^{mR_{1}^{\prime}}]}\nonumber \\
 & \qquad\mathbb{P}_{\mathcal{C}Z^{n}M_{0}}\Big[\sum_{k=1}^{2^{mR_{K}}}\sum_{m_{c}=1}^{2^{mR_{c}}}\eta_{m,k,\check{s}^{m}}>\tau_{m}^{\prime}2^{-m\lambda'}\Big]+\epsilon_{m}^{\prime},\label{eq:-37}
\end{align}
where $\tau_{m}^{\prime}=\tau_{m}-\epsilon_{m},\lambda'=\lambda+2\epsilon$
and
\begin{align}
 & \eta_{m,k,\check{s}^{m}}\triangleq\Pbb\left[d_{E}(S^{m},\check{s}^{m})\leq D_{E},\mathcal{A}_{1}|\hat{S}^{m}\left(j\left(m,k\right)\right)\right]\label{eq:-20}\\
 & =\sum_{s^{m}\in\Scal^{m}}\prod_{i=1}^{m}P_{S|\hat{S}}\bigl(s_{i}|\hat{S}_{i}\left(j\left(m,k\right)\right)\bigr)\nonumber \\
 & 1\left\{ d_{E}(s^{m},\check{s}^{m})\leq D_{E},\bigl(s^{m},\hat{S}^{m}\left(j\left(m,k\right)\right)\bigr)\in\mathcal{T}_{\delta}^{n}\right\} .\label{eq:-11}
\end{align}
To guarantee $\tau_{m}^{\prime}>0$ for any $m$, we choose the sequence
$\left\{ \tau_{m}\right\} $ such that $\tau_{m}>\epsilon_{m}$.

If we can show that the probability in \eqref{eq:-37} decays doubly
exponentially fast with $m$, then the proof will be complete. Now
we prove this.

Note that $\eta_{m,k,\check{s}^{m}}$ is a quantity depending on $\hat{S}^{m}\left(j\left(m,k\right)\right)$,
and the one-time pad satisfies that given $m_{k}$, different $k$'s
correspond to different $j_{k}$'s. Hence given $\left(m_{0},m_{1}^{\prime}\right)$,
for different $\left(k,m_{c}\right)$'s, $j\left(m,k\right)$'s are
different as well. This guarantees that given $\left(\check{s}^{m},m_{0},m_{1}^{\prime}\right)$,
$\eta_{m,k,\check{s}^{m}},\left(m_{c},k\right)\in[2^{mR_{c}}][2^{mR_{K}}]$
are i.i.d. random variables due to the nature of the random codebook,
with mean
\begin{align}
 & \mathbb{E}_{{\mathcal{C}}}\eta_{m,k,\check{s}^{m}}\nonumber \\
 & =\mathbb{E}_{{\mathcal{C}}}\Pbb\left[d_{E}(S^{m},\check{s}^{m})\leq D_{E},\mathcal{A}_{1}|\hat{S}^{m}\left(j\left(m,k\right)\right)\right]\\
 & \leq\mathbb{E}_{{\mathcal{C}}}\Pbb\left[d_{E}(S^{m},\check{s}^{m})\leq D_{E},S^{m}\in\mathcal{T}_{\delta}^{m}|\hat{S}^{m}\left(j\left(m,k\right)\right)\right]\\
 & =\Pbb\left[d_{E}(S^{m},\check{s}^{m})\leq D_{E},S^{m}\in\mathcal{T}_{\delta}^{m}\right].
\end{align}
Now we need the following lemmas.
\begin{lem}
\label{lem:typebound}\cite{Schieler} If $S^{m}$ is i.i.d. according
to $P_{S}$, then for any $\check{s}^{m}$,
\begin{equation}
\Pbb[d(S^{m},\check{s}^{m})\leq D_{E},S^{m}\in\Tcal_{\delta}^{m}]\leq2^{-m(R_{S}(D_{E})-\epsilon_{m,\delta})},\label{eq:rd-1-1}
\end{equation}
where $\epsilon_{m,\delta}$ is a term that vanishes as $\delta\rightarrow0$
and $m\rightarrow\infty$.
\end{lem}
\begin{lem}
\label{lem:typebound-rd-si} \cite{Schieler} Fix $P_{S\hat{S}}$
and $\hat{s}^{m}\in\hat{\mathcal{S}}{}^{m}$. If $S^{m}$ is distributed
according to $\prod_{i=1}^{m}P_{S|\hat{S}=\hat{s}_{i}}$, then for
any $\check{s}^{m}$,
\begin{align}
 & \Pbb[d(S^{m},\check{s}^{m})\leq D_{E},\left(S^{m},\hat{s}^{m}\right)\in\Tcal_{\delta}^{m}|\hat{S}^{m}=\hat{s}^{m}]\nonumber \\
 & \qquad\leq2^{-m(R_{S|\hat{S}}(D_{E})-\epsilon_{m,\delta})},
\end{align}
where $\epsilon_{m,\delta}$ is a term that vanishes as $\delta\rightarrow0$
and $m\rightarrow\infty$.
\end{lem}
\begin{lem}
\label{lem:chernoff-1} \cite{Schieler} If $X^{l}$ is a sequence
of i.i.d. random variables on the interval $[0,a]$ with $\mathbb{E}[X_{i}]=p$,
then
\begin{equation}
\Pbb\Big[\sum_{i=1}^{l}X_{i}>k\Big]\leq\left(\frac{e\!\cdot\!l\!\cdot\!p}{k}\right)^{k/a}.
\end{equation}
\end{lem}
From Lemmas~\ref{lem:typebound} and \ref{lem:typebound-rd-si},
we see that
\begin{align}
\mathbb{E}_{{\mathcal{C}}}\eta_{m,k,\check{s}^{m}} & \leq2^{-m(R_{S}(D_{E})-\epsilon_{m,\delta})},\\
\eta_{m,k,\check{s}^{m}} & \leq2^{-m(R_{S|\hat{S}}(D_{E})-\epsilon_{m,\delta})}.
\end{align}

Using these bounds, we can apply Lemma~\ref{lem:chernoff-1}
to the probability in \eqref{eq:-37} by identifying
\begin{align}
 & l\leq2^{m\left(R_{K}+R_{c}\right)}\label{eq:-2-1-1-1}\\
 & a=2^{-m(R_{S|\hat{S}}(D_{E})-\epsilon_{m,\delta})}\\
 & p\leq2^{-m(R_{S}(D_{E})-\epsilon_{m,\delta})}\\
 & k=\tau_{m}^{\prime}2^{-m\lambda'}=\tau_{m}^{\prime}2^{-m\left(R_{m}-R_{K}-R_{c}+\epsilon_{\delta}+2\epsilon\right)}.
\end{align}
Set $\left\{ \tau_{m}^{\prime}\right\} $ with $\tau_{m}^{\prime}>0$
to be a sub-exponential sequence (i.e., $\tau_{m}^{\prime}=2^{-o(m)}$)
by choosing a proper $\left\{ \tau_{m}\right\} $. Then we have
\begin{equation}
\Pbb\Big[\sum_{k=1}^{2^{mR_{K}}}\sum_{m_{c}=1}^{2^{mR_{c}}}\eta_{m,k,\check{s}^{m}}>\tau_{m}^{\prime}2^{-m\lambda'}\Big]\leq2^{-m\alpha2^{m\beta}},\label{doubleexp-2-2}
\end{equation}
where
\begin{align}
\alpha & =-\left(R_{K}+R_{c}\right)+R_{S}(D_{E})-\epsilon_{m,\delta}\nonumber \\
 & \qquad-\left(R_{m}-R_{K}-R_{c}+\epsilon_{\delta}+2\epsilon\right)\nonumber \\
 & =R_{S}(D_{E})-R_{m}-2\epsilon-\epsilon_{\delta}-\epsilon_{m,\delta}\nonumber \\
 & \geq3\epsilon-\epsilon_{\delta}-\epsilon_{m,\delta},\\
\beta & =R_{S|\hat{S}}(D_{E})-\epsilon_{m,\delta}-\left(R_{m}-R_{K}-R_{c}+\epsilon_{\delta}+2\epsilon\right)\nonumber \\
 & =R_{S|\hat{S}}(D_{E})+R_{K}+\gamma\Gamma_{1}\bigl(\frac{1}{\gamma}(I(S;\hat{S})+3\epsilon)\bigr)\nonumber \\
 & \qquad-R_{m}-\epsilon_{\delta}-4\epsilon-\epsilon_{m,\delta}\nonumber \\
 & \geq\epsilon-\epsilon_{\delta}-\epsilon_{m,\delta}.
\end{align}

For any fixed $\epsilon$, large enough $m$ and small enough $\delta$,
both $\alpha$ and $\beta$ are positive and bounded away from zero.
Then \eqref{doubleexp-2-2} vanishes doubly exponentially fast, and
it in turn implies \eqref{eq:-37} vanishes. This completes the proof
of Theorem \ref{thm:lossy2}.
\end{IEEEproof}

\subsection{\label{sub:Likelihood-encoder}Likelihood encoder}

Consider the codebook defined in the above subproblem, and define
a likelihood encoder by $P_{J|S^{m}}(j|s^{m})\propto\prod_{i=1}^{m}P_{S|\hat{S}}(s_{i}|\hat{s}_{i}(j)),$
where $\propto$ indicates that appropriate normalization is required.
Now we consider Node A concatenates this likelihood encoder, the one-time
pad $M_{k}=J_{k}\wedge K$, the bijective function $(m_{0},m_{1}^{\prime})=g(m_{k},j_{p})$,
and the stochastic channel encoder $P_{X^{n}|M,\mathcal{C}}=\prod_{i=1}^{n}P_{X|V}(x_{i}|v_{i}\left(M\right))$
as described in the subproblem above. For such cascaded encoder, the
induced overall distribution is
\begin{align}
 & P_{KS^{m}JM\hat{S}^{m}U^{n}V^{n}Y^{n}Z^{n}}\nonumber \\
 & =P_{K}P_{S^{m}J}P_{\hat{S}^{m}|J}P_{M|KJ}P_{U^{n}V^{n}|M}P_{Y^{n}Z^{n}|V^{n}}\nonumber \\
 & \triangleq2^{-mR_{K}}P_{S^{m}J}1\{\hat{s}^{m}=\hat{S}^{m}(j)\}P_{M|KJ}\nonumber \\
 & \qquad1\left\{ u^{n}=U^{n}(m_{0}),v^{n}=V^{n}(m)\right\} \prod P_{YZ|V}.
\end{align}
Furthermore, $P_{KS^{m}JM\hat{S}^{m}U^{n}V^{n}Y^{n}Z^{n}}$ is intimately
related to the idealized distribution $Q_{KS^{m}JM\hat{S}^{m}U^{n}V^{n}Y^{n}Z^{n}}$
which is defined in previous subsection.

Schieler and Cuff \cite[Sec. VIII-B]{Schieler} showed that if $R_{t}>I(S;\hat{S})$,
then
\begin{equation}
\lim_{m\to\infty}\Ebb_{\Ccal}\big\lVert P_{S^{m}J}-Q_{S^{m}J}\big\rVert_{TV}=0,
\end{equation}
where $R_{t}$ given in \eqref{eq:-23-1} denotes the exponent of
the size of codebook. Using the property \eqref{eq:PP-QP}, we further
have
\begin{align}
 & \lim_{m\to\infty}\Ebb_{\Ccal}\big\lVert P_{KS^{m}JM\hat{S}^{m}U^{n}V^{n}Y^{n}Z^{n}}-Q_{KS^{m}JM\hat{S}^{m}U^{n}V^{n}Y^{n}Z^{n}}\big\rVert_{TV}\nonumber \\
 & =\lim_{m\to\infty}\Ebb_{\Ccal}\big\lVert P_{S^{m}J}-Q_{S^{m}J}\big\rVert_{TV}=0.
\end{align}
Combine this with Theorem \ref{thm:lossy2}, utilize the property
\eqref{eq:tvcontinuous}, and let $\epsilon\rightarrow0$, then we
have if $R_{L}\leq\min\Bigl\{ R_{K}+\gamma\Gamma_{1}\bigl(\frac{1}{\gamma}\bigl(I(S;\hat{S})\bigr)\bigr)+R_{S|\hat{S}}(D_{E}),R_{S}(D_{E})\Bigr\}$,
the cascaded encoder above satisfies
\begin{align}
 & \lim\limits _{m\to\infty}\mathbb{E}_{\mathcal{C}Z^{n}M_{0}}\left[\max_{R_{m}Hcodes}\mathbb{P}\left[d_{E}(S^{m},\check{S}^{m})\le D_{E}|\mathcal{C}Z^{n}M_{0}\right]\right]\nonumber \\
 & \qquad=0.
\end{align}
It implies that the wiretapper cannot decrypt source even when both
$Z^{n}$ and $M_{0}$ are revealed to him. Hence as a weaker version,
he cannot decrypt source as well when only $Z^{n}$ is revealed to
him. This completes the proof of the secrecy constraint.

To complete the proof of the achievability part, we now show that
the cascaded encoder above can achieve the distortion $\mathbb{E}d_{B}(S,\hat{S})$
for the legitimate receiver. Instead of considering the induced distribution
$P_{KS^{m}JM\hat{S}^{m}U^{n}V^{n}Y^{n}Z^{n}}$ directly, we prove
this holds under the desired distribution $Q_{KS^{m}JM\hat{S}^{m}U^{n}V^{n}Y^{n}Z^{n}}$.
It is easy to verify that $M$ is uniformly distributed under $Q$.
Furthermore, the codebook is randomly generated, and given $M$, the
signal $Y^{n}$  follows $\prod_{i=1}^{n}P_{Y|V}(y_{i}|v_{i}(M))$.
It is well known that a good channel decoder (e.g., joint typicality
decoder) with respect to the memoryless channel $P_{Y|V}$ will drive
the error probability to zero as $m$ goes to infinity, if $R_{t}<\gamma I\left(V;Y\right)$.
That is
\begin{equation}
\lim_{m\to\infty}\Ebb_{\mathcal{C}}\Pbb_{Q}[\hat{M}\neq M]=0.
\end{equation}
Then using the secret key, the legitimate user could recover $J$
with high probability. That is
\begin{equation}
\lim_{m\to\infty}\Ebb_{\mathcal{C}}\Pbb_{Q}[\hat{J}\neq J]=0.
\end{equation}
Furthermore, \eqref{eq:Qsshat} implies $(S^{m},\hat{S}^{m}\left(J\right))$
is an i.i.d sequence under $Q$, hence by law of large numbers, we
have for any $\tau>0$,
\begin{align}
\lim_{m\to\infty}\Pbb_{Q}\Big[d_{B}(S^{m},\hat{S}^{m}\left(J\right))\le\mathbb{E}d_{B}(S,\hat{S})+\tau\Big] & =1.\label{eq:-61}
\end{align}
This implies under distribution $Q$, $\mathbb{E}d_{B}(S,\hat{S})$
is achieved by the legitimate user. Further, since the total variance
between $Q$ and $P$ vanishes as $m\to\infty$, \eqref{eq:-61} also
holds under the distribution $P$. That is, the cascaded encoder above
achieves distortion $\mathbb{E}d_{B}(S,\hat{S})$ for the legitimate
user. This completes the proof of $\mathcal{R}_{\textrm{sep}}^{(i)}$.

\section{\label{sec:Proof-of-Lemma-prob}Proof of Lemma \ref{lem:probability}}

Observe for any $\alpha>0$, any random variable $X$, and any event
$\mathcal{A}$,
\begin{align}
\mathbb{P}\left(\mathcal{A}\right) & =\mathbb{E}_{X}\mathbb{P}\left(\mathcal{A}|X\right)\nonumber \\
 & \geq\mathbb{E}_{X}\left[\mathbb{P}\left(\mathcal{A}|X\right)1\left\{ \mathbb{P}\left(\mathcal{A}|X\right)\geq\alpha\right\} \right]\nonumber \\
 & \geq\alpha\mathbb{E}_{X}\left[1\left\{ \mathbb{P}\left(\mathcal{A}|X\right)\geq\alpha\right\} \right]\nonumber \\
 & =\alpha\mathbb{P}\left[\mathbb{P}\left(\mathcal{A}|X\right)\geq\alpha\right]
\end{align}
and
\begin{align}
\mathbb{P}\left(\mathcal{A}\right) & =\mathbb{E}_{X}\mathbb{P}\left(\mathcal{A}|X\right)\nonumber \\
 & =\mathbb{E}_{X}\left[\mathbb{P}\left(\mathcal{A}|X\right)1\left\{ \mathbb{P}\left(\mathcal{A}|X\right)\geq\alpha\right\} \right]\nonumber \\
 & \qquad+\mathbb{E}_{X}\left[\mathbb{P}\left(\mathcal{A}|X\right)1\left\{ \mathbb{P}\left(\mathcal{A}|X\right)<\alpha\right\} \right]\nonumber \\
 & \leq\mathbb{E}_{X}\left[1\left\{ \mathbb{P}\left(\mathcal{A}|X\right)\geq\alpha\right\} \right]\nonumber \\
 & \qquad+\alpha\mathbb{E}_{X}\left[1\left\{ \mathbb{P}\left(\mathcal{A}|X\right)<\alpha\right\} \right]\nonumber \\
 & \leq\mathbb{P}\left[\mathbb{P}\left(\mathcal{A}|X\right)\geq\alpha\right]+\alpha.
\end{align}
Then we have the following lemma.
\begin{lem}
For any $\alpha>0$, any random variable $X$, and any event $\mathcal{A}$,
$\mathbb{P}\left(\mathcal{A}\right)-\alpha\leq\mathbb{P}\left[\mathbb{P}\left(\mathcal{A}|X\right)\geq\alpha\right]\leq\frac{\mathbb{P}\left(\mathcal{A}\right)}{\alpha}$.
\end{lem}
Consider a sequence of random variables $\left\{ X_{n}\right\} $,
and a sequence of events $\left\{ \mathcal{A}_{n}\right\} $. Applying
the lemma above, we have

\noindent
\begin{equation}
\mathbb{P}\left(\mathcal{A}_{n}\right)-\tau_{n}\leq\mathbb{P}\left[\mathbb{P}\left(\mathcal{A}_{n}|X_{n}\right)\geq\tau_{n}\right]\leq\frac{\mathbb{P}\left(\mathcal{A}_{n}\right)}{\tau_{n}}
\end{equation}
for any positive sequence $\left\{ \tau_{n}\right\} $. If $\lim_{n\to\infty}\mathbb{P}\left(\mathcal{A}_{n}\right)=0$,
then from the fact that no a worst convergent series exists \cite{Ash},
we have there exists a sequence of real numbers $\left\{ \tau_{n}\right\} $
such that $\lim_{n\to\infty}\tau_{n}=0$ and $\lim_{n\to\infty}\frac{\mathbb{P}\left(\mathcal{A}_{n}\right)}{\tau_{n}}=0$.
Hence $\lim_{n\to\infty}\mathbb{P}\left[\mathbb{P}\left(\mathcal{A}_{n}|X_{n}\right)\geq\tau_{n}\right]=0$.
On the other hand, if $\lim_{n\to\infty}\mathbb{P}\left[\mathbb{P}\left(\mathcal{A}_{n}|X_{n}\right)\geq\tau_{n}\right]=0$
for some sequence $\tau_{n}$ with $\lim_{n\to\infty}\tau_{n}=0$,
then $\lim_{n\to\infty}\mathbb{P}\left(\mathcal{A}_{n}\right)\leq\lim_{n\to\infty}\left\{ \mathbb{P}\left[\mathbb{P}\left(\mathcal{A}_{n}|X_{n}\right)\geq\tau_{n}\right]+\tau_{n}\right\} =0$.

\section{\label{sec:Proof-of-LemmaB}Proof of Lemma \ref{lem:B}}

By using a union bound,
\begin{align}
 & \mathbb{P}_{{\mathcal{C}}}\left(\mathcal{B}^{c}\right)\nonumber \\
 & \le\mathbb{P}_{{\mathcal{C}}}\Big[\mathop{\max}\limits _{(m_{0},m_{c},z^{n}):\left(U^{n}\left(m_{0}\right),z^{n}\right)\in\mathcal{T}_{\delta'}^{n}}\phi_{{\mathcal{C}}}\left(m_{0},m_{c},z^{n}\right)>2^{2m\epsilon}\Big]\nonumber \\
 & \qquad+\mathbb{P}_{{\mathcal{C}}}\Big[\mathop{\min}\limits _{(m_{0},m_{c},z^{n}):\left(U^{n}\left(m_{0}\right),z^{n}\right)\in\mathcal{T}_{\delta'}^{n}}\phi_{{\mathcal{C}}}\left(m_{0},m_{c},z^{n}\right)<1\Big].\label{eq:-9-1}
\end{align}

In the following, we prove that both the terms of \eqref{eq:-9-1}
vanish as $m\rightarrow\infty$. Observe
\begin{align}
 & \mathbb{P}_{{\mathcal{C}}}\Big[\mathop{\max}\limits _{(m_{0},m_{c},z^{n}):\left(U^{n}\left(m_{0}\right),z^{n}\right)\in\mathcal{T}_{\delta'}^{n}}\phi_{{\mathcal{C}}}\left(m_{0},m_{c},z^{n}\right)>2^{2m\epsilon}\Big]\nonumber \\
 & =\mathbb{E}_{\mathcal{C}_{0}}\mathbb{P}_{\mathcal{C}_{1}}\Big[\mathop{\max}\limits _{(m_{0},m_{c},z^{n}):\left(U^{n}\left(m_{0}\right),z^{n}\right)\in\mathcal{T}_{\delta'}^{n}}\nonumber \\
 & \qquad\qquad\qquad\qquad\phi_{{\mathcal{C}}}\left(m_{0},m_{c},z^{n}\right)>2^{2m\epsilon}|\mathcal{C}_{0}\Big].\label{eq:-62}
\end{align}
Hence we only need to show the probability in \eqref{eq:-62} vanishes
for any $\mathcal{C}_{0}$ as $m\rightarrow\infty$. Using union bound,
we have
\begin{align}
 & \mathbb{P}_{\mathcal{C}_{1}}\Big[\mathop{\max}\limits _{(m_{0},m_{c},z^{n}):\left(U^{n}\left(m_{0}\right),z^{n}\right)\in\mathcal{T}_{\delta'}^{n}}\phi_{{\mathcal{C}}}\left(m_{0},m_{c},z^{n}\right)>2^{2m\epsilon}|\mathcal{C}_{0}\Big]\nonumber \\
 & \le2^{m\left(R_{0}+R_{c}\right)}\left|\mathcal{T}_{\delta'}^{n}\right|\mathop{\max}\limits _{(m_{0},m_{c},z^{n}):\left(U^{n}\left(m_{0}\right),z^{n}\right)\in\mathcal{T}_{\delta'}^{n}}\nonumber \\
 & \qquad\qquad\qquad\qquad\mathbb{P}_{\mathcal{C}_{1}}\bigl[\phi_{{\mathcal{C}}}\left(m_{0},m_{c},z^{n}\right)>2^{2m\epsilon}|\mathcal{C}_{0}\bigr]\label{eq:UnionBound}
\end{align}
Define $\theta_{m_{1}^{\prime}}\left(z^{n}\right)\triangleq1\left\{ \left(U^{n}\left(m_{0}\right),V^{n}\left(m_{0},m_{1}^{\prime},m_{c}\right),z^{n}\right)\in\mathcal{T}_{\delta}^{n}\right\} $,
then $\phi_{{\mathcal{C}}}\left(m_{0},m_{c},z^{n}\right)=\sum_{m_{1}^{\prime}}\theta_{m_{1}^{\prime}}\left(z^{n}\right)$.
Given $(m_{0},m_{c},z^{n})$ and $\mathcal{C}_{0}$ such that $\left(U^{n}\left(m_{0}\right),z^{n}\right)\in\mathcal{T}_{\delta'}^{n}$,
$\theta_{m_{1}^{\prime}}\left(z^{n}\right),m_{1}^{\prime}\in[2^{mR_{1}^{\prime}}]$
are i.i.d. random variables, with mean

\noindent
\begin{align}
 & \mathbb{E}_{\mathcal{C}_{1}}\theta_{m_{1}^{\prime}}\left(z^{n}\right)\nonumber \\
 & =\mathbb{P}\Bigl[\left(U^{n}\left(m_{0}\right),V^{n}\left(m_{0},m_{1}^{\prime},m_{c}\right),z^{n}\right)\in\mathcal{T}_{\delta}^{n}|U^{n}\left(m_{0}\right)\Bigr]\nonumber \\
 & =\mathbb{P}\bigl[\left(U^{n},V^{n},z^{n}\right)\in\mathcal{T}_{\delta}^{n}|U^{n}\bigr]\nonumber \\
 & \leq2^{-n\left(I\left(V;Z|U\right)-\epsilon_{\delta}\right)},\label{eq:thetaMean}
\end{align}

\noindent where $\epsilon_{\delta}$ tends to zero as $\delta\rightarrow0$,
and \eqref{eq:thetaMean} follows from the joint typicality lemma
\cite{Gamal}. On the other hand, $\left|\mathcal{T}_{\delta'}^{n}\right|\leq2^{n\left(H\left(Z\right)+\epsilon_{\delta'}\right)}$
for some $\epsilon_{\delta'}$ that tends to zero as $\delta'\rightarrow0$.
Hence if we can show that the probability in \eqref{eq:UnionBound}
decays doubly exponentially fast with $m$, then the proof will be
complete. To that end, we first introduce the following lemma on Chernoff
bounds.
\begin{lem}
\label{lem:chernoff} \cite{Schieler}, \cite{Mitzenmacher} If $X^{l}$
is a sequence of i.i.d. $\text{Bern}(p)$ random variables, then for
any $k>0$,
\begin{equation}
\Pbb\Big[\sum_{i=1}^{l}X_{i}>k\Big]\leq\left(\frac{e\!\cdot\!l\!\cdot\!p}{k}\right)^{k},
\end{equation}
and for any $0\le\delta\le1$,
\begin{equation}
\Pbb\Big[\sum_{i=1}^{l}X_{i}\leq(1-\delta)lp\Big]\leq e^{-\frac{\delta^{2}lp}{2}}.
\end{equation}
\end{lem}
By identifying that
\begin{align}
l & =2^{mR_{1}^{\prime}}\label{eq:-60}\\
p & \leq2^{-n\left(I\left(V;Z|U\right)-\epsilon_{\delta}\right)}\\
k & =2^{2m\epsilon}
\end{align}
and applying Lemma \ref{lem:chernoff}, we have
\begin{equation}
\mathbb{P}_{\mathcal{C}_{1}}\left[\phi_{{\mathcal{C}}}\left(m_{0},m_{c},z^{n}\right)>2^{2m\epsilon}|\mathcal{C}_{0}\right]\le2^{-m\alpha2^{m\beta}},\label{eq:DoubleExp}
\end{equation}

\noindent where
\begin{align}
 & \alpha=\gamma I\left(V;Z|U\right)-\gamma\epsilon_{\delta}-R_{1}^{\prime}+2\epsilon=\epsilon-\gamma\epsilon_{\delta},\label{eq:-57}\\
 & \beta=2\epsilon.\label{eq:-28}
\end{align}

For fixed $\epsilon$ and sufficiently small $\delta$, $\alpha>0$
and $\beta>0$. Hence \eqref{eq:DoubleExp} vanishes doubly exponentially
fast. This means that the first term of \eqref{eq:-9-1} vanishes
as $m\rightarrow\infty$.

In the same way, by utilizing Lemma \ref{lem:chernoff} again, we
can prove that for small enough $\delta$, the second term of \eqref{eq:-9-1}
also vanishes as $m\rightarrow\infty$. This completes the proof of
Lemma \ref{lem:B}.

\section{\label{sec:Proof-of-LemmaB-1}Proof of Lemma \ref{lem:B-1}}

Consider
\begin{align}
 & \Pbb\left(M_{1}=m_{1},Z^{n}=z^{n},M_{0}=m_{0},\mathcal{A}_{3}|\mathcal{C}\right)\nonumber \\
= & 2^{-mR_{t}}\Pbb\left(Z^{n}=z^{n},\mathcal{A}_{3}|U^{n}\left(m_{0}\right),V^{n}\left(m_{0},m_{1}\right)\right)\nonumber \\
\leq & 2^{-mR_{t}}2^{-n\left(H\left(Z|V\right)-\epsilon_{\delta}\right)}\nonumber \\
 & \qquad1\left\{ \left(U^{n}\left(m_{0}\right),V^{n}\left(m_{0},m_{1}\right),z^{n}\right)\in\mathcal{T}_{\delta}^{n}\right\} ,\label{eq:-48}
\end{align}
where \eqref{eq:-48} follows from the property of typical sequence
\cite{Gamal}: for any $\left(v^{n},z^{n}\right)\in\mathcal{T}_{\delta}^{n}$,
\begin{equation}
2^{-n\left(H\left(Z|V\right)+\epsilon_{\delta}\right)}\leq\Pbb\left(Z^{n}=z^{n}|V^{n}=v^{n}\right)\leq2^{-n\left(H\left(Z|V\right)-\epsilon_{\delta}\right)},
\end{equation}
with a term $\epsilon_{\delta}$ that vanishes as $m\to\infty$.

Similarly, we have
\begin{align}
 & \Pbb\left(M_{1}=m_{1},Z^{n}=z^{n},M_{0}=m_{0},\mathcal{A}_{3}|\mathcal{C}\right)\nonumber \\
\geq & 2^{-mR_{t}}2^{-n\left(H\left(Z|V\right)+\epsilon_{\delta}\right)}\nonumber \\
 & \qquad1\left\{ \left(U^{n}\left(m_{0}\right),V^{n}\left(m_{0},m_{1}\right),z^{n}\right)\in\mathcal{T}_{\delta}^{n}\right\} .
\end{align}

Therefore,
\begin{align}
 & \Pbb\left(M_{1}=m_{1},\mathcal{A}_{3}|\mathcal{C}Z^{n}M_{0}\right)\nonumber \\
= & \frac{\Pbb\left(M_{1}=m_{1},Z^{n}=z^{n},M_{0}=m_{0},\mathcal{A}_{3}|\mathcal{C}\right)}{\sum_{m_{1}\in\left[2^{mR_{1}}\right]}\Pbb\left(M_{1}=m_{1},Z^{n}=z^{n},M_{0}=m_{0}|\mathcal{C}\right)}\nonumber \\
\leq & \frac{\Pbb\left(M_{1}=m_{1},Z^{n}=z^{n},M_{0}=m_{0},\mathcal{A}_{3}|\mathcal{C}\right)}{\sum_{m_{1}\in\left[2^{mR_{1}}\right]}\Pbb\left(M_{1}=m_{1},Z^{n}=z^{n},M_{0}=m_{0},\mathcal{A}_{3}|\mathcal{C}\right)}\nonumber \\
\leq & \frac{2^{-m\left(R_{0}+R_{1}\right)}2^{-n\left(H\left(Z|V\right)-\epsilon_{\delta}\right)}}{2^{-m\left(R_{0}+R_{1}\right)}2^{-n\left(H\left(Z|V\right)+\epsilon_{\delta}\right)}}\nonumber \\
 & \times\frac{1\left\{ \left(U^{n}\left(m_{0}\right),V^{n}\left(m_{0},m_{1}\right),z^{n}\right)\in\mathcal{T}_{\delta}^{n}\right\} }{\sum_{m_{1}\in\left[2^{mR_{1}}\right]}1\left\{ \left(U^{n}\left(m_{0}\right),V^{n}\left(m_{0},m_{1}\right),z^{n}\right)\in\mathcal{T}_{\delta}^{n}\right\} }\nonumber \\
\leq & \frac{2^{2n\epsilon_{\delta}}1\left\{ \left(U^{n}\left(m_{0}\right),V^{n}\left(m_{0},m_{1}\right),z^{n}\right)\in\mathcal{T}_{\delta}^{n}\right\} }{\sum_{m_{1}\in\left[2^{mR_{1}}\right]}1\left\{ \left(U^{n}\left(m_{0}\right),V^{n}\left(m_{0},m_{1}\right),z^{n}\right)\in\mathcal{T}_{\delta}^{n}\right\} }\nonumber \\
\leq & 2^{-m\left(R_{c}-2\gamma\epsilon_{\delta}\right)}1\left\{ \left(U^{n}\left(m_{0}\right),V^{n}\left(m_{0},m_{1}\right),z^{n}\right)\in\mathcal{T}_{\delta}^{n}\right\} ,\label{eq:-38}
\end{align}
where \eqref{eq:-38} follows from \eqref{eq:-80}.

\section{\label{sec:Proof-of-Theorem-gaussian}Proof of Converse of Theorem
\ref{thm:gaussian}}

From source-channel coding theorem for Gaussian communication, we
have $\frac{\gamma}{2}\log\left(1+\frac{P}{N_{B}}\right)\geq\frac{1}{2}\log^{+}\frac{N_{S}}{D_{B}}$.
Besides, similar to the discrete memoryless case, if $R_{L}>\frac{1}{2}\log^{+}\frac{N_{S}}{D_{E}}$,
then no matter what scheme Nodes A and B use, the henchman and the
wiretapper can ignore $Z^{n}$ altogether and simply use a point-to-point
rate-distortion code to describe $S^{m}$ within distortion $D_{E}$
(with probability 1). Hence we only need prove

\begin{equation}
R_{L}\leq R_{K}+\frac{\gamma}{2}\log^{+}\left(\frac{1+\nicefrac{P}{N_{B}}}{1+\nicefrac{P}{N_{E}}}\right)+\frac{1}{2}\log^{+}\frac{D_{B}}{D_{E}}.\label{eq:-50}
\end{equation}

To that end, we follow similar steps to the proof of the outer bound
of Theorem \ref{thm:lossy}. Since the admissible region of $\left(R_{K},R_{L},D_{B},D_{E}\right)$
only depends on the margin distribution of the wiretap channel, it
is sufficient to consider the channel to be physically degraded. Furthermore,
observe that the derivation up to \eqref{eq:-5} holds verbatim. Hence
to complete the proof, we only need to prove
\begin{align}
\overline{I}(\mathbf{Y};\mathbf{X}|\mathbf{Z}) & \le\frac{1}{2}\log^{+}\left(\frac{1+\nicefrac{P}{N_{B}}}{1+\nicefrac{P}{N_{E}}}\right),\label{eq:-6}\\
\underline{R_{\mathbf{S}|\mathbf{\hat{S}}}}(D_{E}) & \leq\frac{1}{2}\log^{+}\frac{D_{B}}{D_{E}}.\label{eq:-10}
\end{align}

Actually, for physically degraded Gaussian wiretap channel (as considered
here), \eqref{eq:-6} was proven by Tan \cite[Thm. 5]{Tan}. Next
we will show \eqref{eq:-10} also holds. From Theorem \ref{thm:optimistic},
it is equivalent to showing that upon the two-sided information $\hat{S}^{m}$
(within the distortion $D_{B}$), there exists a source code with
rate $\frac{1}{2}\log^{+}\frac{D_{B}}{D_{E}}$ achieving the distortion
$D_{E}$ with positive probability. The corresponding part of the
proof of Theorem \ref{thm:lossy} requires that the alphabet of the
source or its reproduction is finite, hence it cannot be applied to
the case of continuous alphabets, such as the Gaussian case. From
the assumption, $S^{m}$ is in the balls of center $\hat{S}^{m}$
and radius $\sqrt{mD_{B}}$ with high probability. Hence, we use a
sphere covering lemma to prove \eqref{eq:-10}.
\begin{lem}
\cite{Rogers,Verger-Gaugry} \label{lem:ballcovering}Let $R>1$ and
let $\nu_{R,l}$ be the minimal number of (closed) balls of radius
$1$ which can cover a (closed) ball of radius $R$ in $\mathbb{R}^{l}$.
If $l\geq9$, then we have
\begin{equation}
1<\nu_{R,l}\leq\frac{4eR^{l}l\sqrt{l}}{\ln l\text{\textminus}2}(l\ln l+l\ln(\ln l)+l\ln R+12\ln(144l))
\end{equation}
for all $1<R<\frac{l}{2\ln l}$.
\end{lem}
Observe that for any fixed $R$, $\frac{4eR^{l}l\sqrt{l}}{\ln l\text{\textminus}2}(l\ln l+l\ln(\ln l)+l\ln R+12\ln(144l))=2^{l\left(\log R+o\left(1\right)\right)}$.
Hence from Lemma \ref{lem:ballcovering}, we can easily get that for
large enough $m$, it suffices to cover a ball of radius $\sqrt{mD_{B}}$
using $2^{m\left(\frac{1}{2}\log^{+}\frac{D_{B}}{D_{E}}+o\left(1\right)\right)}$
balls of radius $\sqrt{mD_{E}}$. This implies upon $\hat{S}^{m}$,
there exists a source code with rate $\frac{1}{2}\log^{+}\frac{D_{B}}{D_{E}}$
achieving the distortion $D_{E}$ (with high probability). Hence $\underline{R_{\mathbf{S}|\mathbf{\hat{S}}}}(D_{E})\leq\frac{1}{2}\log^{+}\frac{D_{B}}{D_{E}}$
holds. This completes the proof of the converse part.

\section{\label{sec:Proof-of-Theorem-gaussian-1}Proof of Achievability of
Theorem \ref{thm:gaussian}}

The proof of $\mathcal{R}_{\textrm{sep}}^{(i)}$ of Theorem \ref{thm:lossy}
requires that the alphabets of the channel input and output, and the
alphabets of the source and its reproduction, are all finite, hence
it cannot be applied to the Gaussian case directly. Now, we prove
the achievability part for the Gaussian case by exploiting the techniques
of $\mathsf{d}-$tilted information, weak typicality, and discretization.

\subsection{Weak typicality and $\mathsf{d}-$tilted information }

Before proving the achievability part of Theorem \ref{thm:gaussian},
we need introduce some preliminaries. To extend $\mathcal{R}_{\textrm{sep}}^{(i)}$
of Theorem \ref{thm:lossy} to the Gaussian case, we need to replace
the strong typicality \eqref{eq:typicality} with weak typicality.
The $\delta$-typical set and the $\delta$-jointly-typical set are
defined according to the notion of weak typicality\footnote{For ease in writing, we replace $\delta$ with $\frac{\delta}{2}\log e$
in the definitions of $\delta$-typical set and the $\delta$-jointly-typical
set.}, see \cite{Cover}:

\begin{align*}
 & \Tcal_{\delta}^{n}(X)\triangleq\\
 & \Bigl\{ x^{n}\in\Xcal^{n}:\Bigl|-\frac{1}{n}\log\prod_{i=1}^{n}f_{X}\left(x_{i}\right)-h\left(X\right)\Bigr|\leq\frac{\delta}{2}\log e\Bigr\},
\end{align*}
and
\begin{align*}
 & \Tcal_{\delta}^{n}(X,Z)\triangleq\\
 & \Bigl\{\left(x^{n},z^{n}\right)\in\Xcal^{n}\times\Zcal^{n}:\\
 & \Bigl|-\frac{1}{n}\log\prod_{i=1}^{n}f_{X}\left(x_{i}\right)-h\left(X\right)\Bigr|\leq\frac{\delta}{2}\log e,\\
 & \Bigl|-\frac{1}{n}\log\prod_{i=1}^{n}f_{Z}\left(z_{i}\right)-h\left(Z\right)\Bigr|\leq\frac{\delta}{2}\log e,\\
 & \Bigl|-\frac{1}{n}\log\prod_{i=1}^{n}f_{XZ}\left(x_{i},z_{i}\right)-h\left(XZ\right)\Bigr|\leq\frac{\delta}{2}\log e\Bigr\},
\end{align*}
For jointly Gaussian variables $X$ and $Z$, where $Z=X+U$ and $U$
is independent of $X$, the $\delta$-typical set and the $\delta$-jointly-typical
set become

\begin{equation}
\Tcal_{\delta}^{n}(X)\triangleq\Bigl\{ x^{n}\in\mathbb{R}^{n}:\Bigl|\frac{\left\Vert x^{n}\right\Vert ^{2}}{nN_{X}}-1\Bigr|\leq\delta\Bigr\},
\end{equation}
and
\begin{align}
\Tcal_{\delta}^{n}(X,Z)\triangleq\Bigl\{ & \left(x^{n},z^{n}\right)\in\mathbb{R}^{2n}:\Bigl|\frac{\left\Vert x^{n}\right\Vert ^{2}}{nN_{X}}-1\Bigr|\leq\delta,\nonumber \\
 & \Bigl|\frac{\left\Vert z^{n}\right\Vert ^{2}}{nN_{Z}}-1\Bigr|\leq\delta,\nonumber \\
 & \Bigl|\frac{\left\Vert x^{n}\right\Vert ^{2}}{nN_{X}}+\frac{\left\Vert z^{n}-x^{n}\right\Vert ^{2}}{nN_{U}}-2\Bigr|\leq\delta\Bigr\},\label{eq:-58}
\end{align}
respectively, where $\left\Vert x^{n}\right\Vert =\sqrt{\sum_{i=1}^{n}x_{i}^{2}}$
denotes Euclidean norm of $x^{n}$, and $N_{Z},N_{X}$ and $N_{U}$
denote the variances of $Z$, $X$ and $U$.

Similar to \eqref{eq:rd}, the rate-distortion function for continuous
source is defined as
\begin{align}
 & R_{S}(D)=\min_{f_{\check{S}|S}:\mathbb{E}d(S,\check{S})\leq D}I(S;\check{S}).\label{eq:rd-1}
\end{align}

We impose the following basic restrictions on pdf $f_{S}$ and the
distortion measure $d$:
\begin{itemize}
\item Restriction 1: \label{item:a} $R_{S}(D)$ is finite for some $D$,
i.e. $D_{\min}<\infty$, where
\begin{equation}
D_{\min}=\inf\left\{ D\colon~R_{S}(D)<\infty\right\} ;\label{eq:dmin}
\end{equation}
\item Restriction 2: \label{item:b} The minimum in \eqref{eq:rd-1} is
achieved by a pdf $f_{\check{S}^{\star}|S}$, which is unique up to
$f_{\check{S}^{\star}S}$-null sets, that is, if $g_{\check{S}^{\star}|S}$
is another pdf achieving the minimum in \eqref{eq:rd-1}, then $f_{\check{S}^{\star}|S}=g_{\check{S}^{\star}|S}$,\LyXThinSpace{}
$f_{\check{S}^{\star}S}$-almost everywhere.
\end{itemize}
\begin{defn}[$\mathsf{d}-$tilted information \cite{Kostina}]
\label{defn:id} For $D>D_{\min}$, the $\mathsf{d}-$tilted information
in $s$ is defined as
\begin{equation}
\jmath_{S}(s,D)\triangleq\log\frac{1}{\mathbb{E}\bigl[\exp\bigl(\lambda^{\star}D-\lambda^{\star}d(s,\check{S}^{\star})\bigr)\bigr]},\label{eq:idball}
\end{equation}
where the expectation is with respect to $f_{\check{S}^{\star}}$,
i.e. the unconditional distribution of the reproduction random variable
that achieves the minimum in \eqref{eq:rd-1}, and
\begin{equation}
\lambda^{\star}=-R_{S}^{\prime}(D).\label{eq:lambdastar}
\end{equation}
\end{defn}
The following properties of $\mathsf{d}-$tilted information, proven
in \cite{Csiszar74}, are used in the sequel.
\begin{align}
 & \jmath_{S}(s,D)=\imath_{S;\check{S}^{\star}}(s;\check{s})+\lambda^{\star}d(s,\check{s})-\lambda^{\star}D\label{eq:jddensity}\\
 & \mathbb{E}[\jmath_{S}(s,D)]=R_{S}(D)\label{eq:Ejd}\\
 & \mathbb{E}[\exp\left(\lambda^{\star}D-\lambda^{\star}d(S,\check{s})+\jmath_{S}(S,D)\right)]\leq1\label{eq:csiszar}
\end{align}
where \eqref{eq:jddensity} holds for $f_{\check{S}^{\star}}$-almost
every $\check{s}$, while \eqref{eq:csiszar} holds for all $\check{s}\in\check{\mathcal{S}}$,
and
\begin{equation}
\imath_{S;\check{S}}(s;\check{s})=\log\frac{f_{\check{S}|S}(\check{s}|s)}{f_{\check{S}}(\check{s})}\label{eq:isz}
\end{equation}
denotes the information density of the joint distribution $f_{S\check{S}}$
at $(s,\check{s})$ (c.f. \eqref{eq:infdens}).

Similarly, we can define conditional $\mathsf{d}-$tilted information
and prove the corresponding properties. Similar to \eqref{eq:rd-si},
the conditional rate-distortion function for continuous source is
defined as

\noindent
\begin{align}
 & R_{S|\hat{S}}(D)=\min_{f_{\check{S}|S\hat{S}}:\mathbb{E}d(S,\check{S})\leq D}I(S;\check{S}|\hat{S}).\label{eq:-64}
\end{align}

\noindent Now we can establish the following lemma, the proof of which
is given in Appendix \ref{sec:Proof-of-Lemma-rd-si-equi}.
\begin{lem}
\noindent \label{lem:rd-si-equivalence} The minimization in \eqref{eq:-64}
can be divided into two optimization subproblems:
\begin{equation}
R_{S|\hat{S}}(D)=\min_{b(\hat{s}):\mathbb{E}_{\hat{S}}\bigl[b(\hat{S})\bigr]\leq D}\mathbb{E}_{\hat{S}}R_{S|\hat{S}=\hat{s}}(b(\hat{S})),\label{eq:-65}
\end{equation}
where
\begin{equation}
R_{S|\hat{S}=\hat{s}}(\beta)=\min_{f_{\check{S}|S,\hat{S}=\hat{s}}:\mathbb{E}\bigl[d(S,\check{S})|\hat{S}=\hat{s}\bigr]\leq\beta}I(S;\check{S}|\hat{S}=\hat{s})\label{eq:-26}
\end{equation}
denotes the rate-distortion function of source $S$ under condition
that $\hat{S}=\hat{s}$. Moreover, assume $b^{\star}\left(\hat{s}\right)$
achieves the minimum in \eqref{eq:-65}, and $f_{\check{S}^{\star}|S,\hat{S}=\hat{s}}$
achieves the minimum in \eqref{eq:-26} with $\beta=b^{\star}\left(\hat{s}\right)$,
then $f_{\check{S}^{\star}|S,\hat{S}}$ minimizes \eqref{eq:-64}
as well, i.e., $f_{\check{S}^{\star}|S,\hat{S}}$ is a solution to
the minimization in \eqref{eq:-64}.
\end{lem}
Similar to Restrictions 1 and 2, for conditional rate-distortion function,
we impose the following basic restrictions on $f_{S\hat{S}}$ and
the distortion measure:
\begin{itemize}
\item Restriction 3: \label{item:a-1} For all $\hat{s}\in\hat{\mathcal{S}}$,
$R_{S|\hat{S}=\hat{s}}(\beta)$ is finite for some $\beta$, i.e.
$\beta_{\min}\left(\hat{s}\right)<\infty$, where
\begin{equation}
\beta_{\min}\left(\hat{s}\right)=\inf\left\{ \beta\colon~R_{S|\hat{S}=\hat{s}}(\beta)<\infty\right\} ;\label{eq:dmin-1}
\end{equation}
\item Restriction 4: \label{item:b-1} For all $\hat{s}\in\hat{\mathcal{S}}$,
the minimum in \eqref{eq:-26} is achieved by a pdf $f_{\check{S}^{\star}|S,\hat{S}=\hat{s}}$
which is unique up to $f_{\check{S}^{\star}S|\hat{S}=\hat{s}}$-null
sets, that is, if $g_{\check{S}^{\star}|S,\hat{S}=\hat{s}}$ is another
pdf achieving the minimum in \eqref{eq:rd}, then $f_{\check{S}^{\star}|S,\hat{S}=\hat{s}}=g_{\check{S}^{\star}|S,\hat{S}=\hat{s}}$,
$f_{\check{S}^{\star}S|\hat{S}=\hat{s}}$-almost everywhere;
\item Restriction 5: \label{item:a-1-1} $R_{S|\hat{S}}(D)$ is strictly
decreasing in $\left(D_{\min},D_{\textrm{max}}\right)$ where
\begin{equation}
D_{\min}=\inf\left\{ D\colon~R_{S|\hat{S}}(D)<\infty\right\} \label{eq:dmin-2}
\end{equation}
and
\begin{equation}
D_{\textrm{max}}=\inf\left\{ D\colon~R_{S|\hat{S}}(D)=0\right\} .\label{eq:dmin-1-1}
\end{equation}
\end{itemize}
Assume $b^{\star}\left(\hat{s}\right)$ achieves the minimum in \eqref{eq:-65},
and $f_{\check{S}^{\star}|S,\hat{S}=\hat{s}}$ achieves the minimum
in \eqref{eq:-26} with $\beta=b^{\star}\left(\hat{s}\right)$, as
assumed in Lemma \ref{lem:rd-si-equivalence}, then the following
lemma holds. The proof is given in Appendix \ref{sec:Proof-of-Lemma-rd-si-equi-1}.
\begin{lem}
\noindent \label{lem:properties}
\begin{align}
 & R_{S|\hat{S}}(D)=\mathbb{E}_{\hat{S}}R_{S|\hat{S}=\hat{s}}(b^{\star}\left(\hat{s}\right)),\label{eq:ERs_shat}\\
 & \mathbb{E}_{\hat{S}}\bigl[b^{\star}(\hat{S})\bigr]=D,\label{eq:expected-d-star}\\
 & \mathbb{E}\bigl[d(S,\check{S}^{\star})|\hat{S}=\hat{s}\bigr]=b^{\star}\left(\hat{s}\right),\label{eq:d-star}\\
 & R_{S|\hat{S}=\hat{s}}^{\prime}(b^{\star}\left(\hat{s}\right))=-\lambda,\forall\hat{s}\in\hat{\mathcal{S}}\label{eq:lambdastar-1-1-1}
\end{align}
for some constant $\lambda\geq0$.
\end{lem}
\begin{defn}[Conditional $\mathsf{d}-$tilted information]
\label{defn:id-1} For $b^{\star}\left(\hat{s}\right)>\beta_{\min}\left(\hat{s}\right)$,
the conditional $\mathsf{d}-$tilted information in $s$ under condition
$\hat{S}=\hat{s}$ is defined as
\begin{align}
 & \jmath_{S|\hat{S}=\hat{s}}(s,b^{\star}\left(\hat{s}\right))\nonumber \\
\triangleq & \log\frac{1}{\mathbb{E}_{\check{S}^{\star}|\hat{S}=\hat{s}}\left[\exp\left(\lambda^{\star}\left(\hat{s}\right)b^{\star}\left(\hat{s}\right)-\lambda^{\star}\left(\hat{s}\right)d(s,\check{S}^{\star})\right)\right]},
\end{align}
where the expectation is with respect to $f_{\check{S}^{\star}|\hat{S}=\hat{s}}$,
and
\begin{equation}
\lambda^{\star}\left(\hat{s}\right)=-R_{S|\hat{S}=\hat{s}}^{\prime}(b^{\star}\left(\hat{s}\right)).\label{eq:lambdastar-1}
\end{equation}

Combining \eqref{eq:lambdastar-1-1-1} and \eqref{eq:lambdastar-1},
we have for all $\hat{s}$,
\begin{equation}
\lambda^{\star}\left(\hat{s}\right)=\lambda.\label{eq:-7-2}
\end{equation}
Obviously, the conditional $\mathsf{d}-$tilted information for the
distribution $f_{S|\hat{S}=\hat{s}}$ and distortion $b^{\star}\left(\hat{s}\right)$
can be considered as a special unconditional $\mathsf{d}-$tilted
information for $f_{S^{\prime}}$ and $D^{\prime}$ such that $f_{S^{\prime}}=f_{S|\hat{S}=\hat{s}}$
and $D^{\prime}=b^{\star}\left(\hat{s}\right)$. Hence \eqref{eq:jddensity}-\eqref{eq:csiszar}
still hold for $\jmath_{S|\hat{S}=\hat{s}}(s,b^{\star}\left(\hat{s}\right))$,
i.e.,
\begin{align}
 & \jmath_{S|\hat{S}=\hat{s}}(s,b^{\star}\left(\hat{s}\right))\nonumber \\
 & =\imath_{S;\check{S}^{\star}|\hat{S}=\hat{s}}(s;\check{s})+\lambda^{\star}\left(\hat{s}\right)d(s,\check{s})-\lambda^{\star}\left(\hat{s}\right)b^{\star}\left(\hat{s}\right),\label{eq:jddensity-1}\\
 & \mathbb{E}_{S|\hat{S}=\hat{s}}\bigl[\jmath_{S|\hat{S}=\hat{s}}(S,b^{\star}\left(\hat{s}\right))\bigr]=R_{S|\hat{S}=\hat{s}}(b^{\star}\left(\hat{s}\right)),\label{eq:Ejd-1}\\
 & \mathbb{E}_{S|\hat{S}=\hat{s}}\bigl[\exp\Bigl(\lambda^{\star}\left(\hat{s}\right)b^{\star}\left(\hat{s}\right)-\lambda^{\star}\left(\hat{s}\right)d(S,\check{s})+\jmath_{S|\hat{S}=\hat{s}}(S,b^{\star}\left(\hat{s}\right))\Bigr)\bigr]\nonumber \\
 & \leq1,\label{eq:csiszar-1}
\end{align}
where \eqref{eq:jddensity-1} holds for $f_{\check{S}^{\star}|\hat{S}=\hat{s}}$-almost
every $\check{s}$, while \eqref{eq:csiszar-1} holds for all $\check{s}\in\check{\mathcal{S}}$,
and
\begin{equation}
\imath_{S;\check{S}^{\star}|\hat{S}=\hat{s}}(s;\check{s})=\log\frac{f_{\check{S}|S,\hat{S}=\hat{s}}(\check{s}|s)}{f_{\check{S}|\hat{S}=\hat{s}}(\check{s})}\label{eq:isz-1}
\end{equation}
denotes the conditional information density of the joint distribution
$f_{S\check{S}|\hat{S}=\hat{s}}$ at $(s,\check{s})$.
\end{defn}

\subsection{\label{sub:Interlude-1}Subproblem: lossy compression of a codeword
drawn uniformly from a random codebook with side information }

Next we return to proving the achievability part of Theorem \ref{thm:gaussian}.
We follow similar steps to that of the discrete memoryless case. Consider
the subproblem described in Appendix \ref{sub:Interlude} for the
Gaussian source-channel case. Then we can prove that Theorem \ref{thm:lossy2}
still holds. To show the achievability for the Gaussian case, it is
sufficient to consider the case of $U=\emptyset,V=X$.
\begin{thm}
\label{thm:Thm-lossy-1} Theorem \ref{thm:lossy2} with $U=\emptyset,V=X$
holds for Gaussian communication case.
\end{thm}
\begin{IEEEproof}[Proof of Theorem \ref{thm:Thm-lossy-1}]
Since $d_{B}\left(x,y\right)=d_{E}\left(x,y\right)=\left(x-y\right)^{2}$,
in the following, we use $d\left(x,y\right)$ to denote both of them.
Furthermore, for the memoryless Gaussian source $S$, set $\hat{S}$
to be a jointly Gaussian variable with $S$ such that $S=\hat{S}+W$,
$\hat{S}$ and $W$ are independent and $\mathbb{E}\left[W^{2}\right]=D_{B}$.
It is easy to verify that $f_{S\hat{S}}$ and $d(s,\check{s})$ satisfy
Restrictions 1-5.

Next we follow similar steps to the proof of $\mathcal{R}_{\textrm{sep}}^{(i)}$
of Theorem \ref{thm:lossy2}, except for some modifications. First,
we need to replace the strong typicality \eqref{eq:typicality} with
the weak typicality, since strong typicality only works for the variables
with finite alphabets. Second, we need re-define $\mathcal{A}_{1}$
as
\begin{align}
\mathcal{A}_{1}\triangleq & \Bigl\{ S^{m}\in\mathcal{T}_{\delta}^{m},\frac{1}{m}\sum_{i=1}^{m}\jmath_{S}(S_{i},D_{E})\geq R_{S}(D_{E})-\delta,\nonumber \\
 & \frac{1}{m}\sum_{i=1}^{m}\jmath_{S|\hat{S}=\hat{S}_{i}}(S_{i},b^{\star}(\hat{S}_{i}))\geq R_{S|\hat{S}}(D_{E})-\delta,\nonumber \\
 & \frac{1}{m}\sum_{i=1}^{m}b^{\star}(\hat{S}_{i})\geq D_{E}-\delta\Bigr\}
\end{align}
for $\delta>0$. $\mathcal{A}_{2},\mathcal{A}_{3},\mathcal{A}_{123},\mathcal{A}_{23},\mathcal{B}$
remain the same but restricted to the case with $U=\emptyset,V=X$.
Then we have following lemmas, the proofs of which are given in Appendices
\ref{sec:Proof-of-LemmaB-Gaussian} and \ref{sec:Proof-of-LemmaA123-Gaussian},
respectively.
\begin{lem}
\label{lem:B-Gaussian} For Gaussian wiretap channel, $\lim_{m\to\infty}\Pbb\left(\mathcal{B}\right)=1$.
\end{lem}
\begin{lem}
\noindent \label{lem:A123-Gaussian} For any stationary memoryless
source $f_{S\hat{S}}$ with general distortion function $d(s,\check{s})$
satisfying Restrictions 1-5 (not restricted to the quadratic Gaussian
case), $\lim_{m\to\infty}\Pbb\left(\mathcal{A}_{123}\right)=1$.
\end{lem}
Re-define $\eta_{m,k,\check{s}^{m}}$ as
\begin{align}
 & \eta_{m,k,\check{s}^{m}}\left(D_{E}\right)\triangleq\sum_{s^{m}\in\Scal^{m}}\prod_{i=1}^{m}P_{S|\hat{S}}\left(s_{i}|\hat{S}_{i}(j\left(m,k\right))\right)\nonumber \\
 & \times1\Bigl\{ d(s^{m},\check{s}^{m})\leq D_{E},s^{m}\in\mathcal{T}_{\delta}^{m},\nonumber \\
 & \qquad\frac{1}{m}\sum_{i=1}^{m}\jmath_{S}(s_{i},D_{E})\geq R_{S}(D_{E})-\delta,\nonumber \\
 & \qquad\frac{1}{m}\sum_{i=1}^{m}\jmath_{S|\hat{S}=\hat{s}_{i}}(s_{i},b^{\star}(\hat{s}_{i}))\geq R_{S|\hat{S}}(D_{E})-\delta,\nonumber \\
 & \qquad\frac{1}{m}\sum_{i=1}^{m}b^{\star}(\hat{s}_{i})\geq D_{E}-\delta\Bigr\}.
\end{align}
Then the derivation up to \eqref{eq:-4} still holds, i.e.,
\begin{align}
 & \mathbb{P}_{\mathcal{C}Z^{n}M_{0}}\left[\max_{R_{L}codes}\mathbb{P}\Bigl[d(S^{m},\check{S}^{m})\le D_{E}\Bigr]>\tau\right]\nonumber \\
 & \leq\mathbb{P}_{\mathcal{C}Z^{n}M_{0}}\Big[\max_{\check{s}^{m}\in\mathbb{R}{}^{m},m_{1}^{\prime}\in[2^{mR_{1}^{\prime}}]}\sum_{k=1}^{2^{mR_{K}}}\sum_{m_{c}=1}^{2^{mR_{c}}}\nonumber \\
 & \qquad\qquad\qquad\qquad\eta_{m,k,\check{s}^{m}}\left(D_{E}\right)>\tau'2^{-m\lambda'}\Big]+\epsilon'_{m},\label{eq:-69}
\end{align}
where $\eta_{m,k,\check{s}^{m}}\left(D_{E}\right)$ is defined in
\eqref{eq:-20}. Now to take the maximizing operation out of the probability
in \eqref{eq:-69}, we use a discretization technology that is also
used in proof of Lemma \ref{lem:B-Gaussian}. Quantize $\check{s}^{m}$
as $[\check{s}]^{m}\in\mathcal{N}^{m}$ by the following manner, where
$\mathcal{N}$ is defined in \eqref{eq:-68}.
\begin{equation}
[\check{s}_{i}]=\Delta\cdot\mathrm{Round}\left(\frac{\check{s}_{i}}{\Delta}\right).
\end{equation}
Then similar to \eqref{eq:-55}, we can prove
\begin{align}
\frac{\left\Vert [\check{s}]^{m}-s^{m}\right\Vert ^{2}}{m}-\epsilon_{\Delta} & \leq\frac{\left\Vert \check{s}^{m}-s^{m}\right\Vert ^{2}}{m}\nonumber \\
 & \leq\frac{\left\Vert [\check{s}]^{m}-s^{m}\right\Vert ^{2}}{m}+\epsilon_{\Delta}
\end{align}
for some $\epsilon_{\Delta}$ that vanishes as $\Delta\rightarrow0$.
Hence
\begin{equation}
\eta_{m,k,\check{s}^{m}}\left(D_{E}\right)\leq\eta_{m,k,[\check{s}]^{m}}\left(D_{E}+\epsilon_{\Delta}\right).\label{eq:-70}
\end{equation}

Define $\mathcal{F}^{m}\triangleq\left\{ [\check{s}]^{m}\in\mathcal{N}^{m}:\left\Vert [\check{s}]^{m}\right\Vert ^{2}\leq mN_{S}\left(1+\delta\right)\right\} $,
then we have
\begin{align}
 & \mathbb{P}_{\mathcal{C}Z^{n}M_{0}}\left[\max_{R_{L}codes}\mathbb{P}\left[d\left(S^{m},\check{S}^{m}\right)\le D_{E}\right]>\tau\right]\nonumber \\
 & \leq\mathbb{P}_{\mathcal{C}Z^{n}M_{0}}\Big[\max_{[\check{s}]^{m}\in\mathcal{N}^{m},m_{1}^{\prime}\in[2^{mR_{1}^{\prime}}]}\sum_{k=1}^{2^{mR_{K}}}\sum_{m_{c}=1}^{2^{mR_{c}}}\nonumber \\
 & \qquad\qquad\eta_{m,k,[\check{s}]^{m}}\left(D_{E}+\epsilon_{\Delta}\right)>\tau'2^{-m\lambda'}\Big]+\epsilon'_{m}\label{eq:-71}\\
 & =\mathbb{P}_{\mathcal{C}Z^{n}M_{0}}\Big[\max_{[\check{s}]^{m}\in\mathcal{F}^{m},m_{1}^{\prime}\in[2^{mR_{1}^{\prime}}]}\sum_{k=1}^{2^{mR_{K}}}\sum_{m_{c}=1}^{2^{mR_{c}}}\nonumber \\
 & \qquad\qquad\eta_{m,k,[\check{s}]^{m}}\left(D_{E}+\epsilon_{\Delta}\right)>\tau'2^{-m\lambda'}\Big]+\epsilon'_{m}\label{eq:-72}\\
 & \leq2^{mR_{1}^{\prime}}\left|\mathcal{F}^{m}\right|\max_{[\check{s}]^{m}\in\mathcal{F}^{m},m_{1}^{\prime}\in[2^{mR_{1}^{\prime}}]}\nonumber \\
 & \mathbb{P}_{\mathcal{C}Z^{n}M_{0}}\Big[\sum_{k=1}^{2^{mR_{K}}}\sum_{m_{c}=1}^{2^{mR_{c}}}\eta_{m,k,[\check{s}]^{m}}\left(D_{E}+\epsilon_{\Delta}\right)>\tau'2^{-m\lambda'}\Big]+\epsilon'_{m},\label{eq:-37-1}
\end{align}
where \eqref{eq:-71} follows from \eqref{eq:-69} and \eqref{eq:-70},
and \eqref{eq:-72} follows from that $s^{m}$ belongs in the $m$-ball
with radius $mN_{S}\left(1+\delta\right)$ (since $s^{m}\in\mathcal{T}_{\delta}^{m}$),
hence it suffices to use the points in the ball as reconstructions.

Furthermore, similar to \eqref{eq:-59}, we have
\begin{align}
\left|\mathcal{F}^{m}\right| & \leq2^{m\left(\frac{1}{2}\log m+o\left(\log m\right)\right)}.
\end{align}
Hence if we can show that the probability in \eqref{eq:-37-1} decays
doubly exponentially fast with $m$, then the proof will be complete.
To that end, we first introduce the following lemmas which are related
to Lemmas \ref{lem:typebound} and \ref{lem:typebound-rd-si}. The
proof of Lemma \ref{lem:SPO-si} is given in Appendix \ref{sec:Proof-of-Lemma-SPO-si}.
\begin{lem}
\label{lem:SPO}\cite[Thm. 4]{Elkayam} If $f_{S}$ and $d(s,\check{s})$
satisfy Restrictions 1 and 2, and $S^{m}$ is i.i.d. according to
$f_{S}$, then for any $\hat{s}^{m}\in\hat{\mathcal{S}}{}^{m}$,
\begin{align}
 & \Pbb\bigl[d(S^{m},\check{s}^{m})\leq D_{E},\frac{1}{m}\sum_{i=1}^{m}\jmath_{S}(S_{i},D_{E})\geq R_{S}(D_{E})-\delta\bigr]\nonumber \\
 & \leq2^{-m(R_{S}(D_{E})-\epsilon_{m,\delta})},
\end{align}
where $\epsilon_{m,\delta}$ is a term that vanishes as $\delta\rightarrow0$
and $m\rightarrow\infty$.
\end{lem}
\begin{lem}
\label{lem:SPO-si} Fix $f_{S\hat{S}}$ and $\hat{s}^{m}\in\hat{\mathcal{S}}{}^{m}$.
If $f_{S\hat{S}}$ and $d(s,\check{s})$ satisfy Restrictions 1-5,
and $S^{m}$ is distributed according to $\prod_{i=1}^{m}f_{S|\hat{S}=\hat{s}_{i}}$,
then for any $\check{s}^{m}$,
\begin{align}
 & \Pbb\Bigl[d(S^{m},\check{s}^{m})\leq D_{E},\nonumber \\
 & \qquad\frac{1}{m}\sum_{i=1}^{m}\jmath_{S|\hat{S}=\hat{s}_{i}}(S_{i},b^{\star}(\hat{s}_{i}))\geq R_{S|\hat{S}}(D_{E})-\delta,\nonumber \\
 & \qquad\frac{1}{m}\sum_{i=1}^{m}b^{\star}(\hat{s}_{i})\geq D_{E}-\delta|\hat{S}^{m}=\hat{s}^{m}\Bigr]\nonumber \\
 & \leq2^{-m(R_{S|\hat{S}}(D_{E})-\epsilon_{m,\delta})},
\end{align}
where $\epsilon_{m,\delta}$ is a term that vanishes as $\delta\rightarrow0$
and $m\rightarrow\infty$.
\end{lem}
Note that Lemmas \ref{lem:SPO} and \ref{lem:SPO-si} hold for any
stationary memoryless source with general distortion measure (not
restricted to the Gaussian source with quadratic distortion), and
hence they can be used to replace Lemmas \ref{lem:typebound} and
\ref{lem:typebound-rd-si} in the proof of Theorem \ref{thm:lossy2}.

Apply Lemmas \ref{lem:chernoff-1}, \ref{lem:SPO} and \ref{lem:SPO-si},
then we have the probability in \eqref{eq:-37-1} decays doubly exponentially
fast with $m$. This completes the proof of Theorem \ref{thm:Thm-lossy-1}.
\end{IEEEproof}

\subsection{\label{sub:Likelihood-encoder-1-1}Likelihood encoder}

All the derivations in Appendix \ref{sub:Likelihood-encoder} still
hold for the source and channel with continuous alphabets. Hence the
inner bound $\mathcal{R}_{\textrm{sep}}^{(i)}$ (with $U=\emptyset,V=X$)
still holds for the Gaussian communication. It is easy to verify $\mathcal{R}_{\textrm{sep}}^{(i)}$
is just the region of Theorem \ref{thm:gaussian}, hence the achievability
of Theorem \ref{thm:gaussian} holds.

\section{\label{sec:Proof-of-Lemma-rd-si-equi}Proof of Lemma \ref{lem:rd-si-equivalence}}

Proof of ``$\geq$''. Assume $f_{\check{S}'|S\hat{S}}$ achieves
the minimum in \eqref{eq:-64} and define $b'\left(\hat{s}\right)=\mathbb{E}\left[d(S,\check{S}')|\hat{S}=\hat{s}\right]$,
then $\mathbb{E}_{\hat{S}'}\left[b'\left(\hat{S}'\right)\right]\leq D$,
and
\begin{align}
R_{S|\hat{S}}(D) & =I\left(S;\check{S}'|\hat{S}\right)\nonumber \\
 & =\int d\hat{s}f\left(\hat{s}\right)I\left(S;\check{S}'|\hat{S}=\hat{s}\right)\nonumber \\
 & \geq\int d\hat{s}f\left(\hat{s}\right)\nonumber \\
 & \qquad\times\min_{f_{\check{S}|S,\hat{S}=\hat{s}}:\mathbb{E}\left[d(S,\check{S})|\hat{S}=\hat{s}\right]\leq b'\left(\hat{s}\right)}I\left(S;\check{S}|\hat{S}=\hat{s}\right)\nonumber \\
 & =\mathbb{E}_{\hat{S}}R_{S|\hat{S}=\hat{s}}(b'\left(\hat{s}\right))\nonumber \\
 & \geq\min_{b\left(\hat{s}\right):\mathbb{E}_{\hat{S}}\left[b\left(\hat{S}\right)\right]\leq D}\mathbb{E}_{\hat{S}}R_{S|\hat{S}=\hat{s}}(b\left(\hat{s}\right)).\label{eq:-66}
\end{align}

Proof of ``$\leq$''. Observe that
\begin{align}
 & \min_{b\left(\hat{s}\right):\mathbb{E}_{\hat{S}}\left[b\left(\hat{S}\right)\right]\leq D}\mathbb{E}_{\hat{S}}R_{S|\hat{S}=\hat{s}}(b\left(\hat{s}\right))\nonumber \\
 & =\min_{b\left(\hat{s}\right):\mathbb{E}_{\hat{S}}\left[b\left(\hat{S}\right)\right]\leq D}\int d\hat{s}f\left(\hat{s}\right)\nonumber \\
 & \qquad\times\min_{f_{\check{S}|S,\hat{S}=\hat{s}}:\mathbb{E}\left[d\left(S,\check{S}\right)|\hat{S}=\hat{s}\right]\leq b\left(\hat{s}\right)}I\left(S;\check{S}|\hat{S}=\hat{s}\right).\label{eq:-75}
\end{align}
Assume $b''\left(\hat{s}\right)$ and $f_{\check{S}''|S,\hat{S}=\hat{s}}$
achieves the minimum in \eqref{eq:-75}. Then we have
\begin{equation}
\mathbb{E}d\left(S,\check{S}''\right)=\mathbb{E}_{\hat{S}}\mathbb{E}\left[d\left(S,\check{S}''\right)|\hat{S}=\hat{s}\right]\leq\mathbb{E}_{\hat{S}}b''\left(\hat{S}\right)\leq D,\label{eq:-21}
\end{equation}
and
\begin{align}
 & \min_{b\left(\hat{s}\right):\mathbb{E}_{\hat{S}}\left[b\left(\hat{S}\right)\right]\leq D}\mathbb{E}_{\hat{S}}R_{S|\hat{S}=\hat{s}}(b\left(\hat{s}\right))\nonumber \\
 & =\int d\hat{s}f\left(\hat{s}\right)I\left(S;\check{S}''|\hat{S}=\hat{s}\right)\\
 & =I\left(S;\check{S}''|\hat{S}\right)\\
 & \geq\min_{f_{\check{S}|S\hat{S}}:\mathbb{E}d\left(S,\check{S}\right)\leq D}I\left(S;\check{S}|\hat{S}\right)\label{eq:-74}\\
 & =R_{S|\hat{S}}(D),\label{eq:-76}
\end{align}
where \eqref{eq:-74} follows from \eqref{eq:-21}.

Combining \eqref{eq:-66} and \eqref{eq:-76} gives us
\begin{equation}
R_{S|\hat{S}}(D)=\min_{b\left(\hat{s}\right):\mathbb{E}_{\hat{S}}\left[b\left(\hat{S}\right)\right]\leq D}\mathbb{E}_{\hat{S}}R_{S|\hat{S}=\hat{s}}(b\left(\hat{s}\right)).\label{eq:-79}
\end{equation}
Furthermore, it is easy to verify that $f_{\check{S}''|S,\hat{S}=\hat{s}}$
is also a solution to the minimization in \eqref{eq:-64}. Hence the
lemma holds.

\section{\label{sec:Proof-of-Lemma-rd-si-equi-1}Proof of Lemma \ref{lem:properties}}

Equation \eqref{eq:ERs_shat} follows straightforwardly by the assumption
that $b^{\star}\left(\hat{s}\right)$ achieves the minimum in \eqref{eq:-65}.

Furthermore, from the assumptions, we have
\begin{equation}
\mathbb{E}_{\hat{S}}\bigl[b^{\star}(\hat{S})\bigr]\leq D,\label{eq:expected-d-star-1}
\end{equation}
and
\begin{equation}
\mathbb{E}\bigl[d(S,\check{S}^{\star})|\hat{S}=\hat{s}\bigr]\leq b^{\star}\left(\hat{s}\right).\label{eq:d-star-1}
\end{equation}
Then we have
\begin{align}
\mathbb{E}\bigl[d(S,\check{S}^{\star})\bigr] & =\mathbb{E}_{\hat{S}}\mathbb{E}\bigl[d(S,\check{S}^{\star})|\hat{S}=\hat{s}\bigr]\label{eq:-110}\\
 & \leq\mathbb{E}_{\hat{S}}b^{\star}\left(\hat{s}\right)\label{eq:-111}\\
 & \leq D,\label{eq:-112}
\end{align}
where \eqref{eq:-111} follows from \eqref{eq:d-star-1}, and \eqref{eq:-112}
follows from \eqref{eq:expected-d-star-1}.

Since $R_{S|\hat{S}}(D)$ is strictly decreasing (Restriction 5),
$R_{S|\hat{S}}(D)$ is achieved by some $f_{\check{S}|S,\hat{S}}$
only if $\mathbb{E}\bigl[d(S,\check{S})\bigr]=D.$ On the other hand,
Lemma \ref{lem:rd-si-equivalence} tells us $f_{\check{S}^{\star}|S,\hat{S}}$
achieves $R_{S|\hat{S}}(D)$. Hence
\begin{equation}
\mathbb{E}\bigl[d(S,\check{S}^{\star})\bigr]=D.\label{eq:-109}
\end{equation}

Combining \eqref{eq:-112} and \eqref{eq:-109} gives us
\begin{align}
 & \mathbb{E}_{\hat{S}}\bigl[b^{\star}(\hat{S})\bigr]=D,\label{eq:expected-d-star-4}\\
 & \mathbb{E}\bigl[d(S,\check{S}^{\star})|\hat{S}=\hat{s}\bigr]=b^{\star}\left(\hat{s}\right).\label{eq:d-star-4}
\end{align}

Furthermore, since $R_{S|\hat{S}=\hat{s}}(\cdot)$ is convex, minimizing
\eqref{eq:-65} is equivalent to minimizing its Lagrangian function,
i.e.,
\begin{align}
 & \min_{b\left(\hat{s}\right)}\left\{ \mathbb{E}_{\hat{S}}R_{S|\hat{S}=\hat{s}}(b\left(\hat{s}\right))+\lambda\mathbb{E}_{\hat{S}}\bigl[b(\hat{S})\bigr]\right\} ,\label{eq:-113}
\end{align}
where $\lambda\geq0$ is a Lagrangian multiplier. By calculus of variations,
the solution to \eqref{eq:-113} or \eqref{eq:-65} satisfies the
necessary condition
\begin{equation}
f\left(\hat{s}\right)R_{S|\hat{S}=\hat{s}}^{\prime}(b\left(\hat{s}\right))+\lambda f\left(\hat{s}\right)=0.
\end{equation}

\noindent That is
\begin{equation}
\lambda=-R_{S|\hat{S}=\hat{s}}^{\prime}(b\left(\hat{s}\right)).\label{eq:lambdastar-1-1}
\end{equation}
On the other hand, $b^{\star}\left(\hat{s}\right)$ is assumed to
be the solution to \eqref{eq:-65}, hence
\begin{equation}
\lambda=-R_{S|\hat{S}=\hat{s}}^{\prime}(b^{\star}\left(\hat{s}\right)).\label{eq:lambdastar-1-1-1-2}
\end{equation}

\section{\label{sec:Proof-of-LemmaB-Gaussian}Proof of Lemma \ref{lem:B-Gaussian}}

Before proving Lemma \ref{lem:B-Gaussian}, we need introduce the
discretization of Euclidean space, and corresponding properties. Let
\begin{equation}
[Z]\in\mathcal{N}\triangleq\left\{ \cdots,-2\Delta,-\Delta,0,\Delta,2\Delta,\cdots\right\} \label{eq:-68}
\end{equation}
be a quantized version of $Z$, obtained by mapping $Z$ to the closest
quantization point, i.e.,
\begin{equation}
[z_{i}]=\Delta\cdot\mathrm{Round}\left(\frac{z_{i}}{\Delta}\right).
\end{equation}
Then for any $z^{n}\in\mathbb{R}^{n}$,
\begin{equation}
0\leq\frac{\left\Vert z^{n}-[z]^{n}\right\Vert ^{2}}{n}\leq\frac{\Delta^{2}}{4}.
\end{equation}
Furthermore, using triangle inequality we have
\begin{align}
-\frac{\left\Vert z^{n}-[z]^{n}\right\Vert ^{2}}{n} & \leq\frac{\left\Vert z^{n}\right\Vert ^{2}}{n}-\frac{\left\Vert [z]^{n}\right\Vert ^{2}}{n}\leq\frac{\left\Vert z^{n}-[z]^{n}\right\Vert ^{2}}{n}.
\end{align}
Hence
\begin{align}
-\frac{\Delta^{2}}{4} & \leq\frac{\left\Vert z^{n}\right\Vert ^{2}}{n}-\frac{\left\Vert [z]^{n}\right\Vert ^{2}}{n}\leq\frac{\Delta^{2}}{4},
\end{align}
i.e.,
\begin{equation}
\frac{\left\Vert [z]^{n}\right\Vert ^{2}}{nN_{Z}}-\epsilon_{\Delta}\leq\frac{\left\Vert z^{n}\right\Vert ^{2}}{nN_{Z}}\leq\frac{\left\Vert [z]^{n}\right\Vert ^{2}}{nN_{Z}}+\epsilon_{\Delta}\label{eq:-53}
\end{equation}
for some $\epsilon_{\Delta}$ that vanishes as $\Delta\rightarrow0$.
Similarly, it holds that
\begin{equation}
\frac{\left\Vert [z]^{n}-x^{n}\right\Vert ^{2}}{nN_{E}}-\epsilon_{\Delta}\leq\frac{\left\Vert z^{n}-x^{n}\right\Vert ^{2}}{nN_{E}}\leq\frac{\left\Vert [z]^{n}-x^{n}\right\Vert ^{2}}{nN_{E}}+\epsilon_{\Delta}.\label{eq:-55}
\end{equation}
Therefore, combining \eqref{eq:-53} and \eqref{eq:-55}, and according
to definition of weak typicality, we have if $\left(x^{n},[z]^{n}\right)\in\mathcal{T}_{\delta-\epsilon_{\Delta}}^{n}$,
then $\left(x^{n},z^{n}\right)\in\mathcal{T}_{\delta}^{n}$; and in
turn if $\left(x^{n},z^{n}\right)\in\mathcal{T}_{\delta}^{n}$, then
$\left(x^{n},[z]^{n}\right)\in\mathcal{T}_{\delta+\epsilon_{\Delta}}^{n}$,
where $\mathcal{T}_{\delta}^{n}$, $\mathcal{T}_{\delta-\epsilon_{\Delta}}^{n}$
and $\mathcal{T}_{\delta+\epsilon_{\Delta}}^{n}$ correspond to jointly
typical sets of $\left(X,Z\right)$. This implies
\begin{align}
1\left\{ \left(x^{n},[z]^{n}\right)\in\mathcal{T}_{\delta-\epsilon_{\Delta}}^{n}\right\}  & \leq1\left\{ \left(x^{n},z^{n}\right)\in\mathcal{T}_{\delta}^{n}\right\} \nonumber \\
 & \leq1\left\{ \left(x^{n},[z]^{n}\right)\in\mathcal{T}_{\delta+\epsilon_{\Delta}}^{n}\right\} .\label{eq:-56}
\end{align}

Now we start to prove Lemma \ref{lem:B-Gaussian}. From \eqref{eq:-56},
we have \eqref{eq:-52} (given at the top of next page), where $\mathcal{T}_{\delta'+\epsilon_{\Delta}}^{n}$
corresponds to a typical set of $Z$, and $\mathcal{F}^{n}\triangleq\mathcal{N}^{n}\cap\mathcal{T}_{\delta'+\epsilon_{\Delta}}^{n}$.
\begin{figure*}[!t]
\setcounter{mytempeqncnt}{\value{equation}} \setcounter{equation}{329}
\begin{align}
\mathbb{P}_{{\mathcal{C}}}\left(\mathcal{B}^{c}\right) & \leq\mathbb{P}_{{\mathcal{C}}}\Bigl[\mathop{\min}\limits _{m_{c}\in\left[2^{mR_{c}}\right],[z]^{n}\in\mathcal{T}_{\delta'+\epsilon_{\Delta}}^{n}}\sum_{m_{1}^{\prime}}1\left\{ \left(X^{n}\left(m_{c},m_{1}^{\prime}\right),[z]^{n}\right)\in\mathcal{T}_{\delta-\epsilon_{\Delta}}^{n}\right\} <1,\nonumber \\
 & \qquad\textrm{or }\mathop{\max}\limits _{m_{c}\in\left[2^{mR_{c}}\right],[z]^{n}\in\mathcal{T}_{\delta'+\epsilon_{\Delta}}^{n}}\sum_{m_{1}^{\prime}}1\left\{ \left(X^{n}\left(m_{c},m_{1}^{\prime}\right),[z]^{n}\right)\in\mathcal{T}_{\delta+\epsilon_{\Delta}}^{n}\right\} >2^{2m\epsilon}\Bigr]\nonumber \\
 & =\mathbb{P}_{{\mathcal{C}}}\Bigl[\mathop{\min}\limits _{m_{c}\in\left[2^{mR_{c}}\right],[z]^{n}\in\mathcal{F}^{n}}\sum_{m_{1}^{\prime}}1\left\{ \left(X^{n}\left(m_{c},m_{1}^{\prime}\right),[z]^{n}\right)\in\mathcal{T}_{\delta-\epsilon_{\Delta}}^{n}\right\} <1\Bigr]\nonumber \\
 & \qquad+\mathbb{P}_{{\mathcal{C}}}\Bigl[\mathop{\max}\limits _{m_{c}\in\left[2^{mR_{c}}\right],[z]^{n}\in\mathcal{F}^{n}}\sum_{m_{1}^{\prime}}1\left\{ \left(X^{n}\left(m_{c},m_{1}^{\prime}\right),[z]^{n}\right)\in\mathcal{T}_{\delta+\epsilon_{\Delta}}^{n}\right\} >2^{2m\epsilon}\Bigr],\label{eq:-52}
\end{align}
\setcounter{mytempeqncnt}{\value{equation}} \setcounter{equation}{\value{mytempeqncnt}}
\hrulefill{}
\end{figure*}

Using union bound we have
\begin{align}
 & \mathbb{P}_{{\mathcal{C}}}\Bigl[\mathop{\max}\limits _{m_{c}\in\left[2^{mR_{c}}\right],[z]^{n}\in\mathcal{F}^{n}}\nonumber \\
 & \qquad\qquad\sum_{m_{1}^{\prime}}1\left\{ \left(X^{n}\left(m_{c},m_{1}^{\prime}\right),[z]^{n}\right)\in\mathcal{T}_{\delta+\epsilon_{\Delta}}^{n}\right\} >2^{2m\epsilon}\Bigr]\nonumber \\
 & \le2^{mR_{c}}\left|\mathcal{F}^{n}\right|\mathop{\max}\limits _{m_{c}\in\left[2^{mR_{c}}\right],[z]^{n}\in\mathcal{F}^{n}}\nonumber \\
 & \qquad\mathbb{P}_{{\mathcal{C}}}\Bigl(\sum_{m_{1}^{\prime}}1\left\{ \left(X^{n}\left(m_{c},m_{1}^{\prime}\right),[z]^{n}\right)\in\mathcal{T}_{\delta+\epsilon_{\Delta}}^{n}\right\} >2^{2m\epsilon}\Bigr).\label{eq:UnionBound-1}
\end{align}

Furthermore, $\left|\mathcal{F}^{n}\right|$ is upper-bounded by $2^{n\left(\frac{1}{2}\log n+o\left(\log n\right)\right)}$
as shown in \eqref{eq:-59},
\begin{figure*}[!t]
\setcounter{mytempeqncnt}{\value{equation}} \setcounter{equation}{331}
\begin{align}
\left|\mathcal{F}^{n}\right| & \leq\frac{\textrm{Volume of \ensuremath{n-}ball with radius }\sqrt{nN_{Z}\left(1+\delta'+\epsilon_{\Delta}\right)}+\sqrt{n\Delta^{2}}}{\Delta^{n}}\\
 & =\frac{\pi^{n/2}\left(\sqrt{nN_{Z}\left(1+\delta'+\epsilon_{\Delta}\right)}+\sqrt{n\Delta^{2}}\right)^{n}}{\Delta^{n}\Gamma\left(\frac{n}{2}+1\right)}\\
 & \leq\frac{\pi^{n/2}\left(\sqrt{nN_{Z}\left(1+\delta'+\epsilon_{\Delta}\right)}+\sqrt{n\Delta^{2}}\right)^{n}}{\Delta^{n}}\\
 & =2^{n\left(\frac{1}{2}\log\pi-\log\Delta+\log\left(\sqrt{nN_{Z}\left(1+\delta'+\epsilon_{\Delta}\right)}+\sqrt{n\Delta^{2}}\right)\right)}\\
 & \leq2^{n\left(\frac{1}{2}\log n+o\left(\log n\right)\right)}.\label{eq:-59}
\end{align}
\setcounter{mytempeqncnt}{\value{equation}} \setcounter{equation}{\value{mytempeqncnt}}
\hrulefill{}
\end{figure*}
and it is easy to verify that \eqref{eq:thetaMean}-\eqref{eq:-28}
still hold for the Gaussian case. Hence the probability in \eqref{eq:UnionBound-1}
decays doubly exponentially fast with $n$, which further means \eqref{eq:UnionBound-1}
vanishes as $m\to\infty$.

Similarly, we can also prove
\begin{align}
 & \mathbb{P}_{{\mathcal{C}}}\Bigl[\mathop{\min}\limits _{m_{c}\in\left[2^{mR_{c}}\right],[z]^{n}\in\mathcal{F}^{n}}\nonumber \\
 & \qquad\qquad\sum_{m_{1}^{\prime}}1\left\{ \left(X^{n}\left(m_{c},m_{1}^{\prime}\right),[z]^{n}\right)\in\mathcal{T}_{\delta-\epsilon_{\Delta}}^{n}\right\} <1\Bigr]\rightarrow0,
\end{align}
as $m\to\infty$. Hence Lemma \ref{lem:B-Gaussian} holds.

\section{\label{sec:Proof-of-LemmaA123-Gaussian}Proof of Lemma \ref{lem:A123-Gaussian}}

Similar to \eqref{eq:Qsshat}, for continuous random variables the
following holds.
\begin{equation}
\Ebb_{\mathcal{C}}\left[Q_{S^{m}\hat{S}^{m}}\left(s^{m},\hat{s}^{m}\right)\right]=\prod_{i=1}^{m}f_{S\hat{S}}(s_{i},\hat{s}_{i}).\label{eq:-77}
\end{equation}
Hence $(S^{m},\hat{S}^{m})$ is i.i.d. Since (weakly) typical set
has total probability close to one \cite{Cover}, we have
\begin{equation}
\lim_{m\to\infty}\mathbb{P}\left(S^{m}\in\mathcal{T}_{\delta}^{m}\right)=1.\label{eq:-81}
\end{equation}
By the law of large numbers, we also have for any $\delta>0$,
\begin{equation}
\lim_{m\to\infty}\mathbb{P}\Bigl(\frac{1}{m}\sum_{i=1}^{m}\jmath_{S}(S_{i},D_{E})>\mathbb{E}\jmath_{S}(S,D_{E})-\delta\Bigr)=1,\label{eq:-31}
\end{equation}
\begin{equation}
\lim_{m\to\infty}\mathbb{P}\Bigl(\frac{1}{m}\sum_{i=1}^{m}b^{\star}(\hat{S}_{i})\geq\mathbb{E}_{\hat{S}}\left[b^{\star}(\hat{S})\right]-\delta\Bigr)=1,\label{eq:-41}
\end{equation}
and
\begin{align}
 & \lim_{m\to\infty}\mathbb{P}\Bigl(\frac{1}{m}\sum_{i=1}^{m}\jmath_{S|\hat{S}=\hat{S}_{i}}(S_{i},b^{\star}(\hat{S}_{i}))>\nonumber \\
 & \qquad\qquad\qquad\mathbb{E}_{\hat{S}}\mathbb{E}_{S}[\jmath_{S|\hat{S}=\hat{s}}(S,b^{\star}\left(\hat{s}\right))]-\delta\Bigr)=1.\label{eq:-49}
\end{align}

Combining \eqref{eq:-31} with the property \eqref{eq:Ejd}, we have
\begin{equation}
\lim_{m\to\infty}\mathbb{P}\Bigl(\frac{1}{m}\sum_{i=1}^{m}\jmath_{S}(S_{i},D_{E})>R_{S}(D_{E})-\delta\Bigr)=1.\label{eq:-83}
\end{equation}

Combining \eqref{eq:-41} with \eqref{eq:expected-d-star} gives us
\begin{equation}
\lim_{m\to\infty}\mathbb{P}\Bigl(\frac{1}{m}\sum_{i=1}^{m}b^{\star}(\hat{S}_{i})\geq D_{E}-\delta\Bigr)=1.\label{eq:-89}
\end{equation}

Furthermore,
\begin{align}
\mathbb{E}_{\hat{S}}\mathbb{E}_{S}[\jmath_{S|\hat{S}=\hat{s}}(S,b^{\star}\left(\hat{s}\right))] & =\mathbb{E}_{\hat{S}}R_{S|\hat{S}=\hat{s}}(b^{\star}\left(\hat{s}\right))\label{eq:-121}\\
 & =R_{S|\hat{S}}(D_{E}),\label{eq:-8-1}
\end{align}
where \eqref{eq:-121} follows from \eqref{eq:Ejd-1}, and \eqref{eq:-8-1}
follows from \eqref{eq:ERs_shat}.

Combining \eqref{eq:-49} with \eqref{eq:-8-1} gives
\begin{equation}
\lim_{m\to\infty}\mathbb{P}\Bigl(\frac{1}{m}\sum_{i=1}^{m}\jmath_{S|\hat{S}=\hat{S}_{i}}(S_{i},b^{\star}(\hat{S}_{i})))>R_{S|\hat{S}}(D_{E})-\delta\Bigr)=1.\label{eq:-108}
\end{equation}

\eqref{eq:-81}, \eqref{eq:-83}, \eqref{eq:-89} and \eqref{eq:-108}
imply the lemma holds.

\section{\label{sec:Proof-of-Lemma-SPO-si}Proof of Lemma \ref{lem:SPO-si}}

Lemma \ref{lem:SPO-si} is proven in \eqref{eq:-73}-\eqref{eq:-101},
where $\lambda\geq0$ is given in \eqref{eq:lambdastar-1-1-1} (or
\eqref{eq:-7-2}), and \eqref{eq:-101} follows from \eqref{eq:-7-2}
and the property \eqref{eq:csiszar-1}.
\begin{figure*}[!t]
\setcounter{mytempeqncnt}{\value{equation}} \setcounter{equation}{347}
\begin{align}
 & \Pbb\Bigl[d(S^{m},\check{s}^{m})\leq D_{E},\frac{1}{m}\sum_{i=1}^{m}\jmath_{S|\hat{S}=\hat{s}_{i}}(S_{i},b^{\star}\left(\hat{s}_{i}\right))\geq R_{S|\hat{S}}(D_{E})-\delta,\frac{1}{m}\sum_{i=1}^{m}b^{\star}\left(\hat{s}_{i}\right)\geq D_{E}-\delta|\hat{S}^{m}=\hat{s}^{m}\Bigr]\nonumber \\
 & =\sum_{s^{m}:d(s^{m},\check{s}^{m})\leq D_{E},\frac{1}{m}\sum_{i=1}^{m}\jmath_{S|\hat{S}=\hat{s}_{i}}(s_{i},b^{\star}\left(\hat{s}_{i}\right))\geq R_{S|\hat{S}}(D_{E})-\delta}P(s^{m}|\hat{s}^{m})1\Bigl\{\frac{1}{m}\sum_{i=1}^{m}b^{\star}\left(\hat{s}_{i}\right)\geq D_{E}-\delta\Bigr\}\label{eq:-73}\\
 & \leq\sum_{s^{m}:d(s^{m},\check{s}^{m})\leq D_{E},\frac{1}{m}\sum_{i=1}^{m}\jmath_{S|\hat{S}=\hat{s}_{i}}(s_{i},b^{\star}\left(\hat{s}_{i}\right))\geq R_{S|\hat{S}}(D_{E})-\delta}P(s^{m}|\hat{s}^{m})2^{m\lambda\left(D_{E}-d(s^{m},\check{s}^{m})\right)}\nonumber \\
 & \qquad\qquad\qquad\qquad\qquad\qquad\qquad\times2^{\sum_{i=1}^{m}\jmath_{S|\hat{S}=\hat{s}_{i}}(s_{i},b^{\star}\left(\hat{s}_{i}\right))-m\left(R_{S|\hat{S}}(D_{E})-\delta\right)}1\Bigl\{\frac{1}{m}\sum_{i=1}^{m}b^{\star}\left(\hat{s}_{i}\right)\geq D_{E}-\delta\Bigr\}\\
 & \leq2^{-m\left(R_{S|\hat{S}}(D_{E})-\delta\right)}\mathbb{E}\left[2^{m\lambda\left(D_{E}-d(S^{m},\check{s}^{m})\right)+\sum_{i=1}^{m}\jmath_{S|\hat{S}=\hat{s}_{i}}(S_{i},b^{\star}\left(\hat{s}_{i}\right))}|\hat{S}^{m}=\hat{s}^{m}\right]1\Bigl\{\frac{1}{m}\sum_{i=1}^{m}b^{\star}\left(\hat{s}_{i}\right)\geq D_{E}-\delta\Bigr\}\\
 & \leq2^{-m\left(R_{S|\hat{S}}(D_{E})-\delta-\lambda\delta\right)}\mathbb{E}\left[2^{\lambda\left(\sum_{i=1}^{m}b^{\star}\left(\hat{s}_{i}\right)-\sum_{i=1}^{m}d\left(S_{i},\hat{s}_{i}\right))\right)+\sum_{i=1}^{m}\jmath_{S|\hat{S}=\hat{s}_{i}}(S_{i},b^{\star}\left(\hat{s}_{i}\right))}|\hat{S}^{m}=\hat{s}^{m}\right]\\
 & \leq2^{-m\left(R_{S|\hat{S}}(D_{E})-\delta-\lambda\delta\right)},\label{eq:-101}
\end{align}
\setcounter{mytempeqncnt}{\value{equation}} \setcounter{equation}{\value{mytempeqncnt}}
\hrulefill{}
\end{figure*}

\begin{IEEEbiographynophoto}{Lei Yu} received the B.E. and Ph.D. degrees, both in electronic engineering, from University of Science and Technology of China (USTC) in 2010 and 2015, respectively. Currently, he is a postdoctoral researcher at the Department of Electronic Engineering and Information Science (EEIS), USTC. His research interests include information theory, probability theory, and security.\end{IEEEbiographynophoto}
\begin{IEEEbiographynophoto}{Houqiang Li} (S'12) received the B.S., M.Eng., and Ph.D. degrees in electronic engineering from the University of Science and Technology of China, Hefei, China, in 1992, 1997, and 2000, respectively, where he is currently a Professor with the Department of Electronic Engineering and Information Science.
His research interests include video coding and communication, multimedia search, image/video analysis. He served as an Associate Editor of the IEEE TRANSACTIONS ON CIRCUITS AND SYSTEMS FOR VIDEO TECHNOLOGY from 2010 to 2013, and has been with the Editorial Board of the Journal of Multimedia since 2009. He was a recipient of the Best Paper Award for Visual Communications and Image Processing Conference in 2012. \end{IEEEbiographynophoto}
\begin{IEEEbiographynophoto}{Weiping Li} (F'00) received his B.S. degree from University of Science and Technology of China (USTC) in 1982, and his M.S. and Ph.D. degrees from Stanford University in 1983 and 1988 respectively, all in electrical engineering. He was an Assistant Professor, Associate Professor with Tenure, and Professor of Lehigh University from 1987 to 2001. He worked in several high-tech companies in the Silicon Valley with technical and management responsibilities from 1998 to 2010. He has been a Professor in USTC since 2010. He served as the Editor-in-Chief of IEEE Transactions on Circuits and Systems for Video Technology, a founding member of the Board of Directors of MPEG-4 Industry Forum, and several other positions in IEEE and SPIE. He is an IEEE Fellow. \end{IEEEbiographynophoto}


\begin{thebibliography}{10}
\bibitem{Yu-2}L. Yu, H. Li, and W. Li, ``Source-channel secrecy
for Shannon cipher system,'' in \emph{54th Allerton Conference},
Monticello, IL, USA, Sep. 2016.

\bibitem{Schieler} C. Schieler, and P. Cuff, ``The henchman problem:
Measuring secrecy by the minimum distortion in a list,'' \textit{IEEE
Trans. Inf. Theory}, vol. 62, no. 6, pp. 3436\textendash 3450, Jun.
2016.

\bibitem{Shannon49} C. Shannon, ``Communication theory of secrecy
systems,'' \textit{Bell Syst. Tech. J.}, vol. 28, no. 4, pp. 656\textendash 715,
Oct. 1949.

\bibitem{Merhav}N. Merhav and E. Arikan, \textquotedblleft The Shannon
cipher system with a guessing wiretapper,\textquotedblright{} \emph{IEEE
Trans. Inf. Theory,} vol. 45, no. 6, pp. 1860\textendash 1866, Sep.
1999.

\bibitem{Vembu} S. Vembu, S. Verd{ú}, and Y. Steinberg, \textquotedblleft The
source-channel separation theorem revisited,\textquotedblright{} \emph{IEEE
Trans. Inf. Theory,} vol. 41, pp. 44\textendash 54, Jan. 1995.

\bibitem{Verdu} S. Verd{ú} and T. S. Han, \textquotedblleft A general
formula for channel capacity,\textquotedblright{} \emph{IEEE Trans.
Inf. Theory,} vol. 40, pp. 1147\textendash 1157, July 1994.

\bibitem{Chen}P.-N. Chen and F. Alajaji, ``Optimistic Shannon coding
theorems for arbitrary single-user systems,'' \textit{IEEE Trans.
Inf. Theory}, vol. 45, no. 7, pp. 2623-2629, 1999.

\bibitem{Wyner} A. Wyner, ``The wire-tap channel,'' \textit{Bell
Syst. Tech. J.}, vol. 54, no. 8, pp. 1334\textendash 1387, 1975.

\bibitem{Csiszar} I. Csiszár, and J. Körner, ``Broadcast channels
with confidential messages,'' \textit{IEEE Trans. Inf. Theory}, vol.
24, no. 3, pp. 339\textendash 348, May 1978.

\bibitem{Yamamoto} H. Yamamoto, ``Rate-distortion theory for the
Shannon cipher system,'' \textit{IEEE Trans. Inf. Theory}, vol. 43,
no. 3, pp. 827\textendash 835, May 1997.

\bibitem{Han}T. S. Han, Information-Spectrum Methods in Information
Theory, Springer Berlin Heidelberg, Feb 2003.

\bibitem{Wilson}M. P. Wilson, and K. Narayanan, ``Transmitting an
analog Gaussian source over a Gaussian wiretap channel under SNR mismatch,\textquotedblright{}
in \emph{Proc. IEEE Int. Conf. on Telecommunications}, pp.44-47, April
2010.

\bibitem{Bagherikaram} G. Bagherikaram and K. Plataniotis, ``Secure
hybrid digital-analog Wyner-Ziv coding,\textquotedblright{} in \emph{Proc.
of the IEEE 22nd Int. Symp. on Personal Indoor and Mobile Radio Communications},
pp. 1161-1166, Sep. 2011.

\bibitem{Bagherikaram13}G. Bagherikaram and K. Plataniotis, ``Secure
joint source-channel coding with interference known at the transmitter,\textquotedblright{}
\emph{IET Communications}, vol. 6, no. 17, pp. 2796 - 2808, Jan. 2013.

\bibitem{Villard} J. Villard, P. Piantanida, and S. Shamai, ``Secure
transmission of sources over noisy channels with side information
at the receivers,\textquotedblright{} \textit{IEEE Trans. Inf. Theory},
vol. 60, pp. 713\textendash 739, Jan 2014.

\bibitem{Song15} E. Song, P. Cuff, and V. Poor, ``Joint source-channel
secrecy using hybrid coding,\textquotedblright{} in \emph{IEEE Int.
Symp. on Inf. Theory}, Jun. 2015.

\bibitem{Gamal} A. El Gamal and Y.-H. Kim, Network Information Theory.
Cambridge University Press, 2011.

\bibitem{Tan} V Y. F. Tan, and M. R. Bloch, ``Information spectrum
approach to strong converse theorems for degraded wiretap channels,''
\emph{IEEE Trans. on Inf. Forensics and Security}, vol. 10, no. 9,
pp. 1891-1904, Sep. 2015.

\bibitem{Dijk}M. van Dijk, ``On a special class of broadcast channels
with confidential messages,'' \textit{IEEE Trans. Inf. Theory}, vol.
43, no. 2, pp. 712-714, 1997.

\bibitem{Yu}L. Yu, H. Li, and W. Li, ``Joint source-channel secrecy
using uncoded schemes: Towards secure source broadcast,'' Jan. 2017.
{[}Online{]}. Available: https://arxiv.org/abs/1607.07040.

\bibitem{Cover}T. M. Cover and J. A. Thomas, \textit{Elements of
Information Theory}, Wiley, New York, 1991.

\bibitem{Rogers}C. A. Rogers, ``Covering a sphere with spheres,\textquotedblright{}
\emph{Mathematika}, vol. 10, no. 02, pp. 157\textendash 164, 1963.

\bibitem{Verger-Gaugry}J. L. Verger-Gaugry, ``Covering a ball with
smaller equal balls in $\mathbb{R}^{n}$,\textquotedblright{} \emph{Discr.
Comput. Geom.}, vol. 33, no. 1, pp. 143-155, 2005.

\bibitem{Mitzenmacher}M. Mitzenmacher and E. Upfal. Probability and
Computing: Randomized Algorithms and Probabilistic Analysis. Cambridge
University Press, 2005.

\bibitem{Elkayam}N. Elkayam and M. Feder, ``Information spectrum
approach to the source channel separation theorem,\textquotedblright{}
in \emph{IEEE Int. Symp. on Inf. Theory (ISIT)}, Jun. 2014, pp. 1697-1701.

\bibitem{Ash} J. M. Ash, ``Neither a worst convergent series nor
a best divergent series exists,'' \emph{College Mathematics Journal},
vol. 28, no. 4, pp. 296\textendash 297, 1997.

\bibitem{Kostina} V. Kostina and S. Verd{ú}, ``Lossy joint source-channel
coding in the finite blocklength regime,'' \textit{IEEE Trans. Inf.
Theory}, vol. 59, no. 5, pp. 2545\textendash 2575, 2013.

\bibitem{Csiszar74} I. Csiszár, ``On an extremum problem of information
theory,\textquotedblright{} \emph{Studia Scientiarum Mathematicarum
Hungarica}, vol. 9, no. 1, pp. 57-71, 1974.
\end{thebibliography}
\end{document}